\documentclass[12pt]{article}
\usepackage{graphicx}
\usepackage{epstopdf, epsfig}
\pdfoutput=1

\usepackage[a4paper, margin=2.5cm]{geometry}
\usepackage{amsmath} 
\usepackage{multirow} 
\usepackage{amssymb}
\usepackage[round]{natbib}

\usepackage{placeins} 

\usepackage[mathscr]{euscript}
 \let\mathscr\relax
\usepackage[scr]{rsfso}

\usepackage{enumitem}

\usepackage{cleveref}
\crefformat{section}{\S#2#1#3}

\title{Geometry and organization of coherent structures in stably stratified atmospheric boundary layers}

\author{\small Abhishek Harikrishnan$^{1}$, Cedrick Ansorge$^{2}$, Rupert Klein$^{1}$, Nikki Vercauteren$^{3}$}

\date{\small 
$^{1}$Institute of Mathematics, Free University of Berlin\\
$^{2}$Institute of Geophysics and Meteorology, University of Cologne\\
$^{3}$Department of Geosciences, University of Oslo}

\begin{document}


\maketitle


\begin{abstract}

\noindent Global intermittency is observed in the stably stratified Atmospheric Boundary Layer (ABL) and corresponds to having large nonturbulent flow regions to develop in an otherwise turbulent flow. In this paper, the differences between continuous and intermittent turbulence are quantified with the help of coherent structures. Eight classes of coherent structures are identified from literature, most of which are indicated by scalar criteria derived from velocity fields. A method is developed to geometrically classify structures into three categories: blob-like, tube-like or sheet-like. An alternate definition of the intermittency factor $\gamma$ based on coherent structures is introduced to separate turbulent and nonturbulent parts of a flow. Applying this conditioning technique and the geometrical characterization on direct numerical simulations (DNS) of an Ekman flow, we find the following: (i) structures with similar geometries (either tube-like or sheet-like) are found regardless of the strength of stratification; (ii) global intermittency affects all regions of the ABL - viscous sublayer, buffer layer, inner, and outer layer; (iii) for the highly stratified case, sweep/ejection pairs form well-separated clusters within the viscous sublayer which can possibly explain the abundance of hairpin-like vortices with a particular orientation; (iv) nonturbulent regions are occupied with streamwise velocity fluctuations and there is a switch between high- and low-speed streaks at a particular height for all stratified cases. 


\end{abstract}




\newpage

\section{Introduction}
\label{sec:introduction}

The Atmospheric (or Planetary) Boundary Layer (ABL) is the lowest part of the troposphere (up to 2 km), i.e., the region directly responding to changes in surface conditions. It shares many similarities with the turbulent boundary layer in a simulated channel flow but differs in two key aspects: 
First, the structure of this layer is influenced by the presence of clouds and strong surface heating/cooling. Second, the outer layer (also known as the Ekman layer) is affected by the Coriolis effect due to the rotation of the Earth. 
Energy exchange at the surface leads to density variations penetrating throughout the ABL. 
Strong heating, encountered for example in daytime conditions, leads to unstable stratification whereas strong cooling, encountered for example in night-time conditions, yields a stably stratified ABL \citep{garratt1994atmospheric}. The former is characterized by continuous turbulence whereas, in the latter, turbulence may be intermittent or even absent \citep{van2012minimum}. The organization of turbulence in intermittent patches is denoted by \citet{mahrt1989intermittency} as \textit{global intermittency}. Our goal is to study the differences between continuous and intermittent turbulence by comparing the geometry of the coherent structures. 

The demarcation of the flow into turbulent and nonturbulent flow regions was first investigated by \citet{townnsend1948local} in the wake of a circular cylinder. Away from the center of the wake, he noted that regions of fully developed turbulent flow and almost laminar flow were separated by a sharp boundary. This phenomenon was quantified with an \textit{intermittency factor}, denoted by $\gamma$, and it was measured with the kurtosis of velocity derivatives. Similar probe-based measurements were employed by \citet{townsend1949fully, corrsin1955free, fiedler1966intermittency, kovasznay1970large, rao1971bursting, falco1977coherent}. 
Since these techniques have limited spatial resolution, the three-dimensional geometry of the turbulent/nonturbulent patches and the corresponding dynamics are not adequately captured. \citet{da2014characteristics} used a technique based on vorticity magnitude to segregate the turbulent (rotational) and nonturbulent (irrotational) regions on the direct numerical simulations (DNS) of boundary layers, jets and shear-free turbulence. However, this approach inherits the well-known drawbacks of vorticity magnitude: its reliance on a subjective threshold and producing false-positives in shear dominated regions \citep{lugt1979dilemma}. \citet{ansorge2016analyses} addressed the latter problem by applying two horizontal high-pass filters (for streamwise and spanwise directions) on velocity fields close to the wall. 
They found that the high-pass filtered velocity fields were able to correctly detect the absence of turbulence for most of the inner layer. The latter shortcoming of vorticity magnitude can also be overcome by considering another vortex criterion like $Q$ \citep{hunt1988eddies}, which identifies a purely spinning region by subtracting the strain-rate tensor from the vorticity tensor.  




The adoption of such vortex-based partition techniques forms a useful connection to the notion of coherent structures. Numerous types of coherent structures such as streaks, sweeps, ejections, vortical structures have been identified and studied in wall-bounded flows by \citet{robinson1991kinematics}, hereafter referred to as Robinson structures. Previous investigations into coherent structures such as sweeps and ejections attribute these entities to momentum transfer in the boundary layer \citep{katul1997ejection, li2011coherent}. In their study,  \citet{li2011coherent} report a dissimilarity in the transport efficiency of scalars and momentum between the neutral and unstable ABL. They speculate that this is most likely caused by a change in topology of coherent structures and stress the necessity of studies considering their three-dimensional extent. Plenty of field measurements and laboratory studies exist which confirm the presence of hairpin vortices in the neutral ABL \citep{hommema2003packet, carper2004role, huang2009analysis, inagaki2010organized}. However, their role in stratified ABLs is not well known. All of this taken together presents a need to study the role of Robinson structures in the neutrally and stably stratified ABL.






The geometry of coherent structures has also attracted considerable interest. \citet{moisy2004geometry}, hereafter MJ2004, applied the classical technique of box counting to 3D scalar fields of vorticity and strain-rate magnitude. They defined individual structures as a connected set of points satisfying a thresholding condition and studied their geometry by identifying three characteristic lengths for each structure. This enabled them to relate the structures as being close to a sphere, tube, ribbon or a flat sheet. They concluded that the strain-rate structures change from being flat sheets to ribbons with increase in threshold. Vorticity strutcures, on the other hand, shift from ribbons to long tubes with increase in threshold. \citet{bermejo2008non}, hereafter B2008, developed a non-local methodology to study the geometry of structures by applying a curvelet transform on a passive scalar from the DNS of forced, isotropic turbulence. They utilized differential geometry properties such as \textit{Shape Index and Curvedness} to characterize the structures and concluded that blob-like and tube-like structures prevail in the inertial range and sheet-like structures in the dissipation range. In an extension of the study, \citet{bermejo2009geometry} applied the same technique to study the geometry of structures at different resolutions of the numerical simulation and concluded that the coarsest resolution was unable to detect sheet-like structures at small scales in 3D scalar fields of enstrophy and kinetic energy dissipation-rate. Similar characizerization was also carried out for lagrangian structures \citep{yang2010multi, yang2011geometric} and with the adoption of Minkowski functionals \citep{leung2012geometry}. In our current work, we aim to study the geometry of Robinson structures in neutral and stably stratified ABLs by developing a methodology derived from the work of MJ2004 and B2008. This will enable us to establish a detailed comparison on the geometry, distribution and organization of coherent structures. To this end, we pose the following questions,

\noindent (a) How can we extract and study coherent structures in the ABL?

\noindent (b) How different are the properties of coherent structures at increasing levels of stratification?

\noindent (c) How does global intermittency affect the self-organization of structures within the flow in the boundary layer?

%
%
%
%


The paper is organized as follows: We start by discussing the details of our DNS database in section \ref{sec:sim}. Section \ref{sec:coherent_structures} details our methodology to extract and geometrically characterize coherent structures. We start with a brief review of all coherent structures and the indicators used to identify them. Following this, we present a modified neighbor scanning (NS) algorithm to extract structures from 3D scalar fields. We discuss the technique of percolation analysis to objectively choose an appropriate threshold for the entire boundary layer. Finally, we discuss the methodology derived from MJ2004 and B2008 for geometrical characterization of the structures. In section \ref{sec:results}, we detail the results for the different layers of the ABL, namely the viscous, buffer, inner, and outer layers. A significant number of figures produced to support our remarks are presented in a supplementary file. These figures are denoted with C\#, where \# denotes a section number. A special study is also carried out on the geometry of hairpin-like structures. In section \ref{sec:discussion}, we investigate the dataset from the viewpoint of conditional one-point statistics and compare the results to section \ref{sec:results}. All observations are summarized in section \ref{sec:conclusion}.

\newpage


\section{Overview of the dataset}
\label{sec:sim}

\begin{table}
	\begin{center}
		\begin{tabular}{lcccccccc}
			\hline
			Case & Line specification & $Ri_{B}$ & $Fr$ & $Re$ & $L_{x}$& $L_{y}$ & $L_{x}^{+}$ & $L_{y}^{+}$ \\[3pt]
			\hline
			N & -----------  & 0 & $\infty$ & \multirow{4}{*}{1000} & \multirow{4}{*}{20.4$\delta$} & \multirow{4}{*}{20.4$\delta$} & \multirow{4}{*}{27\,040} & \multirow{4}{*}{27\,040}\\
			S\_1 & ............  & 2.64 & 0.02 \\
			S\_2 & -\,-\,-\,-\,-\,-\,-  & 0.76 & 0.07  \\
			S\_3 & --\,---\,--\,---  & 0.58 & 0.09 \\
			\hline
		\end{tabular}
		\caption{Parameters of the numerical simulations used in the work. The cases with the prefix $S$ are stratified and $N$ is the neutrally stratified case. Reynolds number can be defined with $\delta$ and the laminar boundary layer depth. The value of the latter case is shown here. If we define $Re$ with the former, then the value is $26\,450$. Further simulation details can be found in \cite{ansorge2016analyses}.}
		\label{tab:sim_parameters}
	\end{center}
\end{table}

\begin{figure}
	\centerline{\includegraphics[width=0.7\linewidth]{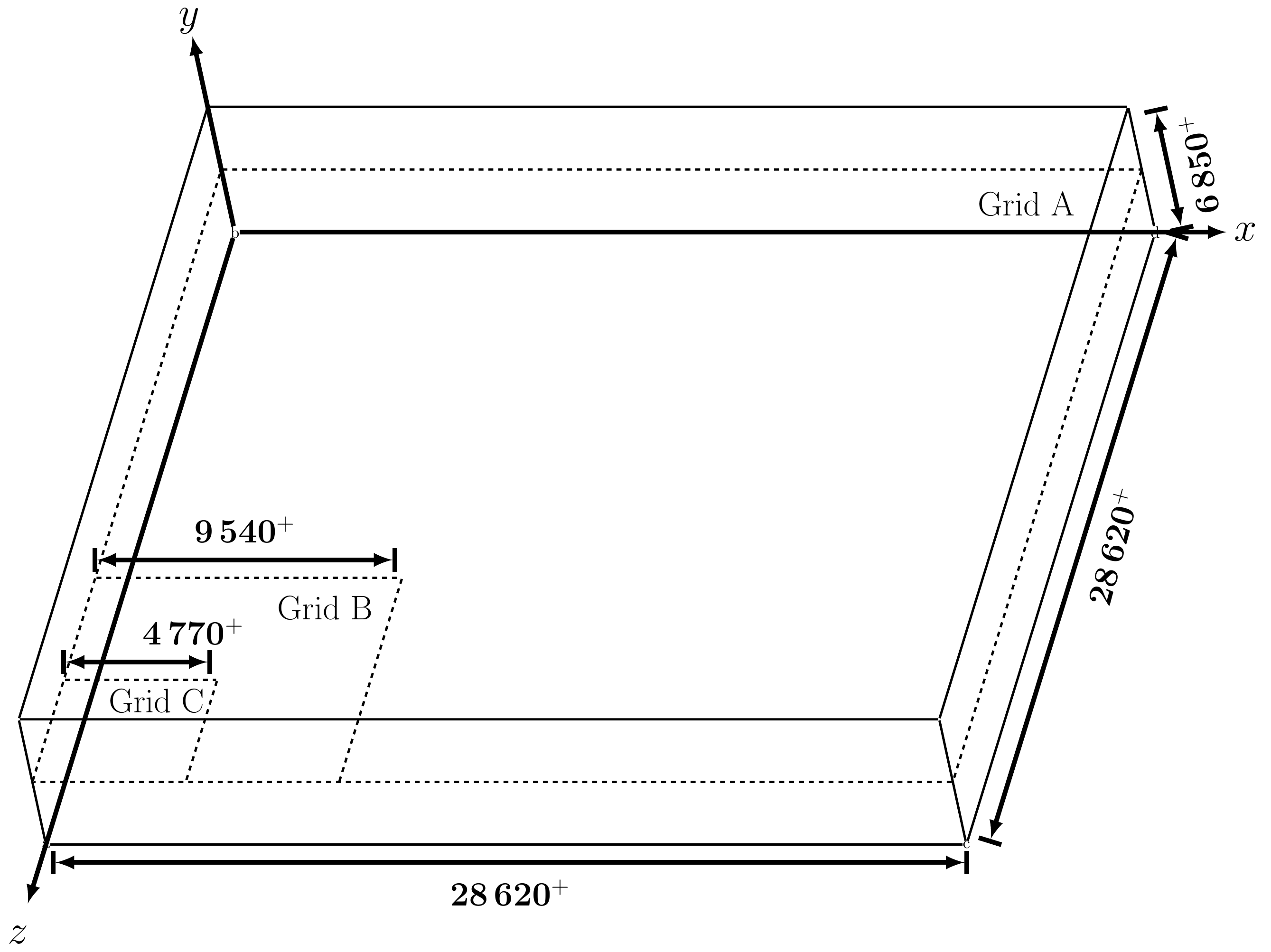}}
	\caption{The schematic shows the three types of subgrids - A, B and C used in the paper. $x, y, z$ correspond to the streamwise, wall-normal and spanwise directions respectively. $(\cdot)^+$ indicates viscous units as described in equation \ref{eq:wallUnits}.}
	\label{fig:computationalDomain}
\end{figure}


We study the ABL with a simplified physical configuration, namely Ekman flow: the flow of a stratified incompressible viscous fluid over a smooth flat plate driven by a uniform pressure gradient which experiences steady rotation around the wall-normal axis. It can be ideally compared to the ABL in the limit of neutral stratification \citep{coleman1990numerical}. We inherit the simulation datasets from the work of \cite{ansorge2016analyses}. Their set-up uses the incompressible (divergence-free) Navier-Stokes equations with the addition of a Coriolis component. It was solved under the Boussinesq approximation, where density variations were neglected except in determining the buoyancy. They also impose the f-plane approximation where the Coriolis parameter acts in a horizontal plane along the vertical axis and is assumed to be constant. A no-slip boundary condition is applied at the wall and the domain is doubly periodic in the horizontal direction. Different levels of density stratification are considered and stratification is imposed via a Dirichlet (fixed-value) boundary condition at the top and bottom boundaries. Further details regarding the simulation can be found in \cite{ansorge2014global, ansorge2016analyses} and \cite{ansorgeThesis}.

\begin{figure}
	\centerline{\includegraphics[width=0.8\linewidth]{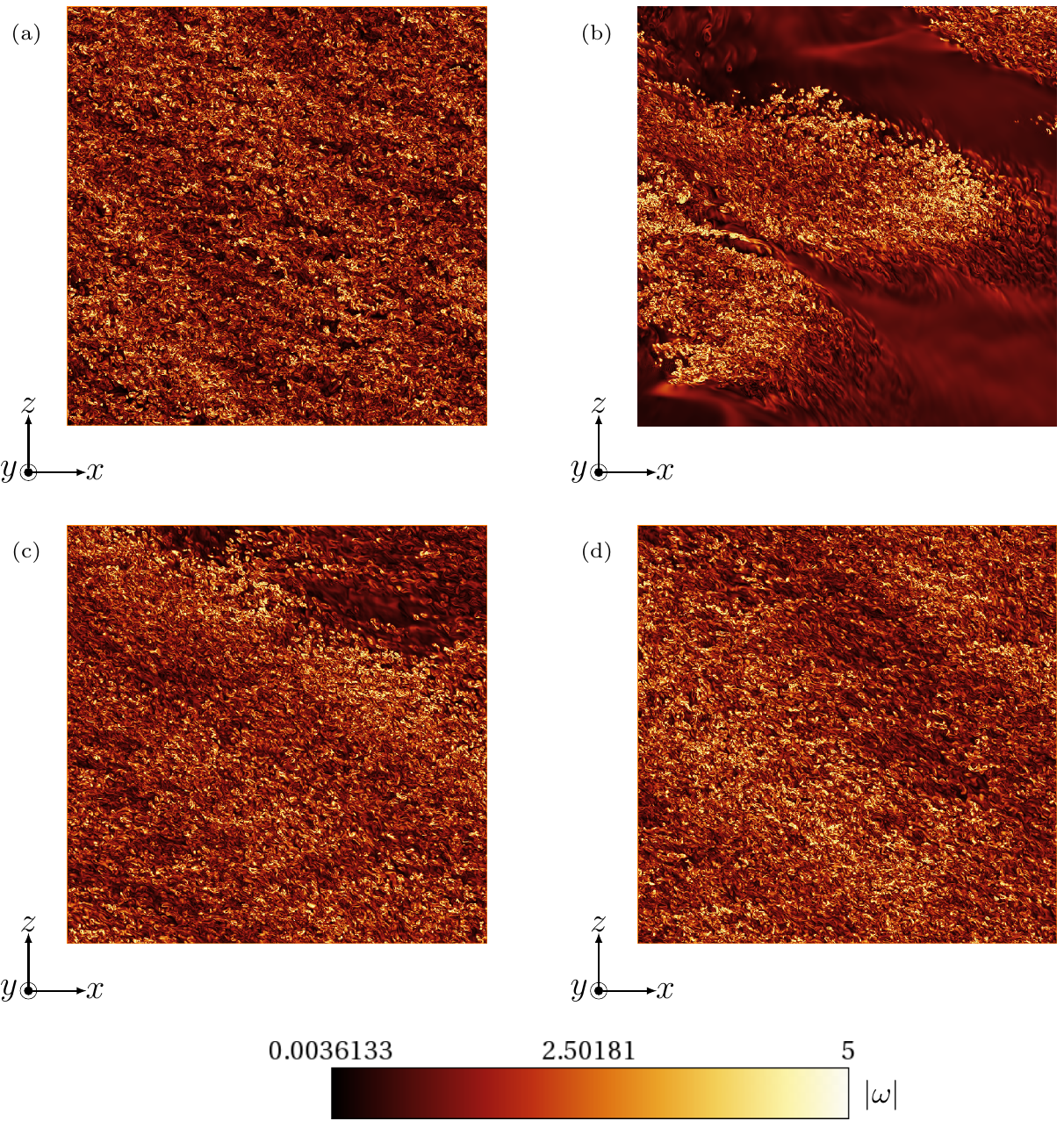}}
	\caption{Horizontal vorticity magnitude slices normalized with its RMS over that slice are shown here at $y^{+} \approx 100$. (a, b, c, d) correspond to cases N, S\_1, S\_2, S\_3 respectively.}
	\label{fig:vortMagSlice}
\end{figure}

Our database is comprised of three stratified cases (S\_1, S\_2, S\_3) along with a neutrally stratified case (N). All simulation parameters are summarized in table \ref{tab:sim_parameters}. The computational domain has $3072 \times 512 \times 6144$ points corresponding to the streamwise ($x$), wall-normal ($y$) and spanwise ($z$) directions respectively. For convenience, we define three horizontal grid sizes to communicate our results. Subgrid A represents the entire computational domain whereas subgrids B and grid C show $1/3$ and $1/6$ of the domain respectively as illustrated in figure \ref{fig:computationalDomain}.

A classification based on the time evolution of vertically integrated turbulence kinetic energy (TKE) suggests that the three cases S\_1, S\_2 and S\_3 are all very stable \citep{ansorge2014global}. In such a regime, turbulence dies out in some regions showing large regions of little to no activity. This is evident from figure \ref{fig:vortMagSlice}(b, c, d) where slices of vorticity magnitude normalized with its root mean square (RMS) over the horizontal slice is shown. 

%

For all cases, the Reynolds number ($Re$) is $26\,450$ in terms of the geostrophic wind velocity $G$, the boundary layer height under neutral conditions $\delta$, and viscosity $\nu$:

\begin{equation}
	Re \equiv G\delta / \nu
\end{equation}

\noindent where $\delta \equiv u_{\tau} / f$, $u_{\tau}$ is the friction velocity and $f$ is the Coriolis parameter. We define the strength of stratification with two dimensionless parameters namely the global bulk Richardson number ($Ri_{B}$) and Froude number ($Fr$). The former is defined as,


\begin{equation}
	Ri_{B} \equiv B_{0} \delta / G^{2}
\end{equation}

\noindent where $B_{0}$ quantifies the buoyancy difference between the top and bottom layer. The Froude number is defined as

\begin{equation}
	Fr = \frac{G^{2}}{B_{0}\Lambda}
\end{equation}

\noindent where $\Lambda$ is the Rossby deformation radius. $Ri_{B}$ is related to $Fr$ as $Ri_{B} = Fr^{-1}\frac{\delta}{\Lambda}$. 
Throughout the paper, we consider the vertical direction in terms of viscous units such that

\begin{equation}
	\label{eq:wallUnits}
	y^{+} \equiv y u_{\tau}/\nu
\end{equation}

\noindent where $u_{\tau}$ is the friction velocity.

\newpage


\section{Characterization of individual structures}
\label{sec:coherent_structures}

\subsection{Identification}
\label{subsec:identificationCS}


\FloatBarrier
\begin{table}
	\begin{center}
		\begin{tabular}{l l c l c}
			\hline
			Type & Coherent structure & Indicator & Region & Color specification \\
			\hline
			&&&&\\
			\textit{Quantitative}
			& High-speed streaks & $u' > 0$ & $y^{+} < 40$ & \includegraphics{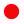}\\
			& Low-speed streaks & $u' < 0$ & $y^{+} < 40$ & \includegraphics{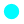}\\
			& Sweeps & $u'>0, v'<0$ & $y^{+} < 1550$ & \includegraphics{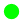}\\
			& Ejections & $u'<0, v'>0$ & $y^{+} < 1550$ & \includegraphics{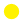}\\
			& Vortices & $\frac{1}{2} (||\Omega||^{2}-||S||^{2}) > 0$ & $y^{+} < 1550$ & \includegraphics{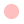}\\
			& Shear layers & $|\omega|$ & $y^{+} < 80$ & \includegraphics{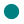}\\
			& Backs & $|\omega|$ & $y^{+} < 1550$ & \includegraphics{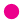}\\
			&&&&\\
			\textit{Qualitative}
			& Pockets & streamlines & $y^{+} < 5$ & - \\
			& Bulges & $|\omega|$ & $y^{+} < 1550$ & - \\
			\hline
		\end{tabular}
		\caption{Types of coherent structures considered in the study. The regions indicate where the dataset has been cut in the $y$-direction. In the case of sweeps, ejections, vortices, backs and bulges where the structures are found throughout the boundary layer, the region is limited to a $y^{+} < 1550$ as it already includes a good portion of the outer layer. The maximum $y^{+}$ value is $6849.54$.}
		\label{tab:structure_classification}
	\end{center}
\end{table}

Even though a precise mathematical formulation for a ``coherent structure'' has not yet surfaced, we adopt the idea from \citet{pope2001turbulent} and understand that it is a  connected region in space which persists in time. We refer to these connected regions as a ``structure'' or ``entity'' throughout the paper, and we will use the taxonomy of boundary layer structures introduced by \citet{robinson1991kinematics}, to study them in an organized manner. 
We divide the Robinson structures into two major categories: (a) quantitative, where structures can be extracted and geometrically characterized and (b) qualitative, where they can only be observed. In this section, we present a brief overview on all Robinson structures, their purpose, which region of the boundary layer they are prominently observed in and how they can be extracted/visualized.

The seminal work of \citet{kline1967structure} has experimentally revealed \textit{streaks} close to the wall of turbulent boundary layers. These regions of relatively slow-moving fluid can be clearly seen until $y^{+} \approx 40$ in flat plate boundary layers \citep{kline1978role}. 
Streaks are extremely helpful in understanding the nature of flow within the viscous sublayer ($y^{+} < 5$). Within the context of ABL, however, these structures are usually studied with Large Eddy Simulations (LES) above the viscous sublayer as the first grid point in LES is relatively far away from the wall \citep{khanna1998three, drobinski2003origin, jayaraman2021transition}. Computing the streamwise velocity fluctuation $u'$, both low-speed ($u' < 0$) and high-speed ($u' > 0$) streaks can be readily observed and used to quantify the effect of global intermittency within this region.


\citet{kim1971production} noted that streaks not only move downstream but also away from the wall. This movement, which is initially slow, extends rapidly outward after a critical height. This process is described as `streak lifting' or \textit{ejection}. Conversely, \citet{corino1969visual} identified a stream of fluid moving towards the wall called \textit{sweep}. The quadrant technique \citep{wallace1972wall} is well suited to detect these entities. It splits the product of streamwise and wall-normal velocity fluctuations into four categories: Q1 $(+u', +v')$, Q2 $(-u', +v')$, Q3 $(-u', -v')$ and Q4 $(+u', -v')$. Out of the four, Q2 and Q4 are used to identify ejection and sweep events respectively. They can be used to quantify momentum transport throughout the ABL \citep{li2011coherent, katul1997ejection, katul2006relative, narasimha2007turbulent}.


\textit{Vortices} were described by \citet{kuchemann1965report} as the `sinews and muscles' of fluid flows, and they have been studied extensively over the years. Despite their immense importance, there exists no universally accepted definition and a vast range of criteria are currently used. For a review on various criteria see \citet{cucitore1999effectiveness, gunther2018state}. 
They fall under two broad categories, one of which does point-wise characterization on the velocity gradient tensor ($\nabla v$) for every instant in time (Eulerian), and the other follows fluid particle trajectories (Lagrangian). While both have their merits, we choose the former technique as the latter would be computationally expensive for our dataset. In choosing an appropriate criterion, we follow the work of  \citet{chakraborty2005relationships}, who found that popular vortex criteria such as the $Q$ \citep{hunt1988eddies}, $\lambda_{2}$ \citep{jeong1995identification} and $\Delta$ criterion \citep{chong1990general} identify very similar regions as vortices. They also define an equivalent threshold which allows for the visualization of similar structures among the criteria at any non-zero threshold. Therefore, a choice among these three will not affect our qualitative results. We choose the $Q$-criterion which is defined as,

\begin{equation}\label{eq:qcrit}
	Q = \frac{1}{2} (||\Omega||^{2} - ||S||^{2}) > 0
\end{equation}

\noindent where $\Omega = \frac{1}{2} [\nabla v - (\nabla v)^{T}]$ is the vorticity tensor, $S = \frac{1}{2} [\nabla v + (\nabla v)^{T}]$ is the strain rate tensor. The $Q$-criterion identifies vortices as regions where vorticity dominates over the strain. They can be visualized only above the viscous sublayer and are thought to play a central role in turbulence production \citep{robinson1991kinematics}.

Closely related to the lift-up of low-speed streaks, the wall-normal \textit{shear layers} are known to exist in the buffer layer ($5 < y^{+} < 30$) and beyond until $y^{+} \approx 80$. They are characterized by high values of the instantaneous velocity gradient ($\partial u' /\partial y$) and are capable of retaining coherence for up to $1000^{+}$ in the streamwise direction \citep{johansson1987shear}. These structures may roll up into vortices \citep{robinson1991kinematics} and may be responsible for near-wall turbulence production \citep{alfredsson1988turbulence}. Since vorticity magnitude ($\omega$) is known to highlight shear regions, it can be used to study the near-wall shear layers and the shear layers on the order of $\delta$-scale termed as \textit{backs}.



\textit{Pockets} have been documented by \citet{falco1977coherent, falco1980production, kim1987turbulence}, but their contribution to turbulence production is not fully understood yet. These structures are understood as `footprints' of outer region motions in the inner layer and are thought to contribute to the generation of hairpin vortices \citep{chu1988vortex}. They are known to exist within the viscous sublayer and are observable with diverging streamlines \citep{robinson1991kinematics}.


Another coherent structure frequently referred to in literature are the \textit{bulges} which are very large scale motions (VLSMs) or superstructures existing between the free stream and the outer edge of the boundary layer. They show numerous narrow incursions, sometimes extending into the buffer layer \citep{corrsin1955free, kovasznay1970large}. They also attenuate small-scale fluctuations close to the wall \citep{marusic2010predictive} are also documented in ABLs \citep{shah2014very, katul2019anatomy} and can instantly be observed with vorticity magnitude \citep{robinson1991kinematics}.

A summary of the Robinson structures, the chosen indicators and the region of identification is presented in table \ref{tab:structure_classification}. Pockets and bulges are qualitatively understood whereas the others can be extracted and geometrically characterized with the technique presented in the following sections.


\subsection{Extraction}
\label{subsec:extraction}

\begin{figure}
	\minipage{\textwidth}
	\centerline{\includegraphics[width=\linewidth]{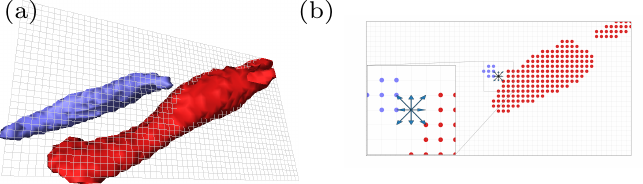}}
	\endminipage\hfill
	\caption{Drawback of the NS approach is illustrated here. An extracted structure is shown in (a). If we look at a plane as shown in (b), we can instantly observe where the issue occurs. Blue and red scatter points belong to two different structures. When the NS algorithm considers the center point indicated by the tail end of arrows as the center of a $3 \times 3$ cube, it assumes that points belonging to both structures are neighbors.}
	\label{fig:close_structures}
\end{figure}

Following MJ2004, we define a structure as a connected set of points for which the indicator is greater than a given threshold. Therefore, a point $x$ belongs to a coherent structure if, 

\begin{equation}\label{eq:definition}
	\alpha(x) > \tau_{p}
\end{equation}

\noindent where $\alpha$ is the indicator for extracting coherent regions and $\tau_{p}$ is an appropriate threshold. The process of selecting $\tau_{p}$ objectively is described in section \ref{subsec:choiceThreshold}. The algorithm used to extract the structures employs the neighbor scanning (NS) approach of MJ2004 where all points in the domain are scanned successively, once $\tau_{p}$ is identified. Every point is considered to be the center of a $3 \times 3$ cube and the neighbors are any of the 26 points (faces, edges and vertices of the cube) that surround it. Although this approach yields good results at high threshold values where the structures are spaced apart, we notice that some structures tend to be grouped together at smaller threshold values. An example is shown in figure \ref{fig:close_structures} where the NS algorithm incorporates all neighbors close to the center point even though it can be clearly seen upon visualization that some neighbors belong to two different structures. This issue arises due to the fact that the NS algorithm doesn't know how the surface mesh is being generated by the visualization algorithm. Therefore, we introduce an extension to the NS approach by correcting the identified neighbors with the marching cubes algorithm \citep{lorensen1987marching} so that it is accurate for visualization purposes. The details of the NS+MC approach is presented in Appendix \ref{appendix:NS+MC}. 

\subsection{On the choice of a threshold}
\label{subsec:choiceThreshold}

As noted by \citet{green2007detection}, the crux when dealing with any Eulerian criterion for structure extraction is its reliance on a user-defined threshold. If the chosen value is too low, a complex, interconnected, sponge-like structure can be seen. On the other hand, a large value can result in too few structures being identified. MJ2004 consider this problem to be analogous to that of a percolation transition, where the complex, interconnected structure only appears above a critical threshold. This threshold, henceforth known as the percolation threshold $(\tau_{p})$, presents a natural, non-subjective way of choosing a global threshold value. \citet{del2006self} have pointed out that $\tau_{p}$ remains unchanged with increasing Reynolds number in channel-flow turbulence.

\begin{figure}
	\minipage{0.33\textwidth}
	\centerline{\includegraphics[width=1\linewidth]{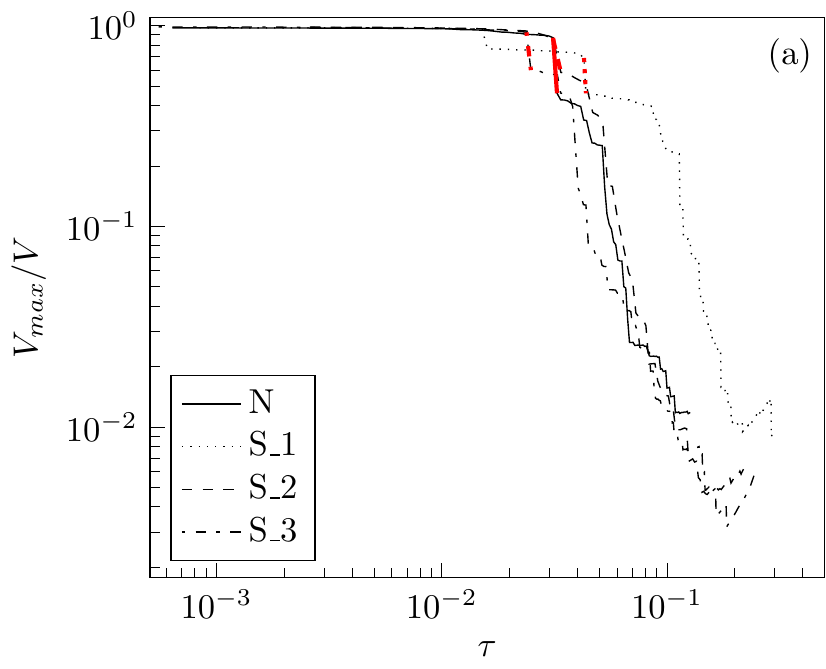}}
	\endminipage\hfill
	\minipage{0.33\textwidth}
	\centerline{\includegraphics[width=1\linewidth]{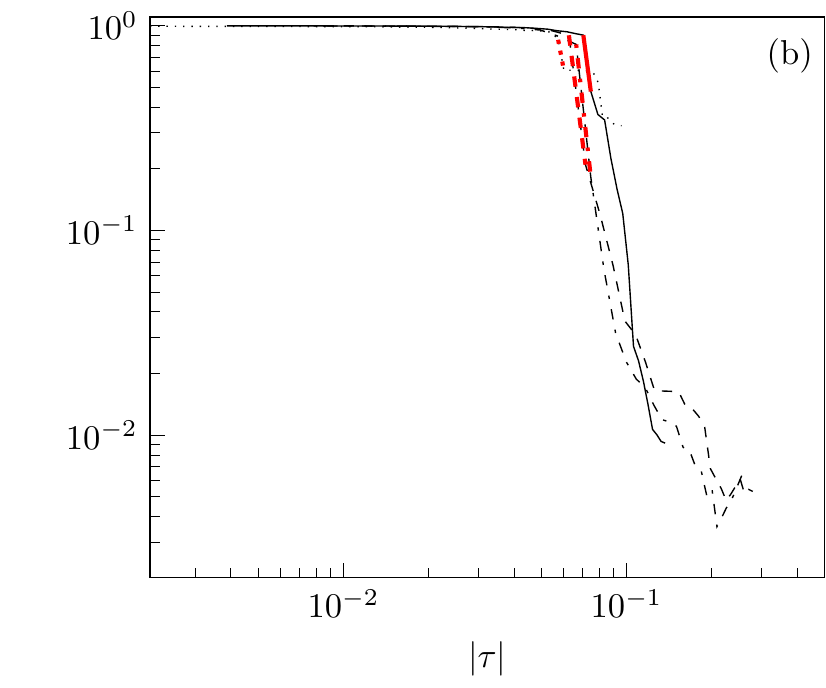}}
	\endminipage\hfill
	\minipage{0.33\textwidth}
	\centerline{\includegraphics[width=1\linewidth]{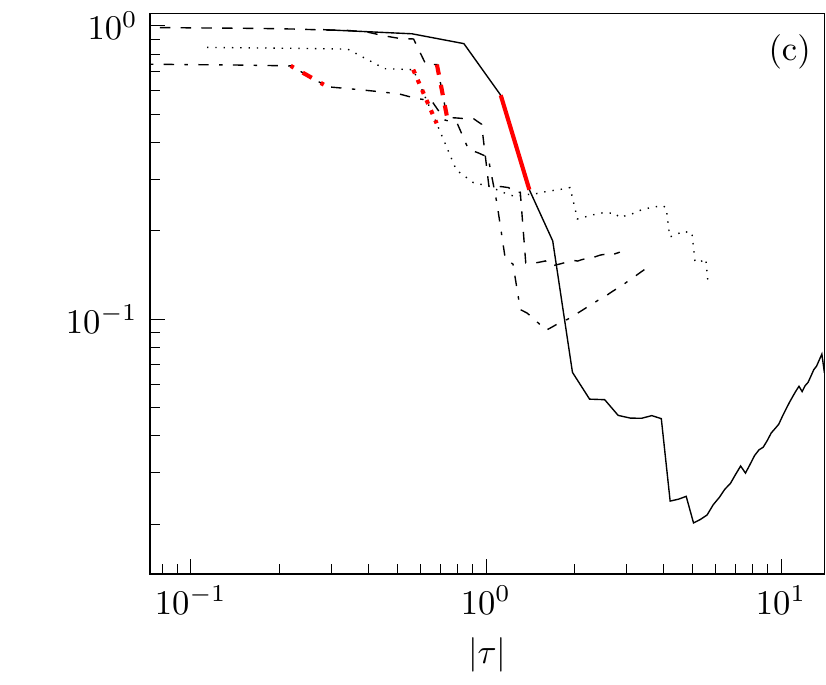}}
	\endminipage\hfill
	\minipage{0.33\textwidth}
	\centerline{\includegraphics[width=1\linewidth]{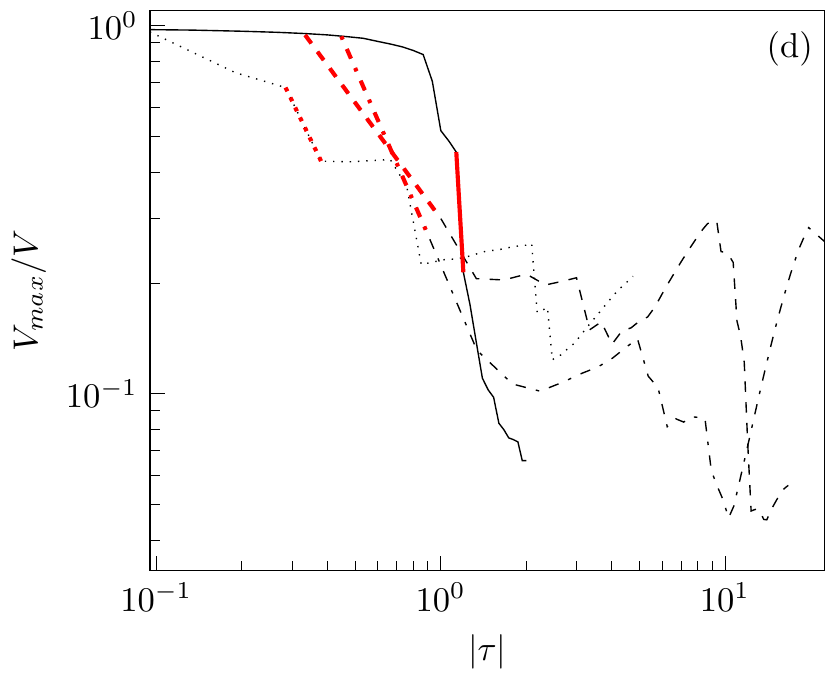}}
	\endminipage\hfill
	\minipage{0.33\textwidth}
	\centerline{\includegraphics[width=1\linewidth]{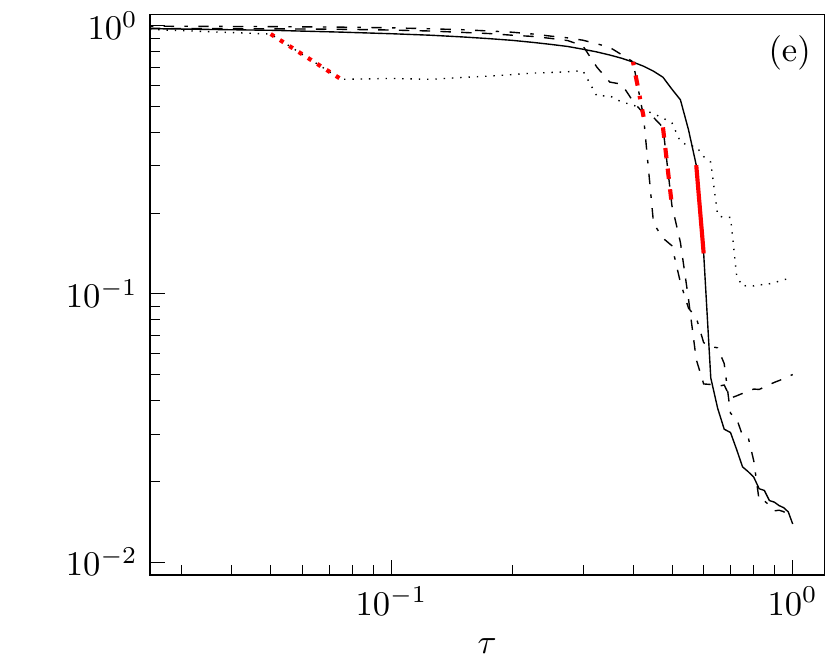}}
	\endminipage\hfill
	\minipage{0.33\textwidth}
	\centerline{\includegraphics[width=1\linewidth]{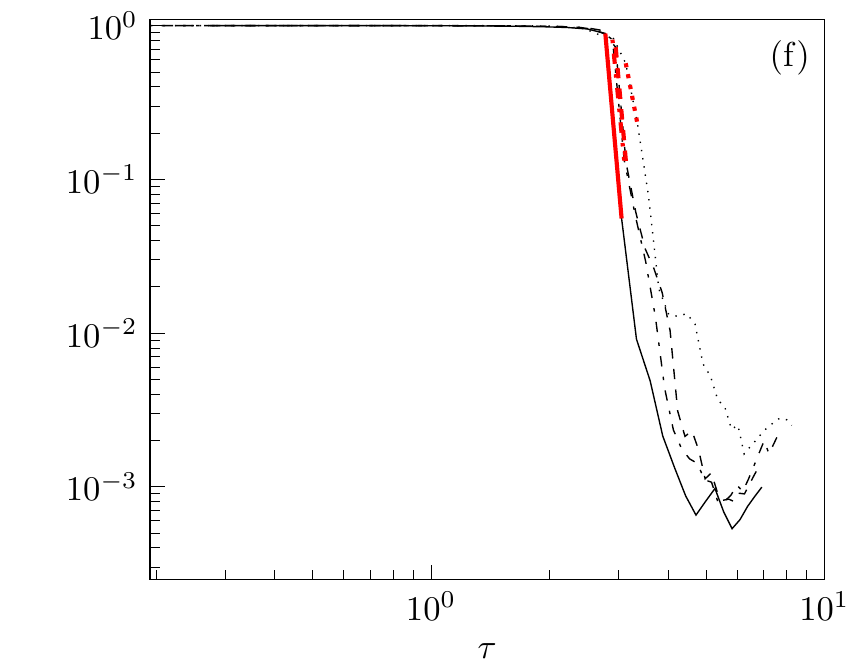}}
	\endminipage\hfill
	\minipage{\textwidth}
	\centerline{\includegraphics[width=.33\linewidth]{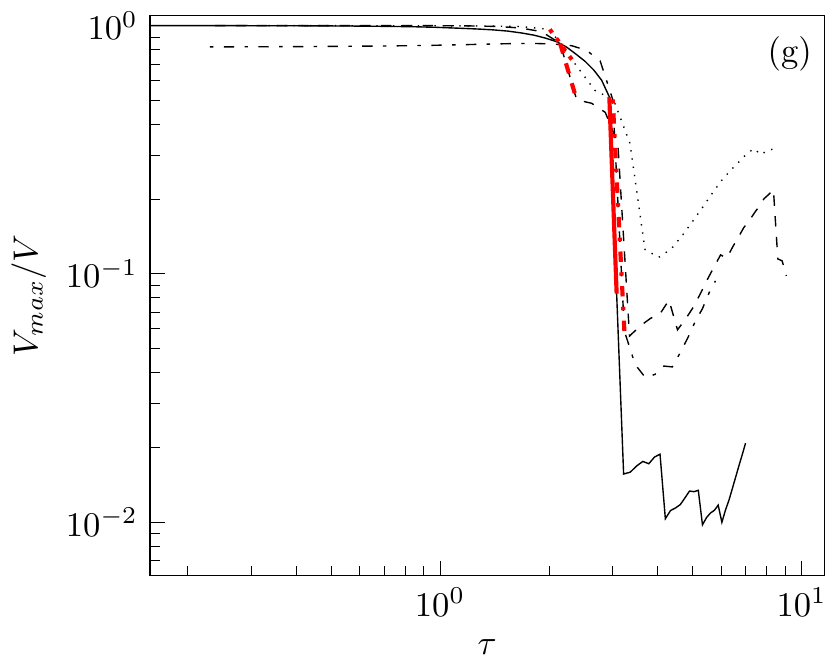}}
	\endminipage\hfill
	\caption{Percolation analysis on seven quantitative indicators as shown in table \ref{tab:structure_classification}. From left to right each row are the (a) high-speed streaks, (b) low-speed streaks, (c) sweeps, (d) ejections, (e) Q-criterion, (f) vorticity magnitude restricted to $y^{+} \approx 80$ and (g) vorticity magnitude until $y^{+} \approx 1550$. $V_{max}$ refers to the volume of the largest structure in the domain and $V$ is the volume of all structures at a particular threshold $\tau$. $|\cdot|$ is the modulus of the threshold value and is used for low-speed streaks, sweeps and ejections (b-d) where all thresholds are negative. A legend corresponding to our different ABL cases in shown in (a). The thick red line highlights the region of maximum slope.}
	\label{fig:percolation_streaks}
\end{figure}

Figure \ref{fig:percolation_streaks} (a-g) shows the computation of the ratio $V_{max}/V$ for increasing values of $\tau$ for all indicators in table \ref{tab:structure_classification}, where $V_{max}$ is the volume of largest structure at that threshold and $V$ is the volume of all structures. When this ratio is $1$, i.e., $V_{max} = V$, we see a large, interconnected structure. As $\tau$ is varied, the magnitude of the slope of the ratio $V_{max}/V$ increases and at $\tau_{p}$, it reaches a maximum. At this point, we see more disconnected structures. 
Due to the high cost of this computation, especially at low values of $\tau$, we had to ensure that this transition was seen with as few values of $\tau$ as possible. The entire range of $\tau$ is split with a chosen number (usually $1000$) of evenly spaced values. According to our experience, a smooth plot can be generated with the first 50 values of $\tau$. If the transition doesn't lie within this region, the order of magnitude of the chosen number can be changed.

For wall-bounded flows, inhomogeneity in the wall-normal direction must be taken into account when choosing a global threshold \citep{del2006self}. For instance, a threshold chosen to specifically highlight structures close to the wall may show an incomprehensible amount of structures away from it and vice versa. This was clearly elucidated by \citet{nagaosa2003statistical} who found that by nondimensionalizing the $Q$-criterion with its RMS for every wall-normal plane, the structures appeared to be more uniformly spread throughout the channel. Therefore, we can rewrite equation \ref{eq:definition} as,

\begin{equation}\label{eq:definition_final}
	\alpha (x) > \tau_{p} \; \overline{\alpha^{'2} (y)}^{1/2}
\end{equation}

\noindent where $\overline{\alpha^{'2} (y)}^{1/2}$ is the RMS of the indicator over wall-normal planes. It is important to note that this normalization is necessary only when dealing with a large domain. If we focus on a specific region as in the case of streaks where we confined ourselves to the region $y^{+} < 40$, then the normalization is no longer useful as first plane and last wall-normal planes are not too far from each other and a single threshold value will highlight structures uniformly. 


This strategy works well at neutral and intermediate levels of stratification i.e., cases N, S\_2 and S\_3. However, for the highly stratified case, we notice that structures tend to aggregate into clusters at non-zero thresholds (see figure \ref{fig:MLPNecessity} (a)). If $\tau_{p}$ is still used to extract structures, then an entire cluster would be identified as a single structure. To circumvent this, we developed a technique where percolation analysis can be applied iteratively to break down a complex cluster into simpler components. This enables every structure to be identified at a unique value of $\tau$. In other words, every structure has its own threshold value. The technique is hence a multilevel percolation threshold approach and is presented in Appendix \ref{appendix:MLP}. 


\subsection{Geometrical characterization}
\label{subsec:geomCharacterization}

Once $\tau_{p}$ is identified, the structures are extracted at this threshold with the NS+MC algorithm described in section \ref{subsec:extraction} and we now proceed to geometrically characterize them with an approach similar to B2008. The main steps along with our modifications are presented below:

\begin{figure}
	\minipage{\textwidth}
	\centerline{\includegraphics[width=1\linewidth]{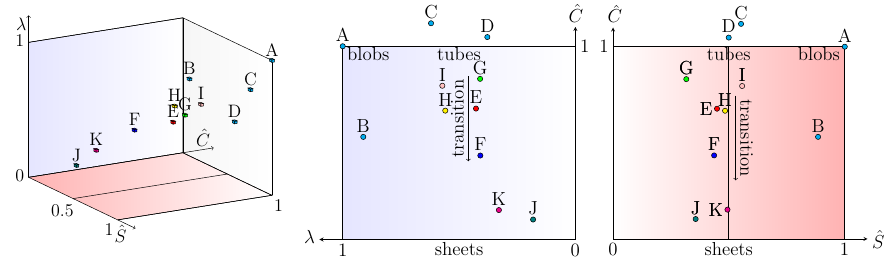}}
	\endminipage\hfill
	\minipage{\textwidth}
	\centerline{\includegraphics[width=\linewidth]{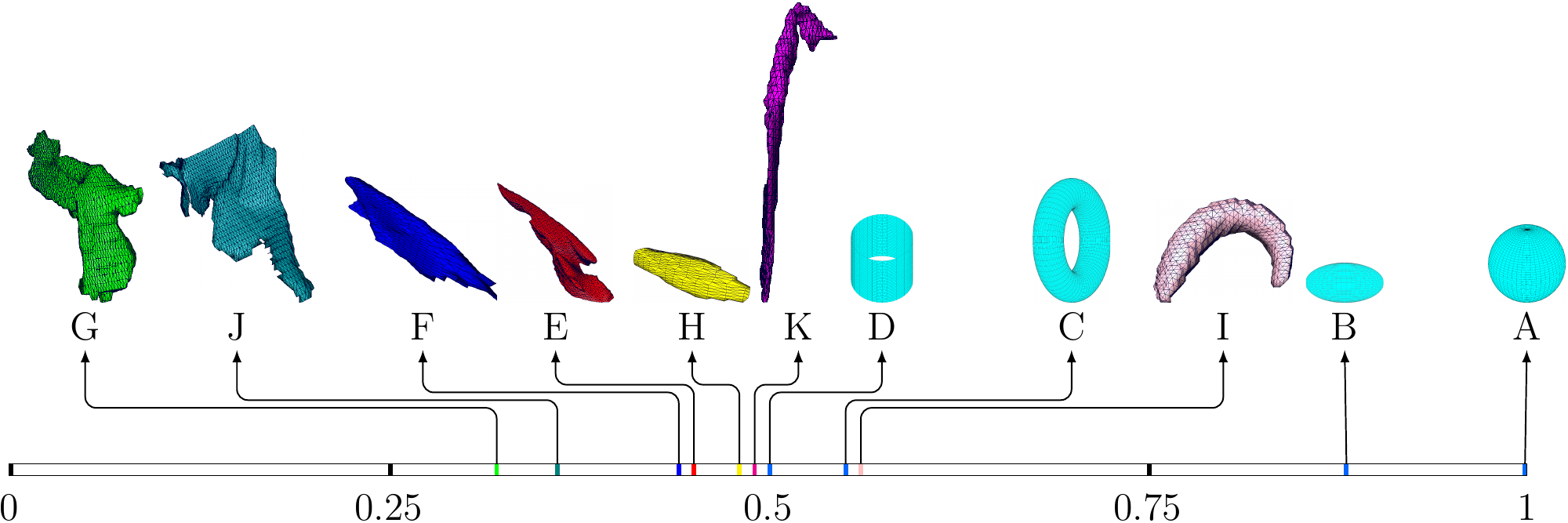}}
	\endminipage\hfill
	\caption{Panel (a) shows the visualization space with a three-dimensional projection and a set of two-dimensional orthogonal projections composed of $\lambda$ and $\hat{C}$ and $\hat{S}$ and $\hat{C}$. Figure adapted from \citet{bermejo2008non}. Panel (b) shows $\hat{S}$ for some commonly observed structures (A - D) and Robinson structures (E - K) extracted from case N. A and B are blob-like structures, C, D, E, G, H and I are tube-like structures, F, J and K are sheet-like structures. Color coding for the Robinson structures is according to table \ref{tab:structure_classification}. }
	\label{fig:shapeIndex}
\end{figure}

\vspace{0.2cm}

\begin{itemize}
	\item[$(a)$] \textit{Fractal dimension: }The first step is to filter structures which are sparser than lines i.e., structures having a mean fractal dimension $\langle D_{\alpha} \rangle < 1$. This is important for two reasons. First, some structures extracted with the NS+MC approach are noise-like consisting of one or a few connected points and cannot be meaningfully interpreted. Second, the computation of curvedness and stretching (described in point b) for geometrical characterization is dependent on the area and volume of a structure and therefore requires that we consider only closed surfaces. Structures having $1 \leq \langle D_{\alpha} \rangle \leq 2$ are excluded because they tend to have complicated geometries and closed, stretched, tube-like surfaces which include a large number of grid points in one dimension and very few in the other two get classified under this category (see C1 for an example where a significant amount of structures are shown to fall under this category). The fractal dimension of a structure is computed with the box-counting approach described in MJ2004 - the extracted structure is placed in the domain which embeds it. Then, we use a fixed grid scan approach where a box of size $r$ is moved across the domain without overlapping and the number of boxes $N_{\alpha}(r; \tau_{p})$ containing a value is counted. This is done for numerous values of $r$ which are chosen as $2^{-n}L$ where $n$ is varied from 0 to 9 and $L$ corresponds to the length of the largest dimension. It should be noted that values of $r$ smaller than the size of a single grid cell are ignored. The fractal dimension is defined as follows,
	
	
	\begin{equation}
		D_{\alpha}(r) = -\frac{\text{d ln } N_{\alpha}(r)}{\text{d ln } r}
		\label{eq: meanFractalDimension}
	\end{equation}
	
	\item[$(b)$] \textit{Shape index, Curvedness and Stretching: }The iso-surfaces of the structures at $\tau_{p}$ which have $\big\langle D_{\alpha} (r)\big\rangle > 1$ are extracted and we proceed to geometrically characterize them with three parameters. The first two, shape index $(S)$ and curvedness $(C)$ are differential-geometry properties which represent the local shape and the intensity of curvature \citep[][]{koenderink1992surface}. In terms of the principal curvatures $\kappa_{1}$ and $\kappa_{2}$ (see \citet{do2016differential} for a description of principal curvatures), they can be written as,
	
	\begin{equation}
		S = \bigg| -\frac{2}{\pi} arctan \bigg(\frac{\kappa_{1} + \kappa_{2}}{\kappa_{1} - \kappa_{2}}\bigg) \bigg|, \quad C = \mu \sqrt{\frac{\kappa_{1}^{2} + \kappa_{2}^{2}}{2}}
	\end{equation}
	
	\noindent where $|.|$ indicates the absolute value. $\mu = 3V/A$ with $V$ and $A$ the volume and area of the structure respectively, is used to nondimensionalize $C$. The area is computed by summing up the area of each triangle in the surface mesh. For accurate volume calculation, the signed volume of tetrahedron as shown in \citet{zhang2001efficient} is used. By introducing a feature center for the joint probability density function (jpdf) of $(S, C)$, B2008 have represented these local properties in a non-local sense. For asymmetrical jpdfs, the feature centers, henceforth denoted as $\hat{S}$ and $\hat{C}$, will lie closer to the region of higher density. The formula to compute the feature center is shown in Appendix C of B2008. A third parameter, denoted by $\lambda$ characterizes the amount of stretching experienced by the structure and is denoted by
	
	\begin{equation}
		\lambda = \sqrt[3]{36\pi}\frac{V^{2/3}}{A}
	\end{equation}
	
	\begin{table}
		\begin{center}
			\begin{tabular}{l}
				\hline
				\textbf{Algorithm 1} Geometrical characterization of structures extracted from scalar fields\\
				\hline
				\vspace{0.25cm}
				\textbf{Input:} Scalar field of the indicator, Percolation threshold $\tau_{p}$\\
				\begin{tabular}{rl}
					(i) & At $\tau_{p}$, extract structures with algorithm \ref{alg:Extraction}.\\
					(ii) & Embed every structure in a box with the maximum extent along every direction.\\
					(iii) & Compute fractal dimension with equation \ref{eq: meanFractalDimension}.\\ 
					(iv) & If $\langle D_{\alpha} \rangle > 1$, compute Shape Index, Curvedness and Stretching.\\
				\end{tabular}\\
				\textbf{Output:} $\hat{S}$, $\hat{C}$ and $\lambda$.\\
				\hline
			\end{tabular}
			\label{alg:MLP}
		\end{center}
	\end{table}
	
	\item[(c)] \textit{Visualization space: }Once the three parameters are computed for every structure, they are plotted onto a visualization space which provides a graphical means of educing clusters of similar structures. All structures can be broadly classified into three categories as blob-like, sheet-like or tube-like depending on where they lie in the visualization space. For instance, a sphere which has $(\hat{S}, \hat{C}, \lambda) = (1, 1, 1)$ is a blob-like structure and any other structures which are close to this region can be thought of as approaching the geometry of a sphere. Similarly, tube-like structures can be identified close to a cylinder which has $(\hat{S}, \hat{C}, \lambda) = (0.5, 1, \lambda)$, where $\lambda$ can indicate the stretching experienced by the tube. The parameter $\hat{C}$ solely determines if a structure is sheet-like ($\hat{C} = 0$) or transitioning towards it ($\hat{C} \rightarrow 0$). Figure \ref{fig:shapeIndex} shows examples of commonly encountered structures and an example of each quantitative coherent structure extracted from case N. Intuitively, one can understand from figure \ref{fig:shapeIndex}(a) that A, B are blob-like, C, D, E, G, H and I are tube-like and the rest are sheet-like structures. This is confirmed with a K-means clustering algorithm initialized with 3 cluster centers. The results are shown in C2. 
\end{itemize}


To summarize, we identify the Robinson structures with indicators as described in section \ref{subsec:identificationCS} and split them into two categories: quantitative and qualitative. For structures belonging to the former, we compute the global percolation threshold $\tau_{p}$ by applying percolation analysis as described in section \ref{subsec:choiceThreshold} and then extract the structures with the NS+MC approach described in section \ref{subsec:extraction}. Finally, we filter all structures with $\big\langle D_{\alpha} (r)\big\rangle < 1$ and compute the stretching and feature centers of shape index and curvedness (section \ref{subsec:geomCharacterization}). This summary is also presented as a four-step algorithm (Algorithm 1) and results of this analysis on the ABL dataset are shown in the subsequent section. 

%
%
%

%

\newpage


\section{Geometry of structures in the ABL}
\label{sec:results}

The results are organized into five sub-sections, four of which correspond to different regions in the boundary layer namely, the viscous sub-layer $(y^{+} < 5)$, buffer layer $(5 < y^{+} < 30)$, inner layer $(y^{+} < 1000)$, and outer layer $(y^{+} > 50)$. This allows us to study the geometrical characteristics and interactions among coherent structures within each region. The final sub-section compares the organization of hairpin structures between the neutral and stably stratified ABL. Although results for all cases are discussed here, the figures for cases S\_2 and S\_3 are shown in the supplementary. 

\subsection{Viscous sub-layer $(y^{+} < 5)$}

\begin{figure}
	\centerline{\includegraphics[width=\linewidth]{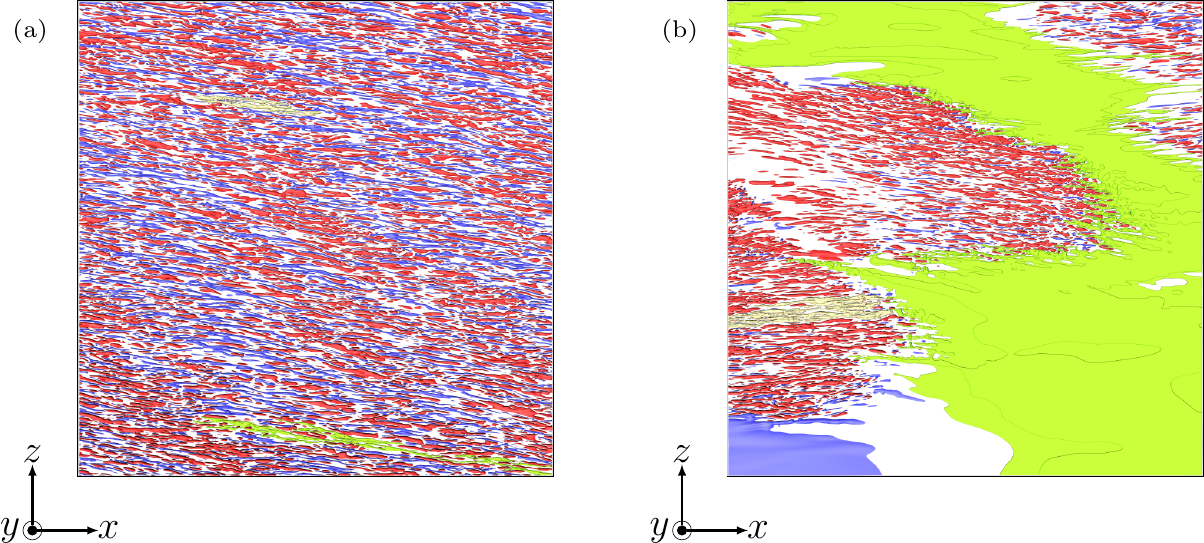}}
	\caption{Isosurfaces of $u'$ in the viscous sublayer for grid B. The structure highlighted in light green is the longest low-speed streak within the domain and the one in pale yellow is the longest high-speed streak. (a) corresponds to case N and (b) to case S\_1. The color specification is according to table \ref{tab:structure_classification} where the regions shaded in red correspond to $u' > 0$ and the ones in blue are $u' < 0$.}
	\label{fig:streaksViscousSublayer}
\end{figure}

The coherent structures of interest in this region are the low-and high-speed streaks, sweeps, ejections and pockets. Other Robinson structures are not observable as the flow in this region is locally laminar. 

First, we discuss the low-and high-speed streaks. Both are identified with the indicator $u' = u - \langle u \rangle$. Here, $\langle \cdot \rangle$ denotes averaging over wall-normal planes for $5$ time steps. Figures \ref{fig:streaksViscousSublayer} and C3.1 show the isosurfaces of $u'$ for all cases at the global percolation threshold $\tau_{p}$. Even qualitatively, one can observe that the distribution and geometry of structures in all cases is different from each other, particularly between cases N and S\_1. 

Case N shown in figure \ref{fig:streaksViscousSublayer}(a) has thin, elongated streaks and their distribution is most comparable to the ones described in flat plate boundary layers (FPBL) \citep{kline1967structure, kline1978role, robinson1991kinematics}. Both low-and high speed streaks tend to be coherent streamwise for several hundred viscous lengths with the longest spanning $6883^{+}$ and $2650^{+}$ units respectively. These numbers are much higher than those reported in R1991. The strength of stratification appears to play an important role in the streamwise and spanwise coherence of low-speed streaks. We note that their streamwise and spanwise coherence tends to increase with an increase in the strength of stratification (from S\_3 to S\_1). In case S\_1, the longest low-speed streak tends to occupy a significant portion of the domain as highlighted in figure \ref{fig:streaksViscousSublayer}(b). If we consider only the high-speed streaks for this case, we would essentially see large empty patches reminiscent of the nonturbulent patches seen in figure \ref{fig:vortMagSlice}(b) at $y^{+} \approx 100$. This suggests that global intermittency causes a deceleration of the flow in the viscous sublayer.
These large low-speed regions force the high speed streaks to form tight clusters. The nature of the high speed streaks, compared to the ones observed in FPBLs, doesn't appear to change much with their mean streamwise and spanwise coherence ranging between $255^{+}$ to $313^{+}$ and $44^{+}$ to $50^{+}$ units respectively. The volume of occupancy of these entities falls with increase in stratification with the converse being true for low speed streaks. For S\_3, low-and high speed streaks occupy $3.1\%$ and $16.3\%$ of grid B respectively, S\_2 with $4.8\%$ and $14.3\%$ respectively and S\_1 with $18.77\%$ and $8.76\%$ respectively. It can be qualitatively observed through figures \ref{fig:streaksViscousSublayer} and C3.1.

\begin{figure}
	\centerline{\includegraphics[width=\linewidth]{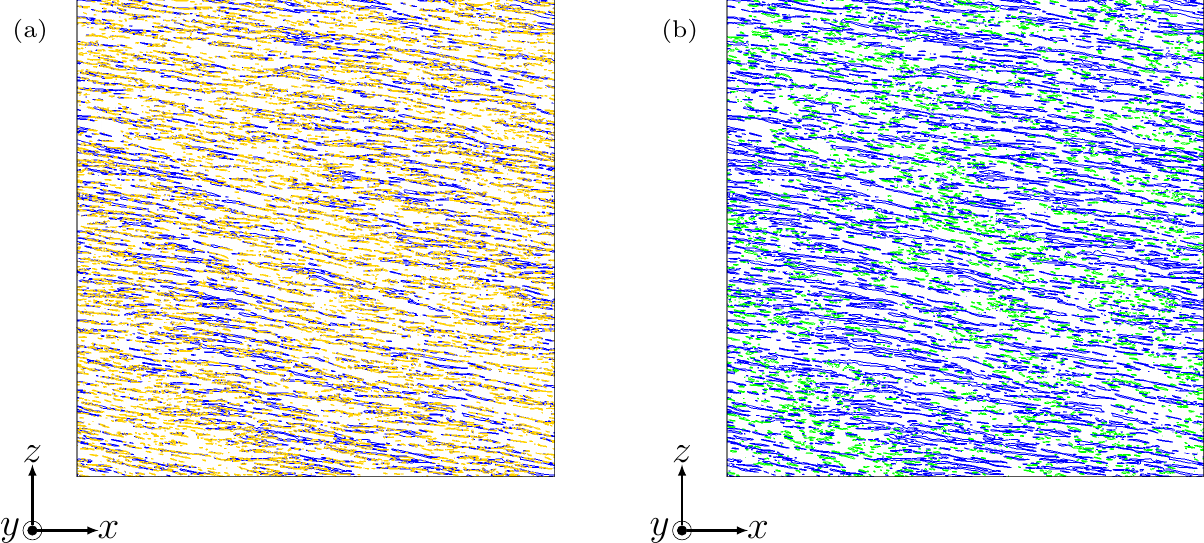}}
	\caption{Contour plots of (a) ejections on top of low-speed streaks and (b) sweeps and low-speed streaks at $y^{+} \approx 3.58$ for case N and grid B. The color specification is according to table \ref{tab:structure_classification} where low-speed streaks are colored blue, sweeps colored green and ejections colored yellow.}
	\label{fig:sweepsEjectionsStreaksViscousSublayerCaseN}
\end{figure}

\begin{figure}
	\centerline{\includegraphics[width=\linewidth]{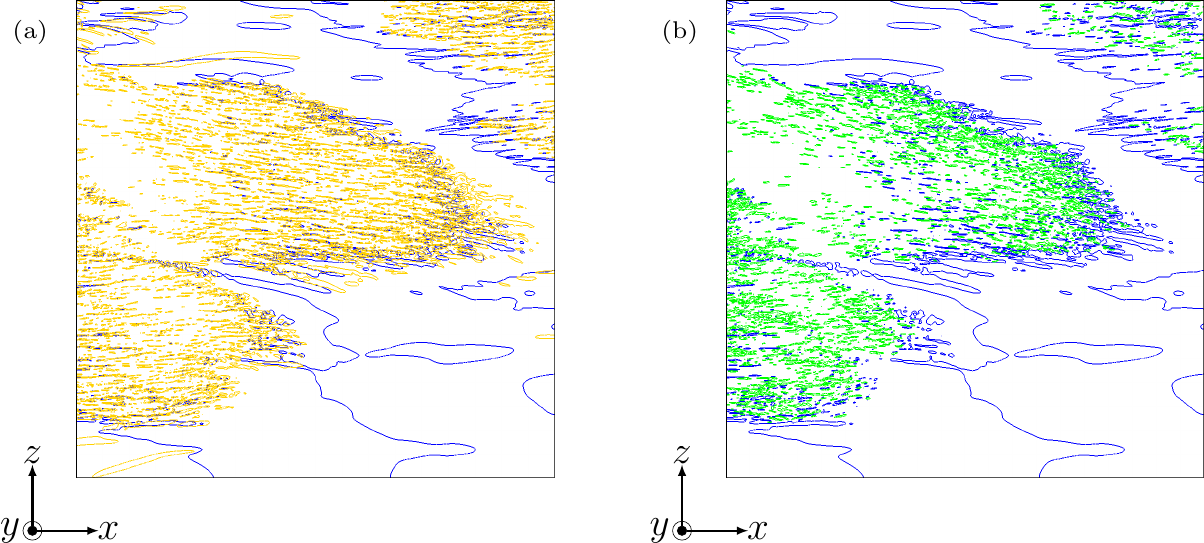}}
	\caption{Similar to figure \ref{fig:sweepsEjectionsStreaksViscousSublayerCaseN}, contour plots of (a) ejections on top of low-speed streaks and (b) sweeps and low-speed streaks at $y^{+} \approx 3.58$ are shown for case S\_1 and grid B. }
	\label{fig:sweepsEjectionsStreaksViscousSublayerCaseS1}
\end{figure}

With figures \ref{fig:sweepsEjectionsStreaksViscousSublayerCaseN}, \ref{fig:sweepsEjectionsStreaksViscousSublayerCaseS1} and C3.2, we study sweeps, ejections and their interactions with low speed streaks at $y^{+} \approx 3.58$. Sweeps and ejections were computed with the quadrant technique and correspond to Q2($-u', +v'$) and Q4($+u', -v'$) events. $v'$ is computed the same way as that of $u'$. Akin to streaks, sweeps and ejections in case N (see figure \ref{fig:sweepsEjectionsStreaksViscousSublayerCaseN}) behave similarly to FPBLs. Confirming previous studies \citep{corino1969visual, bogard1986burst}, we observe that ejection events show less streamwise coherence than low speed streaks and several ejection events may arise from a single low speed streak (see figure \ref{fig:sweepsEjectionsStreaksViscousSublayerCaseN}(a)). This is possible since both are regions of negative $u'$. This character does not appear to change for cases S\_2 and S\_3. For case S\_1 as seen in figure \ref{fig:sweepsEjectionsStreaksViscousSublayerCaseS1}(a), some ejections arise out of the edges of the long low speed streak seen in figure \ref{fig:streaksViscousSublayer}(b). However, most of the structures in the inner region show no interactions. This reinforces the idea that nonturbulent patches may simply be low-speed streaks. 

Sweep events in all cases do not interact with the low speed streaks and appear to be aligned to ejections and low-speed streak. This suggests that sweeps and ejections often exist as a pair and each pair is generally associated with the existence of a vortex above the viscous sublayer \citep{robinson1991kinematics}. With our extraction algorithm, we counted the number of sweep/ejection structures for each case. For case N, the ratio of sweep/ejection structures is $3033/6676 \, (0.45)$, case S\_3 with $3104/4842 \, (0.64)$, case S\_2 with $2798/4980 \, (0.56)$ and case S\_1 with $2194/3054 \, (0.72)$. These numbers illustrate the following: (i) not all sweeps have an ejection associated with it, (ii) although case S\_2 and S\_3 do not show much of a difference, an increase strength of stratification clearly reduces the number of sweeps and ejections. These results together with figures \ref{fig:sweepsEjectionsStreaksViscousSublayerCaseN} and \ref{fig:sweepsEjectionsStreaksViscousSublayerCaseS1} suggest the following: (i) case N will have relatively the least amount of vortices whereas case S\_1 will have the highest close to the wall and (ii) case S\_1 will show that vortices are organized in tight clusters. These views will be revisited in the next sub-section and sub-section \ref{subsec:hairpin}.

\begin{figure}
	\centerline{\includegraphics[width=\linewidth]{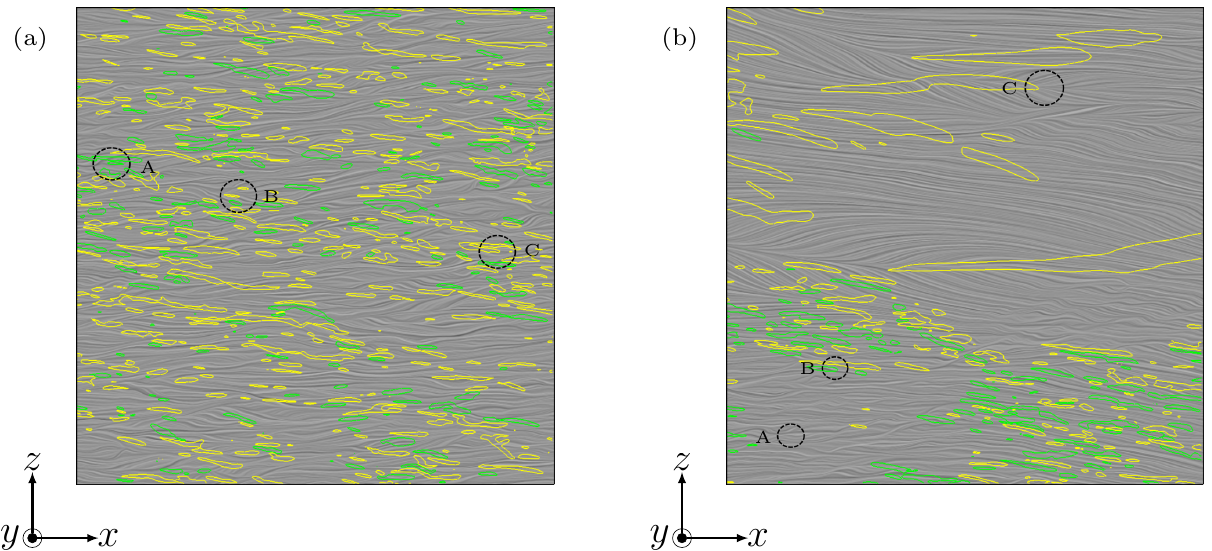}}
	\caption{Pockets are shown with diverging streamlines for (a) case N and (b) case S\_1. Three regions are highlighted in each case which show examples of pocket-like regions. These are overlayed with sweeps and ejections. The combination of pocket-like region and a sweep/ejection can possibly highlight the presence of a vortex above the viscous sublayer. Streamlines are visualized with Line integral convolution for the grid C. The color specification is according to table \ref{tab:structure_classification} where sweeps are colored green and ejections colored yellow.}
	\label{fig:PocketsSweepsCaseNS1}
\end{figure}

Pockets are visualized with streamlines by the line integral convolution technique \citep{cabral1993imaging} in figures \ref{fig:PocketsSweepsCaseNS1} and C3.3. Since pockets are thought to be the `footprints' of a structure aloft the viscous sublayer that induces wallward fluid motion, the outer structure is often associated with vortices. Wallward motion is also known to cause streamlines to diverge, hence we have marked 3 regions of diverging streamlines in each case. Apart from case S\_1, we note that most regions of diverging streamlines can be associated with a sweep/ejection pair thereby suggesting the presence of a vortex. Along with sweeps and ejections, pockets are reviewed again in sub-section \ref{subsec:hairpin}.


\subsection{Buffer layer $(5 < y^{+} < 30)$}

\noindent The reason why we didn't characterize the geometry of structures in the viscous sublayer is because very few of them, which have a fractal dimension greater than 1, are fully formed within this region. Most of these structures extend into the buffer layer and sometimes beyond. In this section, we first establish a comparison between the structures formed within the buffer layer itself i.e., structures which start and end within the buffer layer and structures which extend from the viscous sublayer to this region. Next, we compare the geometry of these structures subject to different levels of stratification. 

\begin{figure}
	\minipage{\textwidth}
	\centerline{\includegraphics[width=0.75\linewidth]{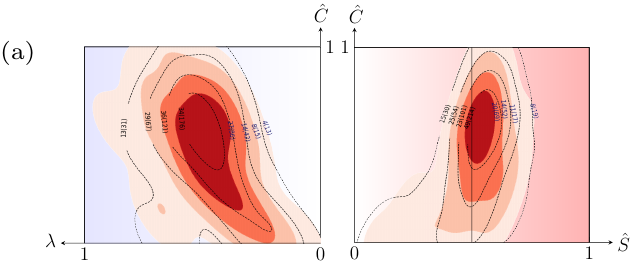}}
	\endminipage\hfill
	\minipage{\textwidth}
	\centerline{\includegraphics[width=0.75\linewidth]{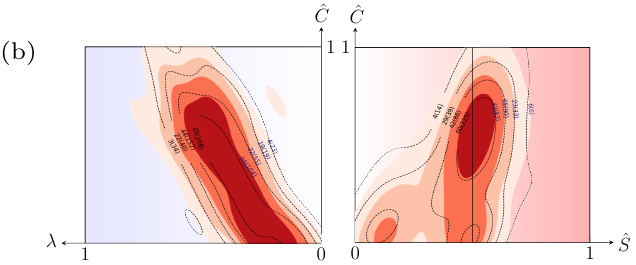}}
	\endminipage\hfill
	\minipage{\textwidth}
	\centerline{\includegraphics[width=0.75\linewidth]{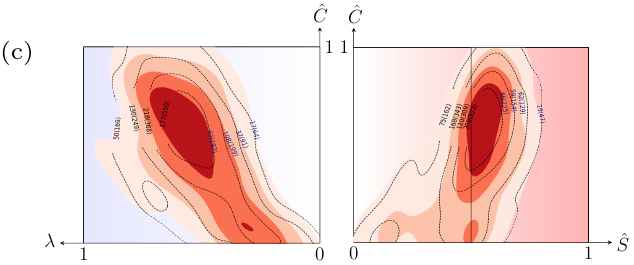}}
	\endminipage\hfill
	\minipage{\textwidth}
	\centerline{\includegraphics[width=0.75\linewidth]{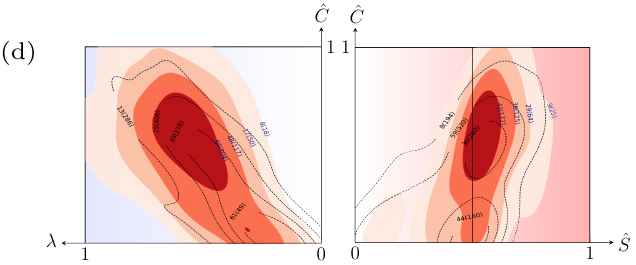}}
	\endminipage\hfill
	\minipage{\textwidth}
	\centerline{\includegraphics[width=0.75\linewidth]{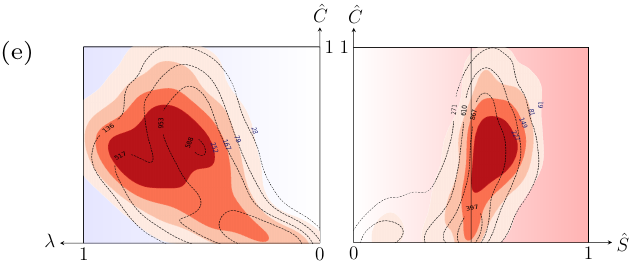}}
	\endminipage\hfill
\end{figure}

\begin{figure}
	\minipage{\textwidth}
	\centerline{\includegraphics[width=0.75\linewidth]{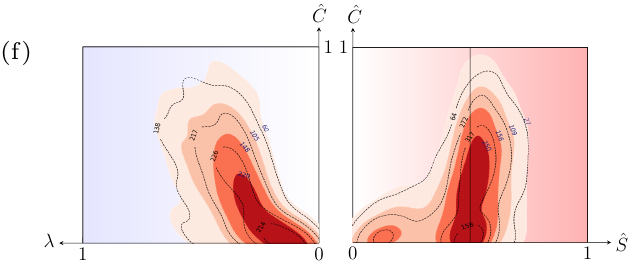}}
	\endminipage\hfill
	\caption{The visualization space for all Robinsion structures except pockets, backs and bulges are shown here with joint pdfs. (a - f) correspond to high-speed streaks, low-speed streaks, sweeps, ejections, vortices and shear layers. Filled contours are used for case S\_1 (with dark shade of red showing the region of high density) whereas unfilled contours with dashed lines are for case N. The number of structures between contours are also indicated - dark blue for case S\_1 and black for case N. Additionally, numbers in parenthesis indicate structures which start within the viscous sublayer and end in the buffer layer. Numbers outside parenthesis indicate structures within the buffer layer itself.}
	\label{fig:GeometryBufferLayerHSSLSS}
\end{figure}

We summarize briefly the procedure to obtain the figures \ref{fig:GeometryBufferLayerHSSLSS} and C3.6. All indicators described in table \ref{tab:structure_classification} are subjected to percolation analysis and an objective threshold $\tau_{p}$ is obtained for each of them. This step is done constrained by grid A for $y^{+} < 1550$. At $\tau_{p}$, all structures are extracted with the algorithm described in sub-section \ref{subsec:extraction}. Extraction is done on grid B to keep the cost of computation low. Next, two filters are applied to ensure that the surfaces are compatible with our geometrical characterization. The first applies the box counting technique (cf. \ref{subsec:geomCharacterization}(a)) to filter all structures with $\langle D_{\alpha} (r) \rangle < 1$. The second one removes all structures close to the side walls as they are incomplete, open surfaces. Finally, the parameters characterizing the geometry of a structure are computed. These are shape index, curvedness and stretching (see sub-section \ref{subsec:geomCharacterization}(b,c)). Since we have characterized a large number of structures, we represent the visualization space with joint pdfs rather than scatter plots as shown in figure \ref{fig:shapeIndex}(a). 

We make a couple of general observations from figures \ref{fig:GeometryBufferLayerHSSLSS} and C3.6: (i) within each category, no obvious distinction can be made in the distribution of geometry of structures across different levels of stratification, i.e., strength of stratification does not appear to impact the geometry, (ii) few structures in the buffer layer are sheet-like ($\hat{C} \rightarrow 0$) and the remaining are in transition ($0 < \hat{C} < 1$). 

Results from figures \ref{fig:GeometryBufferLayerHSSLSS}(a,b) and C3.6(a,b) suggest that high-and low-speed streaks are mostly tube-like structures for all cases. Few high-speed streaks exist in the buffer layer itself whereas low-speed streaks show more activity. This view is supported by \citet{khanna1998three} who found that the intensity of high-speed regions decreases much more rapidly with height than that of the low-speed regions for the near-neutral ABL. Our results suggest that this scenario is true for all cases, regardless of the strength of stratification. Most structures also seem to extend outward from the viscous sublayer (indicated by the number of structures within parenthesis). This is contrary to the view suggested by R1991, who pointed out that high-speed streaks are localized structures extending a limited distance from the wall. 

\citet{willmarth1972structure} state that streaks and ejections are the dominant contributor to Reynolds shear stress $\rho \overline{u' v}$ and show a high activity in the buffer layer of FPBL. They note that sweeps are dominant when $y^{+} < 15$ and ejections dominate when $y^{+} > 15$. This view is supported by \citet{kim1987turbulence} where DNS of channel flow at multiple Reynolds numbers are investigated. We can interpret a similar picture from figures \ref{fig:GeometryBufferLayerHSSLSS}(c,d) and C3.6(c,d). The number of sweep events within the buffer layer itself (indicated by the number of structures outside parenthesis) is much higher than the ejection events for all cases. This suggests a higher sweep activity within the buffer layer. Since sweeps are responsible for wallward flow, these results suggest a high wallward flow in the sweep regions within the buffer layer. 

Vortices and shear layers begin their journey within the buffer layer. We surmised in the previous sub-section that case N will exhibit the least vortex activity and case S\_1 the highest based on the sweep/ejection pair. Results from figure \ref{fig:GeometryBufferLayerHSSLSS}(e) indicate the complete opposite with substantial vortex activity for case N. However, less vortex activity in case S\_1 may be due to the clustering effect we described earlier where a very large region is extracted as a single structure (see figure \ref{fig:MLPNecessity}(a)). Such structures which span the domain are excluded from the geometrical characterization due to computational time. Even if included, they will be characterized as a sheet-like structure since the area is quite large i.e., the ratio $\mu$ becomes very small and $\hat{C} \rightarrow 0$. Apart from this, one can note that vortices identified by the $Q-$criterion are mostly tube-like structures in the buffer layer for all cases. With similar geometry, shear layers are found in less abundence within this region. This is interesting as both $Q-$criterion and $|\omega|$ are vortex indicators suggesting different levels of activity within the buffer layer.

\begin{figure}
	\minipage{\textwidth}
	\centerline{\includegraphics[width=0.75\linewidth]{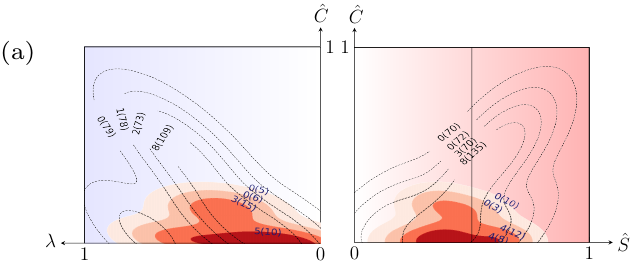}}
	\endminipage\hfill
	\minipage{\textwidth}
	\centerline{\includegraphics[width=0.75\linewidth]{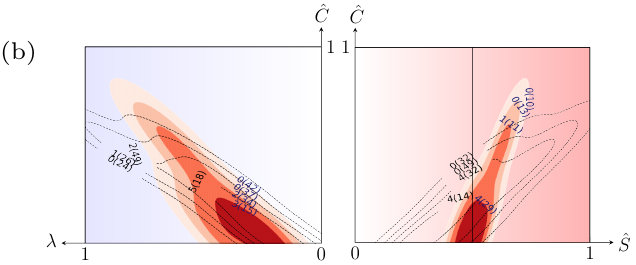}}
	\endminipage\hfill
	\minipage{\textwidth}
	\centerline{\includegraphics[width=0.75\linewidth]{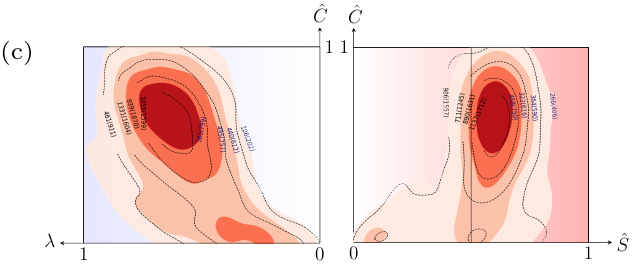}}
	\endminipage\hfill
	\minipage{\textwidth}
	\centerline{\includegraphics[width=0.75\linewidth]{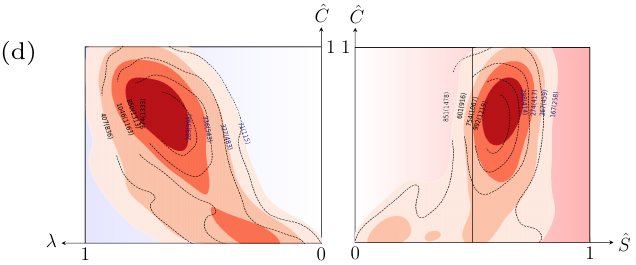}}
	\endminipage\hfill
	\minipage{\textwidth}
	\centerline{\includegraphics[width=0.75\linewidth]{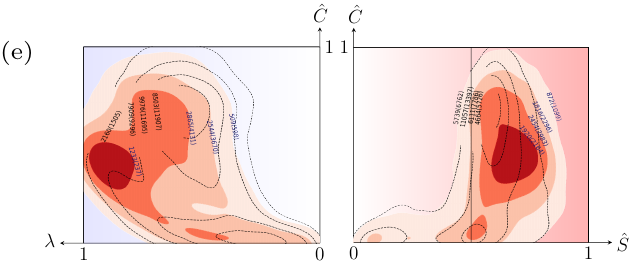}}
	\endminipage\hfill
\end{figure}

\begin{figure}
	\minipage{\textwidth}
	\centerline{\includegraphics[width=0.75\linewidth]{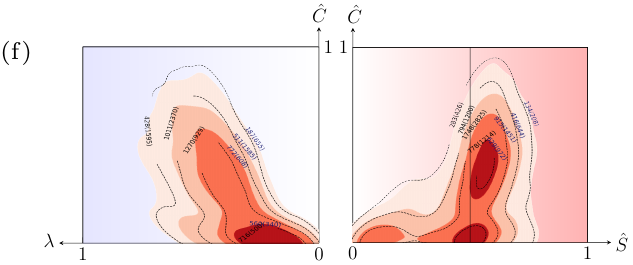}}
	\endminipage\hfill
	\caption{Similar to figure \ref{fig:GeometryBufferLayerHSSLSS}, the visualization space for high-speed streaks, low-speed streaks, sweeps, ejections, vortices and shear layers (a - f) are shown here. In this case, numbers in parenthesis indicate structures which start within the viscous sublayer and end in the inner layer. Numbers outside parenthesis indicate structures beyond the buffer layer.}
	\label{fig:GeometryInnerLayerHSSLSS}
\end{figure}

\subsection{Inner layer $(y^{+} < 1000)$}


\noindent The formation of high-and low-speed streaks within the inner layer itself, i.e., beyond $30 < y^+ < 1000$ is insignificant and at most times nonexistent (see figures \ref{fig:GeometryInnerLayerHSSLSS}(a,b) and C3.7(a,b)). Although we say inner layer, streaks are only studied until a $y^{+} \approx 40$ (as shown in table \ref{tab:structure_classification}). Therefore, we can only conclude that there is very little activity of streaks in the region $30 < y^{+} < 40$. Most of the streaks originate from the viscous sublayer. This further supports the notion that high-speed streaks extend futher outward from the wall than suggested by R1991. While these structures are mostly tube-like, low-speed streaks show a downward shift in both $\hat{S}\hat{C}$ and $\lambda\hat{C}$ planes showing a clear transition towards sheet-like structures. 

Sweeps and ejections in figures \ref{fig:GeometryInnerLayerHSSLSS}(c,d) and C3.7(c,d) exhibit a distinguishing feature from all other structures we have examined so far. When visualized with scatter points, one can observe two distinct clusters (tube-like and sheet-like) for the structures formed beyond the buffer layer. By K-means clustering algorithm, we segregate the points into two clusters (see C3.4). All structrues conforming to each cluster are then extracted and added to an empty 3D scalar field so that they can be reconstructed and visualized (see C3.5). We see that tube-like and sheet-like structures are distributed quite randomly throughout the domain, even along the wall-normal direction.  Although higher activities of sweep and ejection events are evident from the figures, much more sweep events exist than ejection events for all cases except S\_3 where the converse is observed. The ratio of sweep/ejection events is also much more consistent here than in the viscous sublayer. For case N, the ratio of sweep/ejection events is $892/820 (1.09)$, case S\_1 with  $296/201 (1.47)$, case S\_2 with $614/520 (1.18)$ and case S\_3 with $440/635 (0.69)$. If we ignore case S\_3, we can still observe the same trend spotted in the viscous sublayer that an increase in the strength of stratification reduces sweep and ejection events. With all these information, we can now comment on the overall behavior of these events: sweeps dominate the inner and buffer layer suggesting high wallward flow within the sweep regions whereas ejections dominate the viscous sublayer suggesting high outwards flow within the ejection regions from the wall.

A significant number of structures can be seen for figures \ref{fig:GeometryInnerLayerHSSLSS}(e,f) and C3.7(e,f) showing high vortex and shear layer activity in the inner layer. For shear layers, we have limited the region of study to $y^{+} \approx 80$ (see table \ref{tab:structure_classification}). Any structures which extend beyond this wall-normal height are classified as backs and are discussed in section \ref{subsec:outerLayer}. Interestingly, both vortices and shear layers exhibit tube-like and sheet-like geometry with the latter having more sheet-like structures. The reconstructed scalar field for shear layer structures in case S\_1 (see C3.11) traces out a vague outline of the turbulent/non-turbulent interface suggesting that the outer edges of turbulent regions exhibit high amounts of shear.

\begin{figure}
	\minipage{\textwidth}
	\centerline{\includegraphics[width=0.75\linewidth]{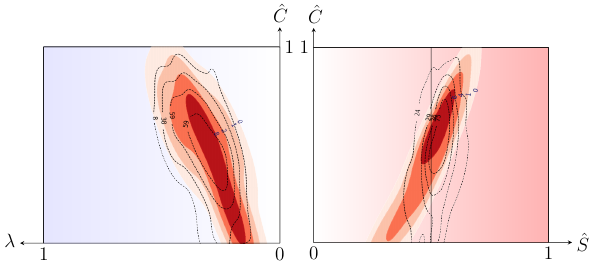}}
	\endminipage\hfill
	\caption{Visualization space for backs for case S\_1 (filled contours) and case N (unfilled contours). The number of structures between contours are also indicated - dark blue for case S\_1 and black for case N.}
	\label{fig:GeometryOuterLayerBacks}
\end{figure}

\subsection{Outer layer $(y^{+} > 50)$}
\label{subsec:outerLayer}

\noindent Although this region presents a lot of overlap with the inner layer, we focus only two Robinson structures namely, extended shear layers also known as backs and bulges which are often referred in literature as very large scale motions (VLSM).

The inability of vorticity magnitude to segregate vortex and shear structures is used to find backs, specifically looking at structures which originate within the buffer layer and extend beyond $y^{+} > 80$. Although plenty of $|\omega|$ structures can be seen in the inner layer (indicated by the number of structures in parenthesis in figures \ref{fig:GeometryInnerLayerHSSLSS}(f) and C3.7(f)), only a handful of structures fulfill the criteria for backs. The highest amount of backs out of the total number of structures $(165/105119)$ is detected for case N, followed by $(35/47391)$, $(11/29342)$, $(8/59906)$ for cases S\_2, S\_1 and S\_3 respectively. 
Figure \ref{fig:GeometryOuterLayerBacks} and C3.8 show the geometrical characterization of backs for all cases. The results suggest no explicit dependence on the strength of stratification and similar geometries exist for all cases. Also, their extent suggests that these structures are not on the order of $\delta-$ scale as indicated in previous works \citep{chen1978large, robinson1991kinematics}. The longest streamwise extent for all cases varies from $26^+$ to $959^+$ units and the longest wall-normal extent varies from $15^+$ to $110^+$ units. Previous research with these structures also show that these structures are inclined in the streamwise direction. We computed the angle of inclination with $\theta = arctan(\Delta y^+ / \Delta x^+)$ for all cases and found that some structures are inclined in the streamwise direction with a mean inclination angle for each case ranging between $50^{\circ}$ to $67^{\circ}$. Although structures in case S\_2, S\_3 and N show a wide distribution of inclination angles ranging between $16^{\circ}$ and $80^{\circ}$, case S\_1 has no structure with an inclination angle smaller than $45^{\circ}$. This can possibly indicate a weak dependence on the strength of stratification. 

\begin{figure}
	\minipage{\textwidth}
	\centerline{\includegraphics[width=1\linewidth]{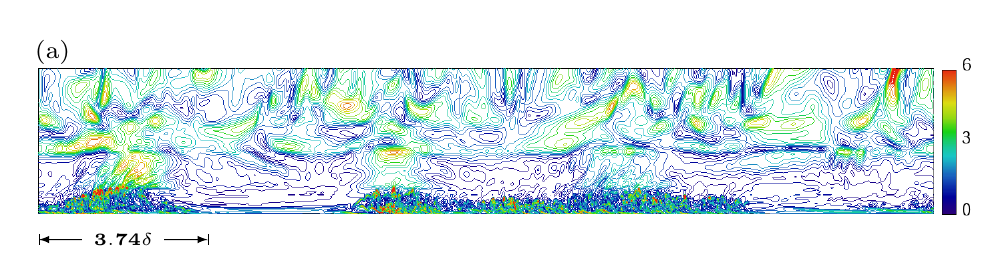}}
	\endminipage\hfill
	\minipage{\textwidth}
	\centerline{\includegraphics[width=1\linewidth]{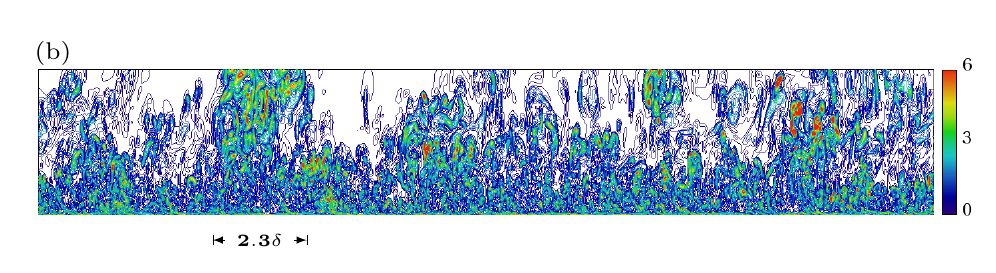}}
	\endminipage\hfill
	\caption{Contours of vorticity magnitude along the XY plane highlighting the presence of $\delta$-scale bulges. Here, the full domain in the streamwise direction is visualized until $y^{+} < 1550$. (a) shows case S\_1 and (b) shows case N. The wall-normal direction, which is $1.75\delta$, has been exaggerated 3 times to show the structures clearly.}
	\label{fig:GeometryOuterLayerBulges}
\end{figure}

The presence of $\delta$-scale structures is most apparent in figure \ref{fig:GeometryOuterLayerBulges}(a), where contours of vorticity magnitude can be seen interacting with the outer layer. These large, $\delta$-scale structures are referred in literature as bulges. We identify these structures by visualizing contours of $|\omega|$. One should note that the $|\omega|$ has been normalized with its RMS as given in equation \ref{eq:definition_final}. This evenly highlights the entire boundary layer including the large scale structures in the outer layer. We also note that bulges exist in all cases regardless of the strength of stratification and an example has been shown for each case in figures \ref{fig:GeometryOuterLayerBulges} and C3.8. The point of interaction of the outer layer and the boundary layer shows a spike in $|\omega|$ suggesting the presence of large scale vortices. 





\subsection{Special study: Hairpin-like structures}
\label{subsec:hairpin}

\begin{figure}
	\minipage{\textwidth}
	\centerline{\includegraphics[width=0.75\linewidth]{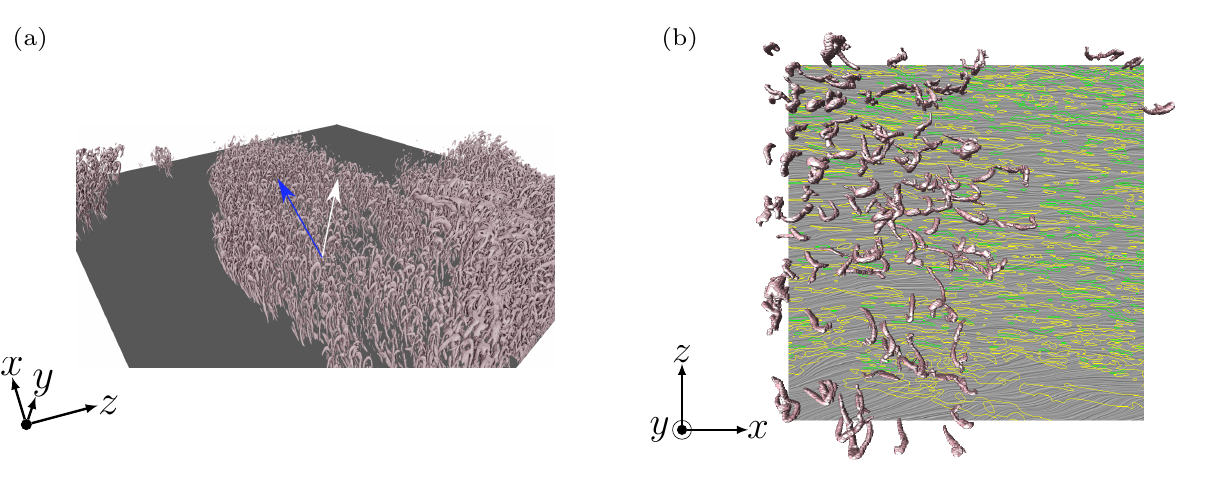}}
	\endminipage\hfill
	\caption{(a) shows a volume rendering of case S\_1 with $Q-$criterion at $\tau_{p}$. The domain is restricted to a wall-normal height of $y^+ \approx 370$ to remove large scale structures. The white arrow indicates the approximate orientation of a majority of hairpin-like structures and the blue arrow indicates the streamwise velocity direction. (b) shows the hairpins chosen for analysis after MLP. It is projected over streamlines showing pockets, sweeps and ejections at a $y^+ \approx 3.58$ in the same region where MLP was performed. }
	\label{fig:HairpinsHS}
\end{figure}

\noindent In this section, we pursue a specific category of vortical structures which are shaped like a hairpin (or horseshoe). This is done for several reasons: (i) upon visualization of case S\_1 (see figure \ref{fig:HairpinsHS}(a)), we see a remarkable amount of vortices shaped like hairpins organized in clusters in the region $30 < y^+ < 200$ \citep{harikrishnan2020curious}, (ii) these hairpin-like structures seem to be oriented in a particular direction and (iii) as pointed out previously, a large region spanning the whole domain in case S\_1 is identified as a single structure and not used for geometrical characterization, thereby restricting our analysis in this region. 

To extract the hairpin-like structures, we apply multi level percolation analysis (MLP) as discussed in Appendix \ref{appendix:MLP} on a small domain of size $300 \times 90 \times 600$ (streamwise $\times$ wall-normal $\times$ spanwise) as shown in figures \ref{fig:MLPNecessity}(a,b). A small domain along with $V_{max} / V = 0.5$ as the stopping criterion is chosen in order to keep the computational costs bearable. As seen in figure \ref{fig:MLPNecessity}(c), this stopping criterion is sufficient to educe individual structures. All structures within this region were extracted and from these 100 typical structures were chosen manually. We also chose $100$ hairpin-like structures from other cases as well.

\begin{figure}
	\minipage{\textwidth}
	\centerline{\includegraphics[width=0.95\linewidth]{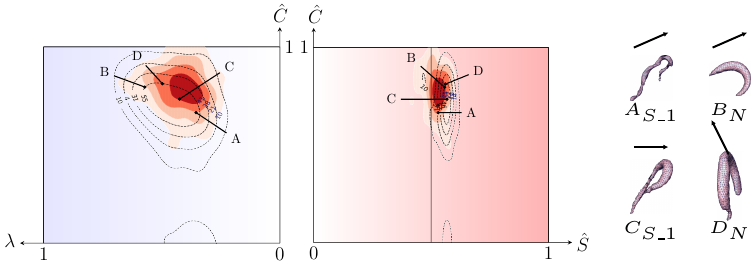}}
	\endminipage\hfill
	\caption{Visualization space with for 100 hairpin-like structures are shown for case S\_1 (filled contours) and case N (unfilled contours). Two examples for each case are visualized on the right hand side. The subscripts indicate the case to which the structure belongs.}
	\label{fig:GeometryHairpins}
\end{figure}

Figures \ref{fig:GeometryHairpins} and C3.9 show the geometrical characterization of hairpin-like structures for all cases. Interestingly, no sheet-like structures can be found for case S\_1 and the curvedness ranges between $0.5$ and $1$. This can be attributed to the application of MLP which produces `simple structures' existing at their own threshold. For other cases, the structures can be complex, sometimes with several structures interacting at once which tends to have an impact on the curvedness value. Example visualizations on the top row of figure \ref{fig:GeometryHairpins} highlight an intriguing aspect of hairpin-like structures - the head of the hairpins seem to be oriented in a similar direction. This view can be further confirmed with figure \ref{fig:HairpinsHS}(a), where a majority of hairpin-like structures are turned at some angle to the streamwise direction. This behavior is not observed in other cases. 

Finally, the importance of pockets and sweep/ejection pairs in vortex generation is reviewed. Figure \ref{fig:HairpinsHS}(b) shows hairpin-like structures, pockets, sweeps and ejections superimposed over one another. Upon close inspection, we find some structures, close to the wall, which can be directly associated with diverging streamlines and a sweep/ejection pair. This reaffirms the idea that pockets are caused by the presence of an outer rotating structure and the presence of a sweep/ejection pair can be associated with the leg of a hairpin-like structure. These results illustrate the geometry and distribution of Robinson structures quite well, despite the analysis being restricted to one-third of the entire domain.





\newpage


\section{Discussion}
\label{sec:discussion}

This section outlines some practical implications of our work: First, in sub-section \ref{subsec: globalIntermittency}, we introduce an alternative quantification of global intermittency that overcomes the known shortcomings of vorticity magnitude. Next, in sub-section \ref{subsec: condStats}, we compare the results from section \ref{sec:results} to the ones obtained with a conditional analysis of one-point statistics that does not make use of feature extraction. 

\subsection{Global intermittency}
\label{subsec: globalIntermittency}

\begin{figure}
	\centerline{\includegraphics[width=\linewidth]{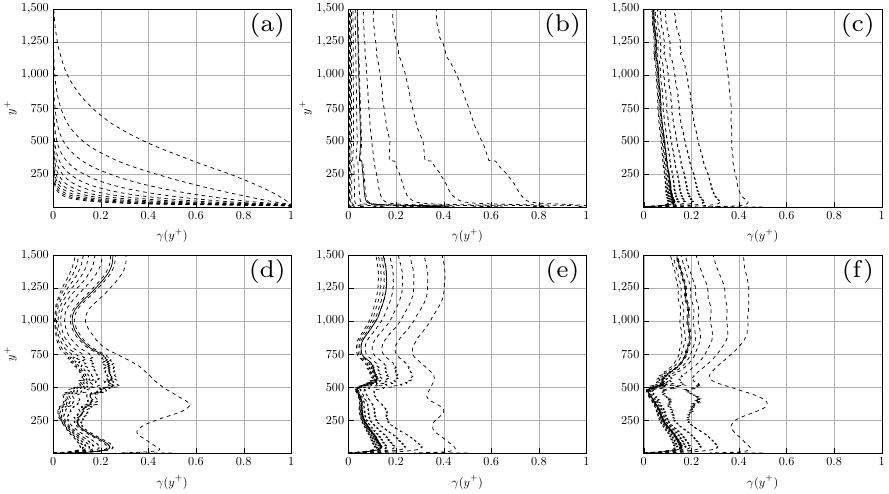}}
	\caption{Top panel shows vertical profiles of $\gamma(y^+)$ for case N: (a) vorticity magnitude without normalization with RMS over every wall-normal plane, (b) vorticity magnitude with normalization with RMS over every wall-normal plane and (c) $Q-$criterion after normalization with RMS for every wall-normal plane. Bottom panel shows vertical profiles of $\gamma(y^+)$ for $Q-$criterion for case (d) S\_1, (e) S\_2 and (f) S\_3. All figures show profiles for increasing values (left to right in each figure) of threshold. The first dashed line corresponds to $0$ threshold whereas the solid line indicates percolation threshold.}
	\label{fig:intermittencyVorticityComparison}
\end{figure}

\noindent The intermittency factor according to \cite{pope2001turbulent} is given by

\begin{equation}\label{eq:popeIntermittencyFactor}
	\gamma(x, t) = \langle H(|\omega(x, t)| - \omega_{\tau}) \rangle
\end{equation}

\noindent where $H$ is a Heaviside function, $|\omega|$ is the vorticity magnitude, $\omega_{\tau}$ is a small threshold, $x$ is a streamwise position in the flow, $t$ is time and $\langle \cdot \rangle$ represents averaging over horizontal planes. The equation \ref{eq:popeIntermittencyFactor} is generalized to include other indicators as follows,

\begin{equation}
	\label{eq: intermittencyEq}
	\gamma(y) \equiv \langle H(\alpha (x, y, z) - \alpha_{\tau_{p}})\rangle
\end{equation}

\noindent where $\alpha$ is an indicator and $\alpha_{\tau_{p}}$ is the global percolation threshold for the indicator. Time $t$ has been excluded since we are dealing with a single timestep in our current work. The choice of $\alpha$ and $\alpha_{\tau_{p}}$ are critical as they differentiate the turbulent and nonturbulent parts. This definition of $\gamma$ is general as $\alpha$ can be any vortex indicator. 

In previous works, \citet{da2014characteristics, ansorge2016analyses} used vorticity magnitude $|\omega|$ as the indicator. 
Figure \ref{fig:intermittencyVorticityComparison}(a, b) show $|\omega|$ applied to the case N. The normalization with RMS over wall-normal planes is not applied for figure \ref{fig:intermittencyVorticityComparison}(a), whereas it is done for figure \ref{fig:intermittencyVorticityComparison}(b). The profiles of $\gamma(y^{+})$ in figure \ref{fig:intermittencyVorticityComparison}(a) is similar to \citet{ansorge2016analyses, kovasznay1970large}. However, this clearly highlights the problems associated with $|\omega|$. First, the profiles indicate that for low values of $\tau$, the region close to the wall, where shear is dominant, is filled with structures. Second, the choice of $\tau_{p}$ is not applicable for the entire boundary layer. The latter issue can be fixed with a normalization of wall-normal planes with its RMS  as shown in Figure \ref{fig:intermittencyVorticityComparison}(b). Here, the profiles are more or less vertical for high values of $\tau$. It should be noted that very close to the wall ($y^{+} < 30$), minor deviations occur. The former issue can be fixed by using an alternate indicator. Figure \ref{fig:intermittencyVorticityComparison}(c) shows case N with $Q-$criterion. The profile correctly shows that regions close to the wall are relatively `less filled' with structures when compared with $|\omega|$. 


Figures \ref{fig:intermittencyVorticityComparison}(d, e, f) show the intermittency factor profiles for the three stratified cases with $Q-$criterion. Unlike case N, the profiles show significant variations beyond $y^{+} = 30$. 
We justify the choice of a single threshold by noting that the standard deviation of the profiles is small. The implication of a modified method to quantify global intermittency on conditional statistics is shown in the next subsection.

\subsection{A conditional statistics perspective}
\label{subsec: condStats}


Sub-section 5.1 suggests that the $Q-$criterion can be used to partition the turbulent flow in analogy to the turbulent/nonturbulent partitioning used for the conditional study of large-scale intermittency in turbulent flows \citep[cf.][]{shah2014very, ansorge2016analyses}. We present the conditional analysis for high-and low speed streaks, sweeps and ejections here. 

\begin{figure}
	\minipage{\textwidth}
	\centerline{\includegraphics[width=\linewidth]{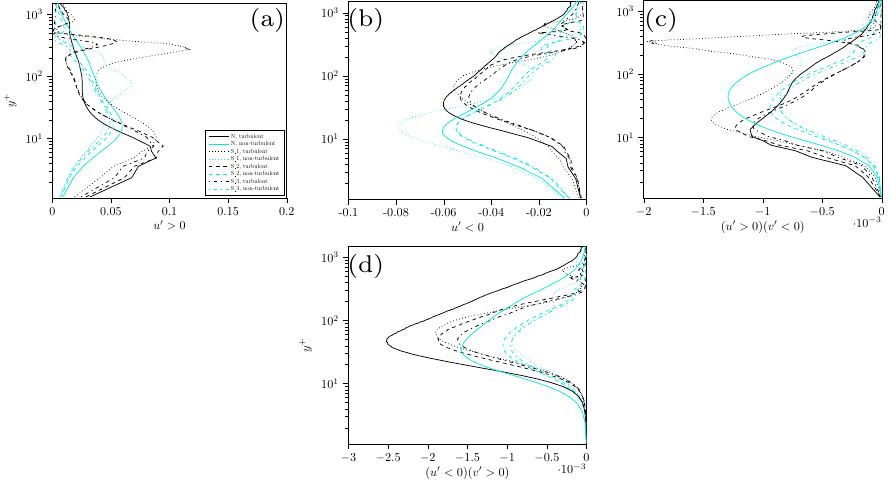}}
	\endminipage\hfill
	\caption{Vertical profiles of streamwise velocity fluctuation for (a) high-speed streaks, (b) low-speed streaks, (c) sweeps and (d) ejections. Line specification is according to table \ref{tab:sim_parameters} for the different cases. Black color indicates turbulent regions and teal shows the nonturbulent regions.}
	\label{fig:conditionalStatistics}
\end{figure}

The flow is split into turbulent and nonturbulent regions with equation \ref{eq: intermittencyEq} using the global percolation threshold $Q_{\tau_{p}}$ as discriminator. The wall-normal planes are averaged such that every height is represented by a single value (cf. figure \ref{fig:conditionalStatistics}). The vertical profiles show that the effect of low-speed streaks (cf. figure \ref{fig:conditionalStatistics}(b)) can extend well over $y^+ > 100$ for all cases indicating that an analysis up to $y^+ \approx 40$ (used in the previous section) may not be sufficient to understand the role of these structures.

When conditioning to high-speed ($u'>0$; figure \ref{fig:conditionalStatistics}(a)) and low-speed streaks ($u'<0$; figure \ref{fig:conditionalStatistics}(b)), we find a distinct
difference between the lowest part of the boundary layer $(y^+<~30)$ and the region aloft. In vicinity of the surface, the
positive fluctuations in the turbulent part are larger than in the laminar one; on the contrary, when looking at the low-
speed streaks, the negative perturbations are larger in the laminar parts. In other words, intense high-speed streaks
tend to belong to the turbulent partition, intense low-speed to the laminar one. This is in accordance with the dominant
origin of turbulence in the buffer layer, namely shear production: given the boundary condition $u(y^+=0)=0$, a larger
velocity immediately implies more shear, thus more intense turbulence production, and larger likelihood of the
corresponding region to be detected as turbulent by the conditioning method introduced in section 5.1 (and vice versa in
low-speed streaks).

This picture is inverted aloft $y^+\approx 30$: There, the nonturbulent partition contains faster moving fluid than the
turbulent one in high-speed streaks and in low-speed streaks (the negative perturbation is smaller, so the fluid is
actually faster in an absolute sense). In this context, we have to be aware that the simple characterization of `streaks'
by the sign of $u'$ bears no geometric characterization whatsoever, i.e., above $y^+=40$, we characterize regions with
positive or negative fluctuation of streamwise velocity which is not necessarily a `wall streak' in the classical sense. In
this context, the non-locality of turbulence above the buffer layer becomes important. Turbulent fluid mostly originates
from below and is thus slower, whereas as non-turbulent fluid originates from above and is therefore faster \cite{ansorge2016analyses}. This is manifested in larger positive velocity fluctuations in the laminar partition of high-speed
streaks and less negative, i.e., also faster in streamwise direction, velocity perturbations in low-speed streaks.

The figures \ref{fig:conditionalStatistics}(c,d) partially confirm the results from section \ref{sec:results} where sweep events appear to dominate over ejection events in the buffer layer with the highest activity in the turbulent regions for case S\_1. Sweeps also show significant activity beyond $y^+ \approx 200$ suggesting strong wallward flow in these regions. Correspondingly, the region beyond the buffer layer and within $y^+ \approx 200$ shows significant ejection activity for all cases in the turbulent regions. This was not seen in section \ref{sec:results}. The results indicate a switching behavior for case S\_1 with sweeps and ejections in turbulent regions. This suggests alternating regions of wallward and outward flow at different heights for case S\_1.

\begin{figure}
	\minipage{\textwidth}
	\centerline{\includegraphics[width=\linewidth]{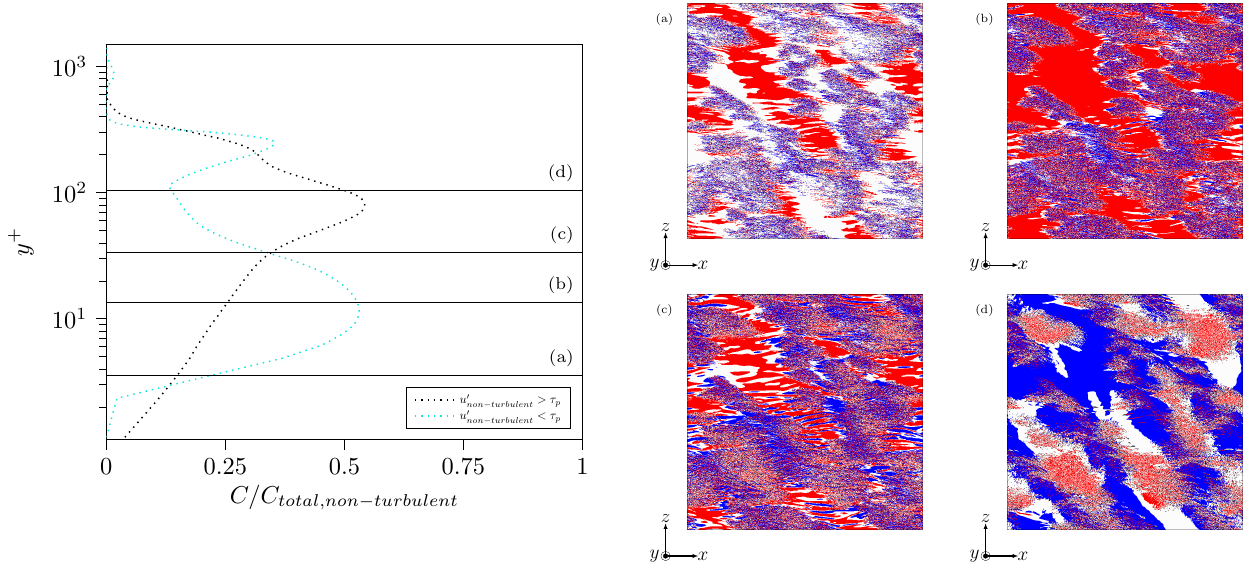}}
	\endminipage\hfill
	\caption{For case S\_1, vertical profiles indicating the number of points $C$ satisfying the condition $\langle u'_{nonturbulent} > 0 \rangle$ and $\langle u'_{nonturbulent} < 0 \rangle$ over the total nonturbulent points $C_{total, nonturbulent}$ are shown on the left plot. Line specification is according to table \ref{tab:sim_parameters}. On the right, horizontal slices of $\langle u'_{nonturbulent} > 0 \rangle$ (red), $\langle u'_{nonturbulent} < 0 \rangle$ (blue) and $Q-$criterion (black) are shown at (a) $y^+ \approx 3.58$, (b) $y^+ \approx 13.45$, (c) $y^+ \approx 31.26$ (d) $y^+ \approx 104$.}
	\label{fig:conditionalStatisticsS1}
\end{figure}

We define the nonturbulent fluctuating regions as $\langle u'_{nonturbulent} \rangle \equiv \langle u' \lessgtr u'_{\tau_{p}} \rangle$ for $Q < Q_{\tau_p}$. Interestingly, for case S\_1, $\langle u'_{nonturbulent} < 0\rangle$ shows significant activity in the buffer layer. We examine this by counting the number of points which satisfy the aforementioned condition over the counts of all nonturbulent points at every wall-normal plane. This result is shown in figure \ref{fig:conditionalStatisticsS1} and C4.1. For case S\_1, we observe that low-speed streaks dominate over the high-speed streaks until a $y^+ \approx 31.26$, after which there is a switch and the high-speed streaks dominate. This behavior is observable for other stratified cases as well but effect is not as significant. However, the switch takes place at a much lower height at $y^+ \approx 6.18$ and $y^+ \approx 7.55$ for cases S\_2 and S\_3 respectively.

\newpage

%
%
\section{Conclusions}
\label{sec:conclusion}

We have established a detailed comparison on the geometry of Robinson structures among three stably stratified cases and a neutrally stratified case of an Ekman flow. For this purpose, we have developed a methodology derived from the works of MJ2004 and B2008. This employs a modified neighbor scanning algorithm to extract structures from 3D scalar fields. The identified neighbors are corrected with the marching cubes visualization algorithm. Performance of this extraction can be further improved by considering an efficient implementation of marching cubes such as the Lewiner algorithm \citep{lewiner2003efficient}. Structures having a fractal dimension less than 1 are not considered. We also exclude structures which are attached to the domain walls. For all remaining structures, three parameters are computed, namely shape index, curvedness and stretching. This allows us to geometrically classify a structure as blob-like, tube-like or sheet-like. Due to the computationally expensive nature of this methodology, the analysis was limited to one-third of the entire domain i.e., grid B. To understand the behavior of structures on the entire domain i.e., grid A, we apply conditional one-point statistics on four Robinson structures namely high-and low-speed streaks, sweeps and ejections. The conditioning is based on a new definition of intermittency factor introduced in equation \ref{eq: intermittencyEq}. The following conclusions are reached:


\vspace{0.2cm}

\begin{enumerate}[]
	\itemsep0.2cm
	\item [(a)] For all Robinson structures, similar geometries (mostly tube-like and sheet-like) are observed across all levels of stratification. Blob-like structures are nonexistent. 
	\item [(b)] Even if structures are clustered, as shown for sweeps and ejections in the inner layer, no relation can be made between a type of geometry such as tube-like or sheet-like and its distribution across the spatial domain.
	\item [(c)] Although the viscous sublayer is locally laminar and characterization with turbulent and nonturbulent subvolumes is not applicable in this region, the effect of global intermittency can be clearly seen here (see figure \ref{fig:streaksViscousSublayer}(b)). The defining characteristic is the association of an ejection with a low speed streak. Large regions where no ejections arise out of a low speed streak (as seen in figure \ref{fig:sweepsEjectionsStreaksViscousSublayerCaseS1}(a)) are reminiscent of the nonturbulent region observed in the buffer layer and beyond. Therefore, global intermittency can be generalized to indicate active/inactive regions rather than turbulent/nonturbulent regions to include its effect in the viscous sublayer. 
	\item [(d)] Sweep/ejection pairs are constrained to the active regions and form tight clusters in the viscous sublayer. We hypothesize that this can possibly lead to the formation of vortical structures shaped like hairpins above the buffer layer. 
\end{enumerate}

\vspace{0.2cm}

In future work, we will focus on the dynamics of the Robinson structures. With a suitable tracking algorithm, the Robinson structures can be followed in time and changes in the geometry can be studied. One can take this further by considering the visualization space as a phase space which hosts all possible geometrical states of the system. This allows for the estimation of two instantaneous metrics namely the local dimension and inverse persistence \citep{faranda2017dynamical} which may offer insight into the preferred geometrical shape of a structure.

The methodology described in section \ref{sec:coherent_structures} requires significant computational power. In total, $10^6$ structures were geometrically characterized with this technique. To ensure reproducibility of the results, all codes used in the course of the work are made available here: https://github.com/Phoenixfire1081/GeometryCharacterization.git.




\vspace{0.25cm}

\noindent \textbf{Acknowledgements}

\noindent The authors are grateful to Elie Bou-Zeid for fruitful discussions during the course of the work. Due to the size of the database, computing was performed on the Jülich Supercomputing Centre under the project hku24 and stadit. The data was stored under the project dns2share. Additionally, data was also stored at the Zuse Institut Berlin (ZIB). This work has been supported by the Deutsche Forschungsgemeinschaft (DFG) through grant CRC 1114 ``Scaling Cascades in Complex Systems'', Project Number 235221301, Project B07: Self-similar structures in turbulent flows and the construction of LES closures. Cedrick Ansorge acknowledges funding through the European Commission's ERC Starting Grant no. 851347.

\newpage
%
%
\appendix
\section{Neighbor Scanning approach with Marching Cubes neighbor correction}
\label{appendix:NS+MC}

For any indicator $\alpha$, we apply the threshold $\tau_{p}$ and scan all resulting points. With the neighbor scanning (NS) approach, all points surrounding the center of a $3 \times 3$ cube are added as neighbors. This process is repeated until no new neighbors can be found. To ensure that the extracted structures are accurate to visualization, we correct the identified neighbors with the marching cubes (MC) algorithm. Although numerous visualization algorithms are available, we choose MC as it is employed by our visualization software Amira \citep{Stalling2005}. This algorithm uses one grid cell (eight vertices of a cube) to determine how the triangles should be constructed. Depending on which of the eight vertices satisfy the threshold, there are $2^{8}$ ways in which the surface mesh can be represented. For further details, the reader is referred to the original paper by \citet{lorensen1987marching}. Out of the 256 combinations, there are 92 cases where the surface mesh is not continuous and in which only 2-4 nonadjacent points of the grid cell satisfy the threshold. When these points belong to different structures, the NS approach is unable to distinguish between them as it doesn't know how the surface mesh is going to be constructed. We, therefore, present the following modification to the NS approach:

\begin{figure}
	\centering
	\minipage{\textwidth}
	\centerline{\includegraphics[width=1\linewidth]{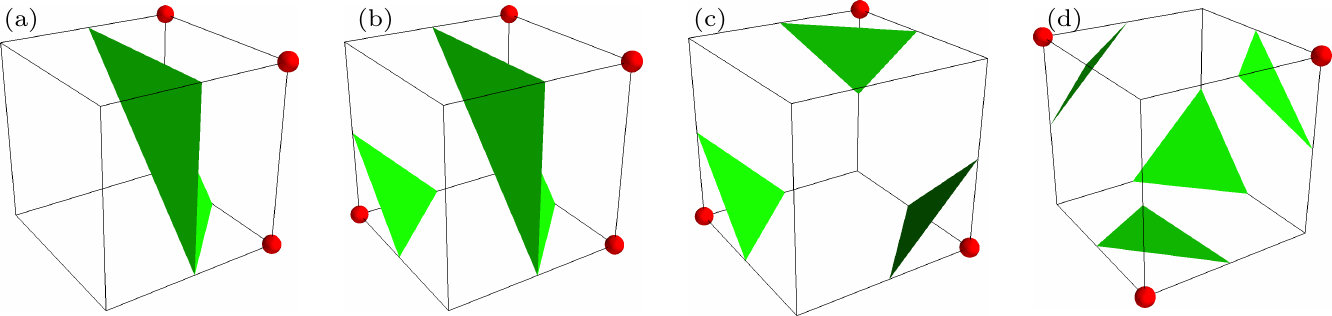}}
	\endminipage\hfill
	\caption{Examples of surface mesh construction with the marching cubes algorithm. (a) shows a continuous mesh which will be a part of only one structure. (b, c, d) show examples where multiple structures can be a part of a single cube.}
	\label{fig:surface_mesh}
\end{figure}

\vspace{0.2cm}

\begin{enumerate}[]
	\itemsep0.2cm
	\item[$(a)$] \textit{Construction of a lookup table: }We visualize every single case out of the 256 combinations and group indices of the cube. For cases with a continuous mesh as in figure \ref{fig:surface_mesh}(a), the lookup table is assigned an empty array. This is also done for the case where all or none of vertices satisfy the threshold as it corresponds to all the points being inside the structure or outside respectively. When more than one mesh element is present as shown in Figure \ref{fig:surface_mesh} (b, c, d), the indices belonging to each mesh element are grouped together. The lookup table is shown in table \ref{tab:GroupingMC}.
	
	\item[$(b)$] \textit{Finding the case in the lookup table: }Once the neighbors are identified with the NS approach, all $2^{3}$ subcubes need to be matched with a case in the lookup table. If $V = \{v_{0}, \ldots, v_{7}\}$ is the set of vertices of the subcube, each vertex is assigned a value $1$ when the neighbor is added and $0$ otherwise. The corresponding case can be found by:
	
	\begin{equation}
		case = \sum_{i = 1}^{8} 2^{i}[V(i) = 1]
		\label{eq:caseMC}
	\end{equation}
	
	\item[$(c)$] \textit{Correcting the neighbors: }Once the appropriate case is found, the grouped indices can be retrieved from the lookup table. No changes are made to the neighbors if the array is empty. For a non-empty array, the `right group' of indices have to be kept as the neighbors and the remaining are discarded. This is chosen as the one which contains the center point of the $3 \times 3$ cube. Since this is our current point of interest and is common to all the subcubes, we can be certain that the mesh element corresponding to this is a part of the structure. For implementation purposes, these steps are shown as an algorithm (see Algorithm A).
\end{enumerate}

\begin{table}
	\begin{center}
		\begin{tabular}{l}
			\hline
			\textbf{Algorithm A} Extracting structures from scalar fields with Neighbor Scanning and Marching \\ 
			Cubes\\
			\hline
			\vspace{0.25cm}
			\textbf{Input:} Scalar field of the indicator, threshold $\tau$\\
			\begin{tabular}{rl}
				(i) & Convert scalar field with floating-point numbers as boolean data type where a point \\
				& $x == True$ when $x > \tau$. \\
				(ii) & For every point in the domain where $x == True$,\\
				& \begin{tabular}{rl}
					(a) & Assign $x$ with a structure label.\\
					(b) & Check neighbors of the point by considering the point to be at the center \\
					& of a $3 \times 3$ cube.\\
					(c) & If neighbors have a $True$ value, label them with the same one as assigned before.\\
					(d) & Split the $3 \times 3$ cube as $8$ subcubes.\\
					(e) & For each cube, determine which Marching Cubes case it belongs to by using \\
					& equation \ref{eq:caseMC}.\\
					(f) & With a table look-up (table \ref{tab:GroupingMC}), the grouping can be seen. If table look-up is\\
					& empty, the labels are not altered. If not, the group with the center point of the\\
					& cube is kept and the remaining labels are altered.\\
					(g) & For all remaining neighbors, repeat from (b).
				\end{tabular}\\
			\end{tabular}\\
			\textbf{Output:} Scalar field with labels for structures, $V_{max}$, $V$.\\
			\hline
		\end{tabular}
		\label{alg:Extraction}
	\end{center}
\end{table}



\begin{table}[h!]
	\begin{center}\scriptsize
		\begin{tabular}{llllllllll}
			\hline
			Case & Indices & Case & Indices & Case & Indices & Case & Indices & Case & Indices\\[3pt]
			\hline
			1 & - & 53 & [2], [4, 5] & 105 & [3], [5, 6] & 157 & - & 209 & - \\
			2 & - & 54 & [2], [0, 4, 5] & 106 & [0, 3], [5, 6] & 158 & - & 210 & - \\
			3 & - & 55 & - & 107 & [3], [1, 5, 6] & 159 & - & 211 & [1], [4, 6, 7]\\
			4 & - & 56 & - & 108 & - & 160 & - & 212 & - \\
			5 & - & 57 & [3], [4, 5] & 109 & - & 161 & [5], [7] & 213 & - \\
			6 & [0], [2] & 58 & - & 110 & - & 162 & [0], [5], [7] & 214 & - \\
			7 & - & 59 & [3], [1, 4, 5] & 111 & - & 163 & [1, 5], [7] & 215 & - \\
			8 & - & 60 & - & 112 & - & 164 & [0, 1, 5], [7] & 216 & [3], [5]\\
			9 & - & 61 & [2, 3], [4, 5] & 113 & - & 165 & [2], [5], [7] & 217 & - \\
			10 & - & 62 & - & 114 & - & 166 & [0], [2], [5], [7] & 218 & - \\
			11 & [1], [3] & 63 & - & 115 & - & 167 & [1, 2, 5], [7] & 219 & [1], [3, 4, 6, 7]\\
			12 & - & 64 & - & 116 & - & 168 & [0, 1, 2, 5], [7] & 220 & - \\
			13 & - & 65 & - & 117 & - & 169 & [3, 7], [5] & 221 & - \\
			14 & - & 66 & [0], [6] & 118 & - & 170 & [0, 3, 7], [5] & 222 & - \\
			15 & - & 67 & [1], [6] & 119 & - & 171 & [1, 5], [3, 7] & 223 & - \\
			16 & - & 68 & [0, 1], [6] & 120 & - & 172 & - & 224 & - \\
			17 & - & 69 & - & 121 & [3], [4, 5, 6] & 173 & [2, 3, 7], [5] & 225 & - \\
			18 & - & 70 & [0], [2, 6] & 122 & - & 174 & [0, 2, 3, 7], [5] & 226 & [0], [5, 6, 7]\\
			19 & [1], [4] & 71 & - & 123 & [3], [1, 4, 5, 6] & 175 & - & 227 & - \\
			20 & - & 72 & - & 124 & - & 176 & - & 228 & - \\
			21 & [2], [4] & 73 & [3], [6] & 125 & - & 177 & - & 229 & - \\
			22 & [0, 4], [2] & 74 & [0, 3], [6] & 126 & [1], [7] & 178 & - & 230 & [0], [2, 5, 6, 7]\\
			23 & [1, 2], [4] & 75 & [1], [3], [6] & 127 & - & 179 & - & 231 & - \\
			24 & - & 76 & [0, 1, 3], [6] & 128 & - & 180 & - & 232 & - \\
			25 & [3], [4] & 77 & - & 129 & - & 181 & [2], [4, 5, 7] & 233 & - \\
			26 & - & 78 & - & 130 & [0], [7] & 182 & [2], [0, 4, 5, 7] & 234 & - \\
			27 & [1], [3], [4] & 79 & - & 131 & [1], [7] & 183 & - & 235 & - \\
			28 & - & 80 & - & 132 & [0, 1], [7] & 184 & - & 236 & [2], [4]\\
			29 & [2, 3], [4] & 81 & [4], [6] & 133 & [2], [7] & 185 & - & 237 & - \\
			30 & - & 82 & [0, 4], [6] & 134 & [0], [2], [7] & 186 & - & 238 & - \\
			31 & [1, 2, 3], [4] & 83 & [1], [4], [6] & 135 & [1, 2], [7] & 187 & - & 239 & - \\
			32 & - & 84 & [0, 1, 4], [6] & 136 & [0, 1, 2], [7] & 188 & - & 240 & - \\
			33 & - & 85 & [4], [2, 6] & 137 & - & 189 & - & 241 & - \\
			34 & [0], [5] & 86 & [0, 4], [2, 6] & 138 & - & 190 & - & 242 & - \\
			35 & - & 87 & [1, 2, 6], [4] & 139 & [1], [3, 7] & 191 & [0], [6] & 243 & - \\
			36 & - & 88 & - & 140 & - & 192 & - & 244 & - \\
			37 & [2], [5] & 89 & [3], [4], [6] & 141 & - & 193 & - & 245 & - \\
			38 & [0], [2], [5] & 90 & [0, 3, 4], [6] & 142 & - & 194 & [0], [6, 7] & 246 & - \\
			39 & - & 91 & [1], [3], [4], [6] & 143 & - & 195 & [1], [6, 7] & 247 & - \\
			40 & - & 92 & [0, 1, 3, 4], [6] & 144 & - & 196 & [0, 1], [6, 7] & 248 & - \\
			41 & [3], [5] & 93 & [2, 3, 6], [4] & 145 & - & 197 & - & 249 & - \\
			42 & [0, 3], [5] & 94 & - & 146 & - & 198 & [0], [2, 6, 7] & 250 & - \\
			43 & [1, 5], [3] & 95 & [1, 2, 3, 6], [4] & 147 & [1], [4, 7] & 199 & - & 251 & - \\
			44 & - & 96 & - & 148 & - & 200 & - & 252 & - \\
			45 & [2, 3], [5] & 97 & - & 149 & [2], [4, 7] & 201 & - & 253 & - \\
			46 & [0, 2, 3], [5] & 98 & [0], [5, 6] & 150 & [2], [0, 4, 7] & 202 & - & 254 & - \\
			47 & - & 99 & - & 151 & [1, 2], [4, 7] & 203 & [1], [3, 6, 7] & 255 & - \\
			48 & - & 100 & - & 152 & - & 204 & - & 256 & - \\
			49 & - & 101 & - & 153 & - & 205 & - & & \\
			50 & - & 102 & [0], [2, 5, 6] & 154 & - & 206 & - & & \\
			51 & - & 103 & - & 155 & [1], [3, 4, 7] & 207 & - & & \\
			52 & - & 104 & - & 156 & - & 208 & - & & \\
			\hline
		\end{tabular}
		\caption{Grouped indices for every case in MC.}
		\label{tab:GroupingMC}
	\end{center}
\end{table}


\newpage




\section{Multilevel percolation analysis}
\label{appendix:MLP}

\begin{figure}
	\minipage{\textwidth}
	\centerline{\includegraphics[width=0.9\linewidth]{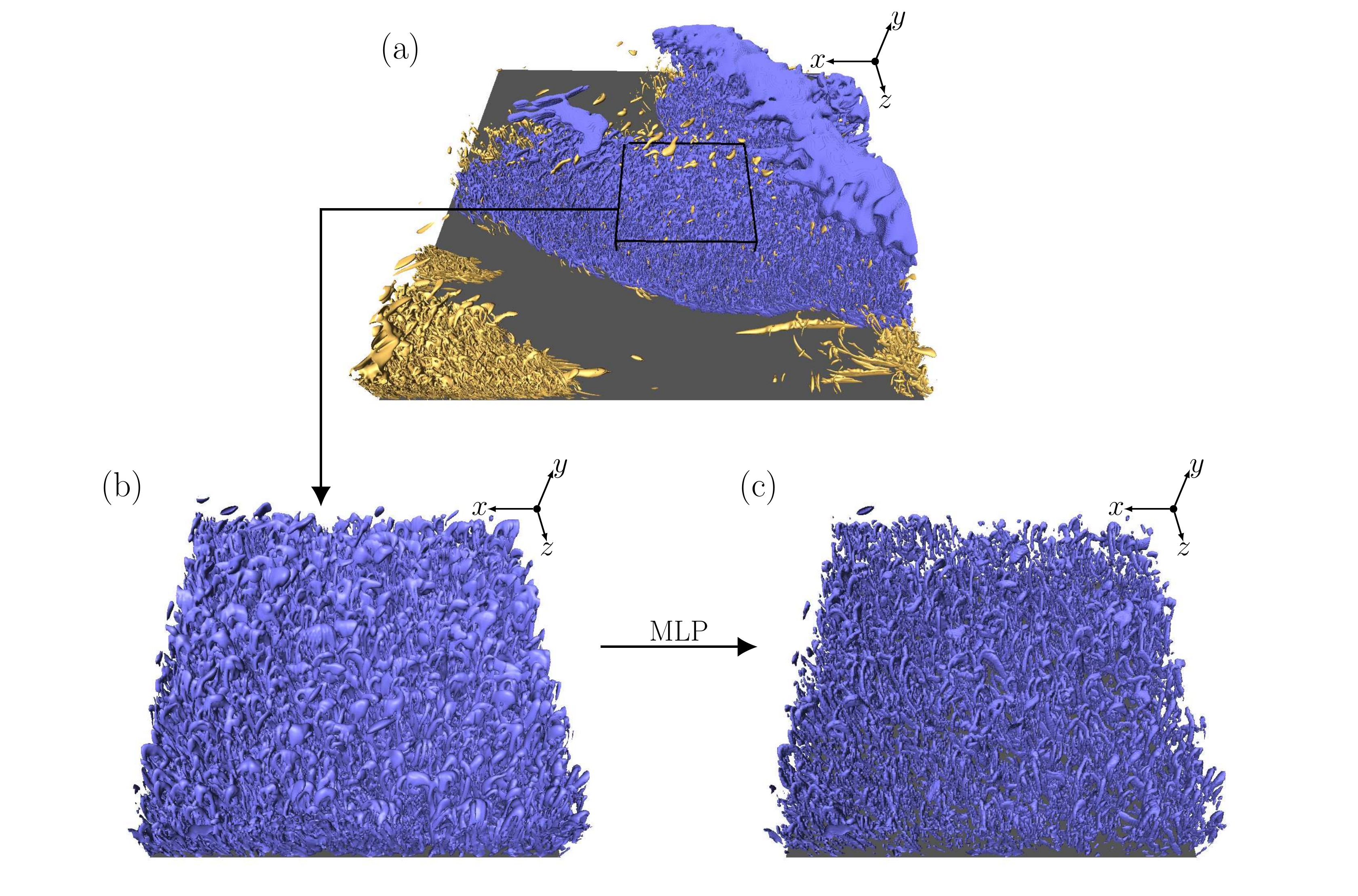}}
	\endminipage
	\caption{$Q-$criterion applied to the case S\_1 is shown here. (a) The highlighted blue region shows a single structure. A small domain from this structure as seen in (b) is subjected to MLP and the result can be seen in (c).}
	\label{fig:MLPNecessity}
\end{figure}

\begin{figure}
	\minipage{\textwidth}
	\centerline{\includegraphics[width=0.9\linewidth]{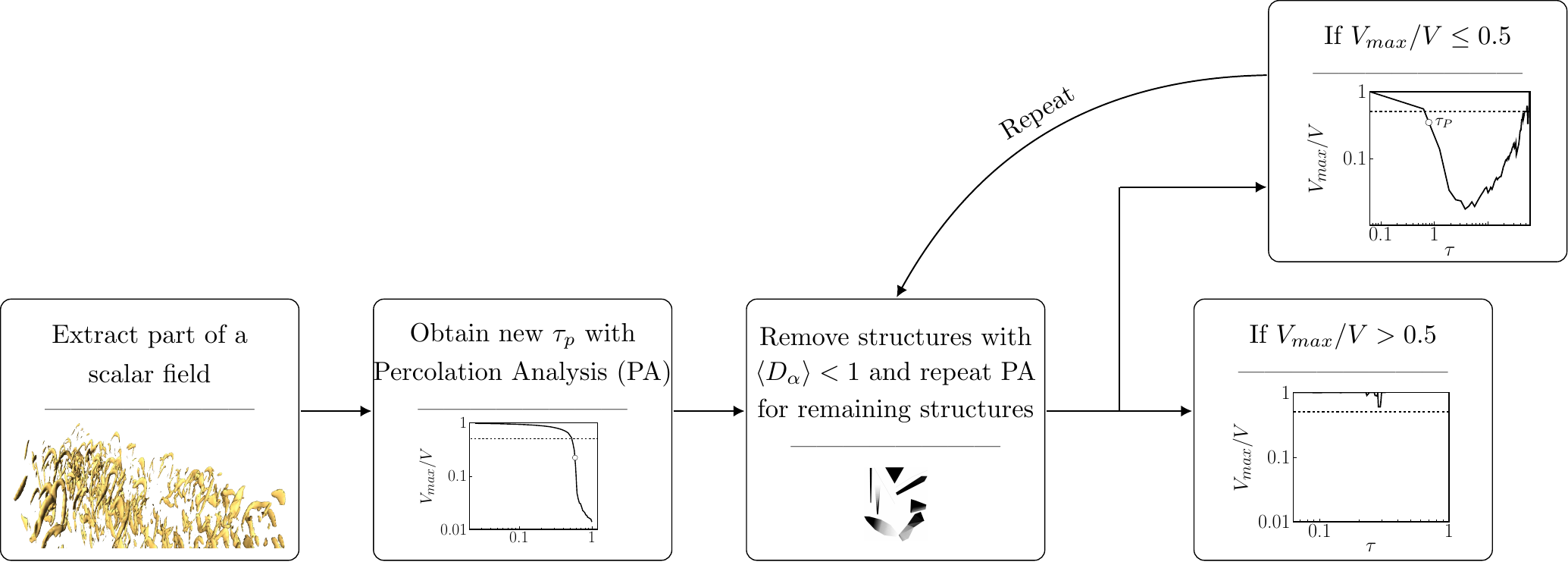}}
	\endminipage
	\caption{A schematic for MLP is shown here. $\tau_{p}$ is the percolation threshold, $\langle D_{\alpha} \rangle$ is the mean fractal dimension, $V_{max}$ is the volume of the largest structure and $V$ is the sum of the volume of all structures in the domain.}
	\label{fig:MLPMethod}
\end{figure}

In this section, we introduce an extension of the percolation analysis technique described in section \ref{subsec:choiceThreshold} to specifically deal with highly stratified flows. Since global intermittency causes large nonturbulent patches to develop within the flow, a global percolation threshold $\tau_{p}$ is not sufficient to educe individual structures. For instance, consider the case S\_1 as shown in Figure \ref{fig:MLPNecessity}(a). Here we are visualizing with the $Q-$criterion indicator and thresholded at $\tau_{p}$. If one proceeds with $\tau_{p}$, the entire region highlighted in blue will be extracted as a single structure and geometrically characterized. We can observe in Figure \ref{fig:MLPNecessity}(b) that this large blue structure is actually composed of numerous smaller structures. In order to appropriately educe these structures, we propose a novel technique called multilevel percolation (MLP) analysis where percolation analysis is applied in a repetitive manner on all structures at $\tau_{p}$ until all large clusters are broken down into individual structures. The procedure for MLP is presented in the schematic \ref{fig:MLPMethod}. It involves the following steps:

\vspace{0.2cm}

\begin{figure}
	\minipage{0.5\textwidth}
	\centerline{\includegraphics[width=\linewidth]{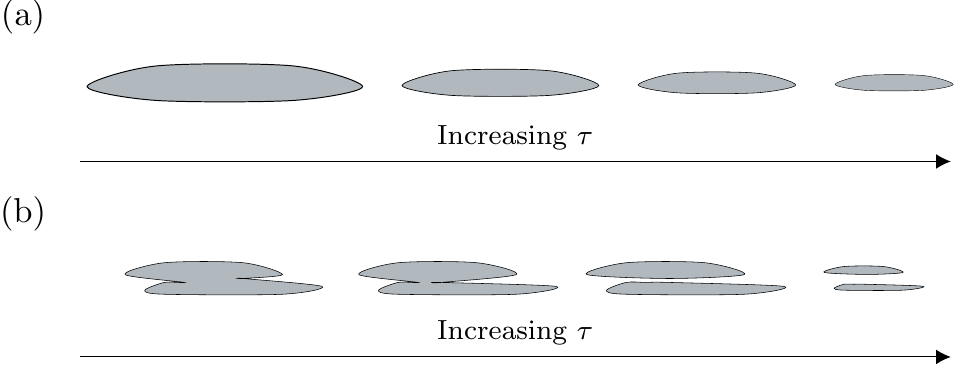}}
	\endminipage
	\minipage{0.5\textwidth}
	\centerline{\includegraphics[width=\linewidth]{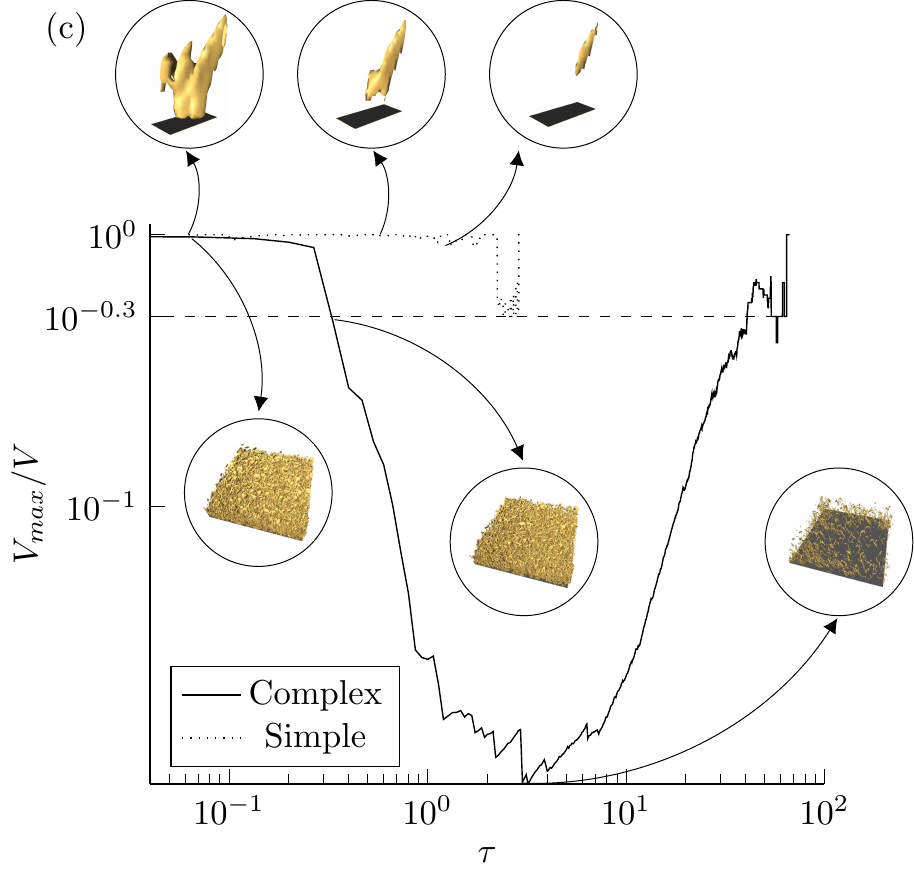}}
	\endminipage
	\caption{(a, b) show the behavior of the structure when the minimum value of the ratio $V_{max}/V$ is chosen as $1$ and $0.5$ respectively. (c) shows a proof-of-concept for MLP applied to a subset of the data from case S\_1. Three points are highlighted here corresponding to the following thresholds $\tau = 0.0625, 0.33455, 3.01098$.}
	\label{fig:MLPIdea}
\end{figure}

\begin{enumerate}
	\itemsep0.2cm
	\item[(a)]\, The global percolation threshold $\tau_{p}$ is computed with the procedure described in section \ref{subsec:choiceThreshold}.
	\item[(b)]\, All structures are extracted. For each structure percolation analysis is repeated and a new value of $\tau_{p}$ is obtained for every structure.
	\item[(c)]\, At the new $\tau_{p}$, all structures are extracted and fractal dimension is computed with the method described in section \ref{subsec:geomCharacterization} (a).
	\item[(d)]\, Structures with $\langle D_{\alpha} \rangle < 1$ are removed. This step is crucial to obtain a noise-free result.
	\item[(e)]\, Finally, the filtered structures are checked against a stopping criterion. This criterion applies percolation analyis on the structure again and checks for the minimum value of the ratio $V_{max}/V$ for the entire range of threshold values. If the minimum value of the ratio $V_{max}/V$ is always above $0.5$, we classify it as a simple structure and it is subjected to no further processing. Essentially the new $\tau_{p}$ calculated in step (c) is associated with the structure i.e. this structure exists at the new $\tau_{p}$. On the other hand, if the minimum value of the ratio $V_{max}/V$ falls below $0.5$ at some threshold, we denote it as a complex structure. A new $\tau_{p}$ is computed and the entire procedure is repeated from step (c).
	\item[(f)]\, This process is repeated on all structures until all complex clusters are broken into individual strutcures. 
\end{enumerate}

%
%
%
%

\vspace{0.2cm}


\noindent \textit{Rationale for the stopping criterion:}

\noindent Adopting a minimum value of the ratio $V_{max}/V$ over the entire threshold range is chosen as a stopping criterion due to its simplicity. One can extract completely individual structures by setting the minimum value of the ratio $V_{max}/V  = 1$. This condition ensures that at any increasing threshold the volume of the largest structure is always equal to the total volume of all structures i.e., there is exactly one structure (see Figure \ref{fig:MLPIdea}(a)).

We note that setting $V_{max}/V  = 1$ is not always practical due to the large computation time requirement. Therefore, we identify an alternate minimum value of $0.5$ for our stopping criterion. When $V_{max}/V  > 0.5$, at most two structures can exist where one has a larger volume than the other (see Figure \ref{fig:MLPIdea}(b)). Two structures with exactly the same volume exist when $V_{max}/V  = 0.5$. The choice of a minimum value other than $1$ is subjective and must be considered only for larger or complex datasets.

A proof-of-concept for a simple and complex structure is illustrated in Figure \ref{fig:MLPIdea}(c). MLP is applied to a small region of the blue structure shown in Figure \ref{fig:MLPNecessity}(a). The size of the domain is $300 \times 90 \times 600$. The minimum value for the ratio was chosen as $0.5$. In its current implementation, the computation took approximately 5 days and the result is shown in Figure \ref{fig:MLPIdea}(c).

\newpage


%
%
\bibliographystyle{unsrtnat} 
\bibliography{referencesFinal} 

\end{document}


\begin{titlepage}
\maketitle
\thispagestyle{empty}
\end{titlepage}

%
%

\noindent \Large C1. Box-counting of 3D scalar fields

\normalsize

\begin{flushleft}
	\noindent C1.1 \textit{Example: Q-criterion}
\end{flushleft}

\begin{figure}[h!]
	\centering
	\includegraphics[width=0.5\linewidth]{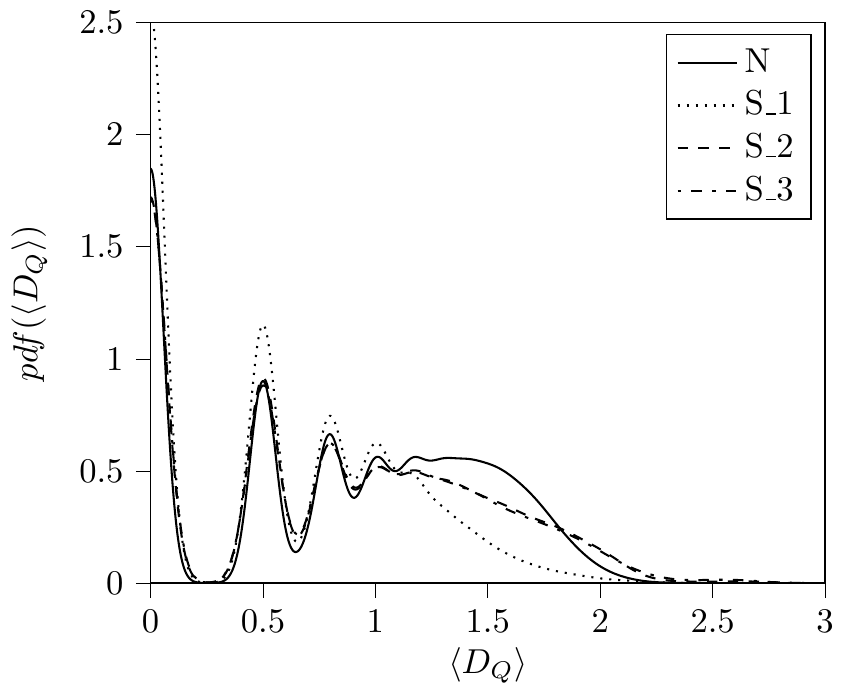}
	\caption{Probability density function of the mean fractal dimension $\langle D_{Q} \rangle$ for $Q-$criterion for all cases. Box-counting is performed on grid B upto $y^{+} < 1550$.}
\end{figure}

\begin{figure}[h!]
	\centering
	\includegraphics[width=\linewidth]{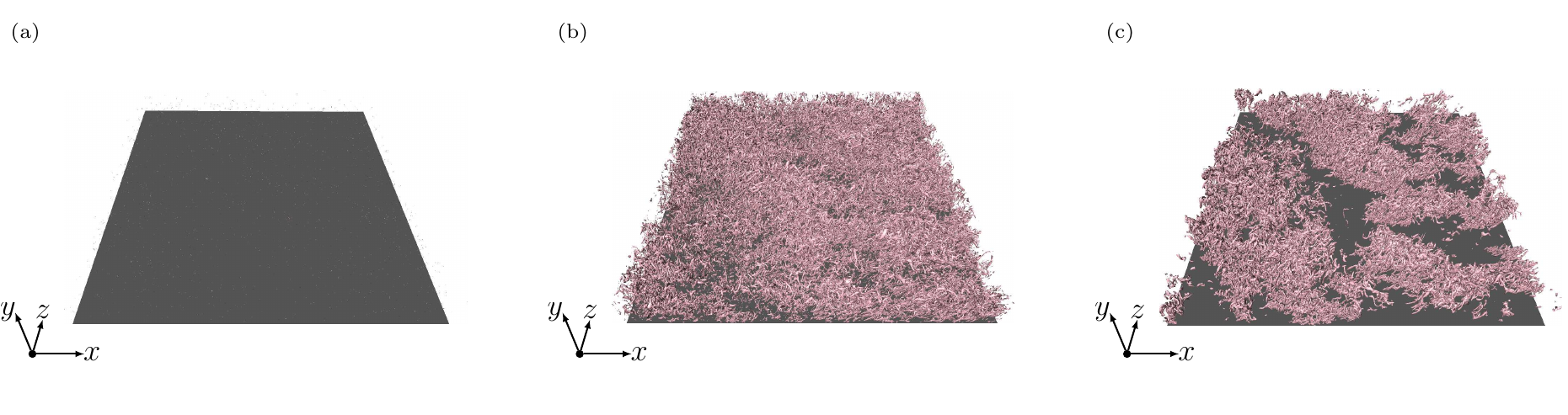}
	\caption{Reconstructed $Q-$criterion scalar fields for (a) $\langle D_{Q} \rangle < 1$, (b) $1 \leq \langle D_{Q} \rangle \leq 2$ and (c) $\langle D_{Q} \rangle > 2$. All figures correspond to case N and visualized for grid B.}
\end{figure}

\begin{figure}[h!]
	\centering
	\includegraphics[width=\linewidth]{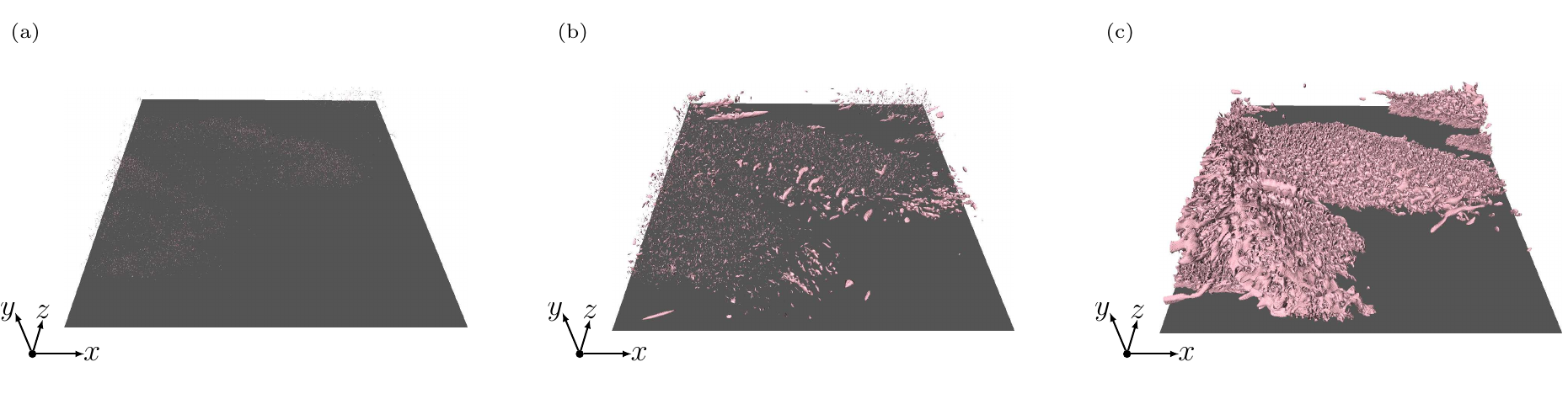}
	\caption{Reconstructed $Q-$criterion scalar fields for (a) $\langle D_{Q} \rangle < 1$, (b) $1 \leq \langle D_{Q} \rangle \leq 2$ and (c) $\langle D_{Q} \rangle > 2$. All figures correspond to case S\_1 and visualized for grid B.}
\end{figure}

\begin{figure}[h!]
	\centering
	\includegraphics[width=\linewidth]{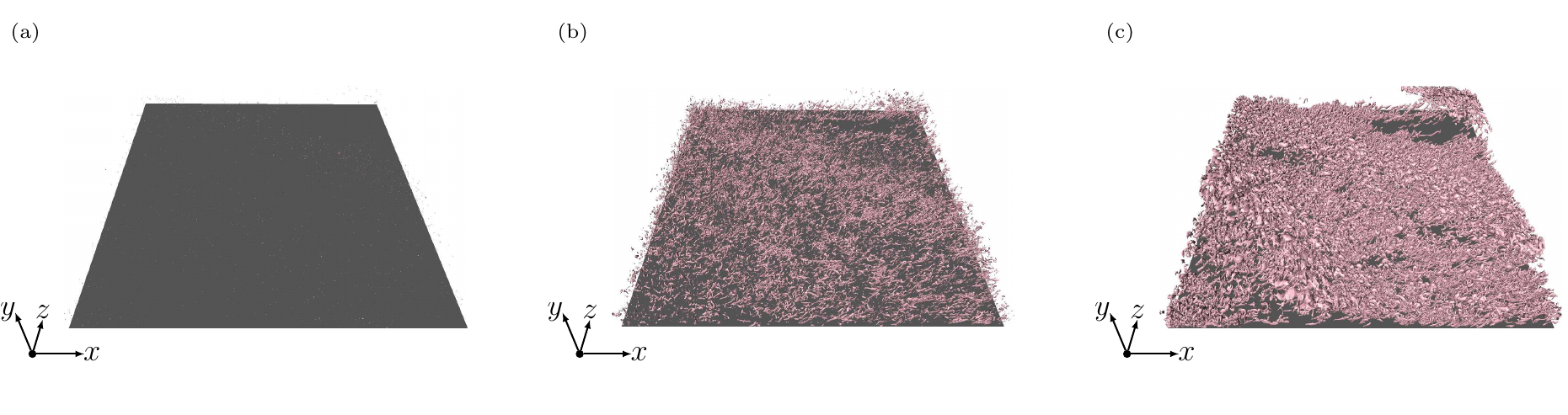}
	\caption{Reconstructed $Q-$criterion scalar fields for (a) $\langle D_{Q} \rangle < 1$, (b) $1 \leq \langle D_{Q} \rangle \leq 2$ and (c) $\langle D_{Q} \rangle > 2$. All figures correspond to case S\_2 and visualized for grid B.}
\end{figure}

\begin{figure}[h!]
	\centering
	\includegraphics[width=\linewidth]{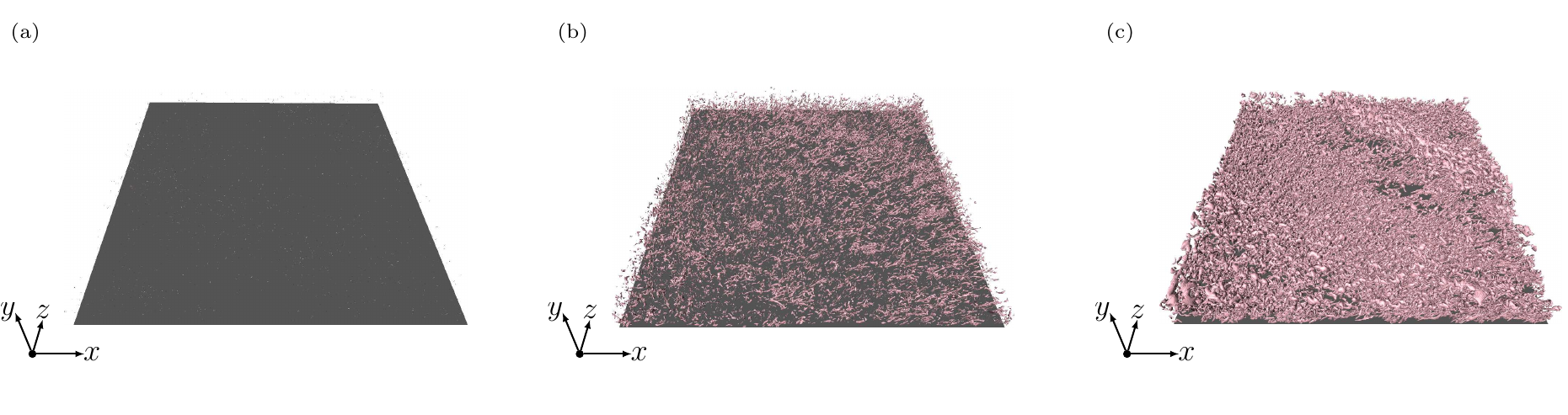}
	\caption{Reconstructed $Q-$criterion scalar fields for (a) $\langle D_{Q} \rangle < 1$, (b) $1 \leq \langle D_{Q} \rangle \leq 2$ and (c) $\langle D_{Q} \rangle > 2$. All figures correspond to case S\_3 and visualized for grid B.}
\end{figure}

\clearpage

\noindent \Large C2. Geometrical characterization

\normalsize

\begin{flushleft}
	\noindent C2.1 \textit{Example: Some common and Robinson structures from case N}
\end{flushleft}

\begin{figure}[h!]
	\minipage{0.33\textwidth}
	\centerline{\includegraphics[width=0.95\linewidth]{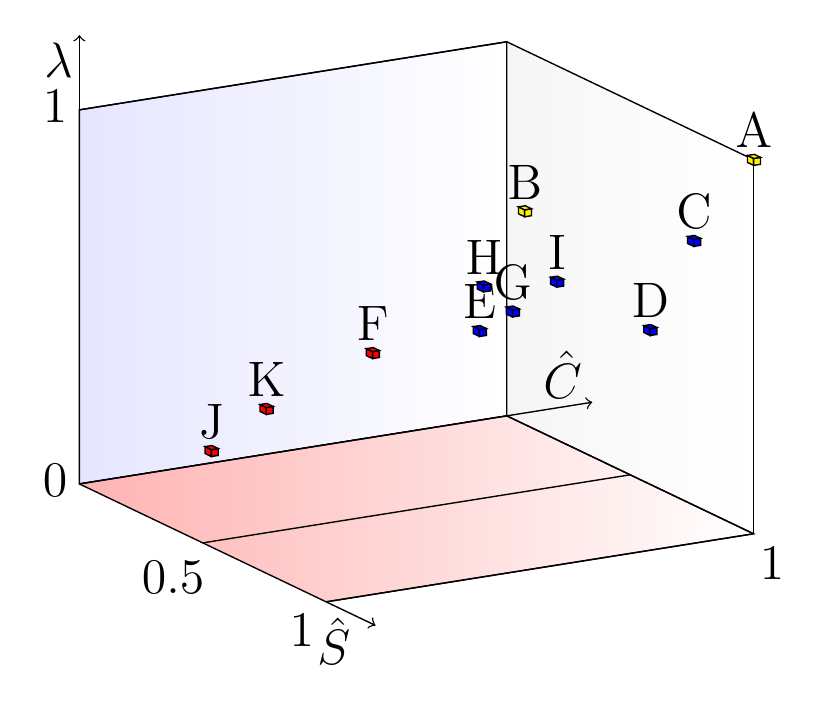}}
	\endminipage\hfill
	\minipage{0.33\textwidth}
	\centerline{\includegraphics[width=0.95\linewidth]{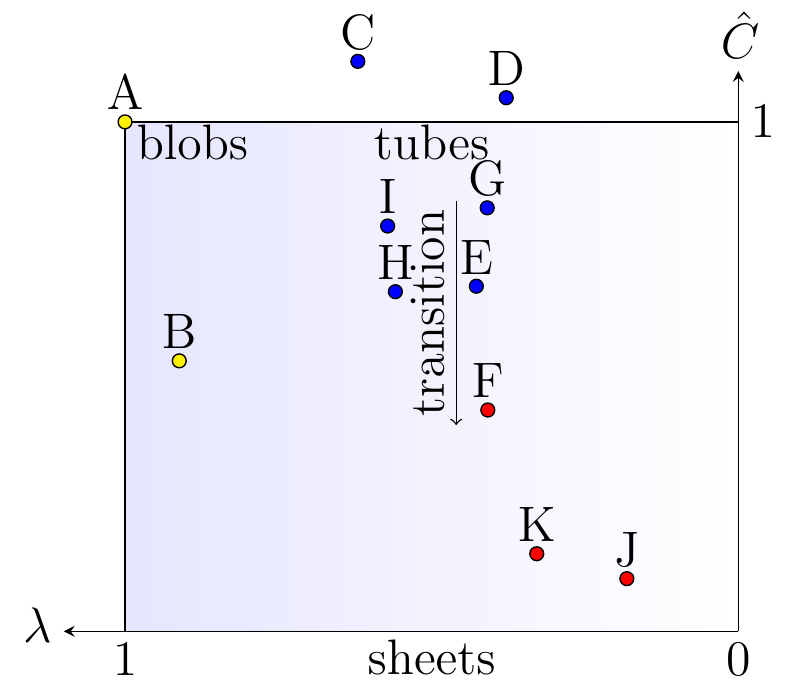}}
	\endminipage\hfill
	\minipage{0.33\textwidth}
	\centerline{\includegraphics[width=0.95\linewidth]{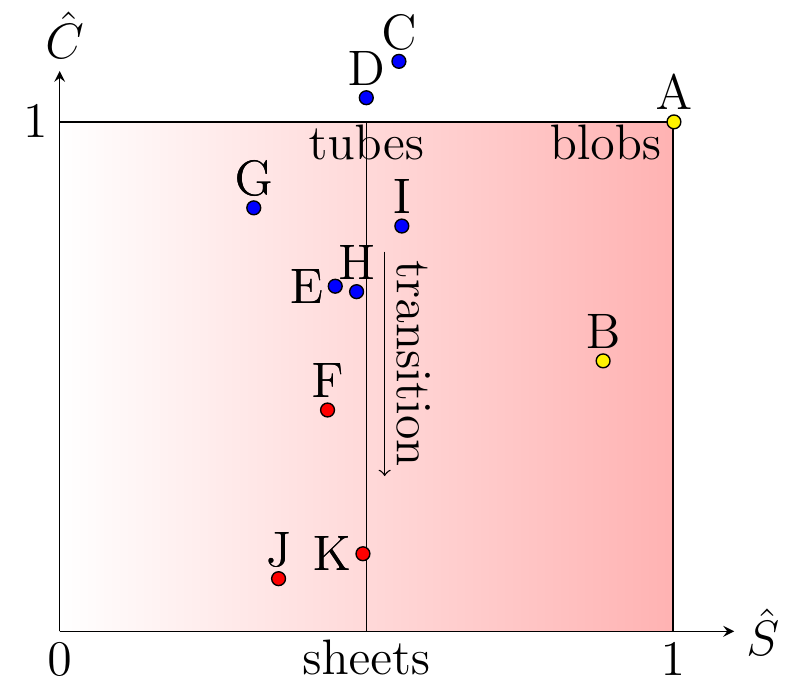}}
	\endminipage\hfill
	\caption{The visualization space for commonly encountered structures and Robinson structures are shown here. A, B, C and D represent a sphere, ellipse, torus and cylinder respectively. E - K are the Robinson structures: high-speed streak, low-speed streak, sweep, ejection, vortex, shear layer and back respectively. The color indicates different clusters as classified by the K-means clustering algorithm. Yellow markers indicate blob-like structures, blue markers indicate tube-like structures and red markers indicate sheet-like structures.}
\end{figure}

\clearpage

\noindent \Large C3. Geometry of structures in the ABL

\normalsize

\begin{flushleft}
	\noindent C3.1 \textit{Low-and high speed streaks in viscous sublayer}
\end{flushleft}

\begin{figure}[h!]
	\centering
	\includegraphics[width=\linewidth]{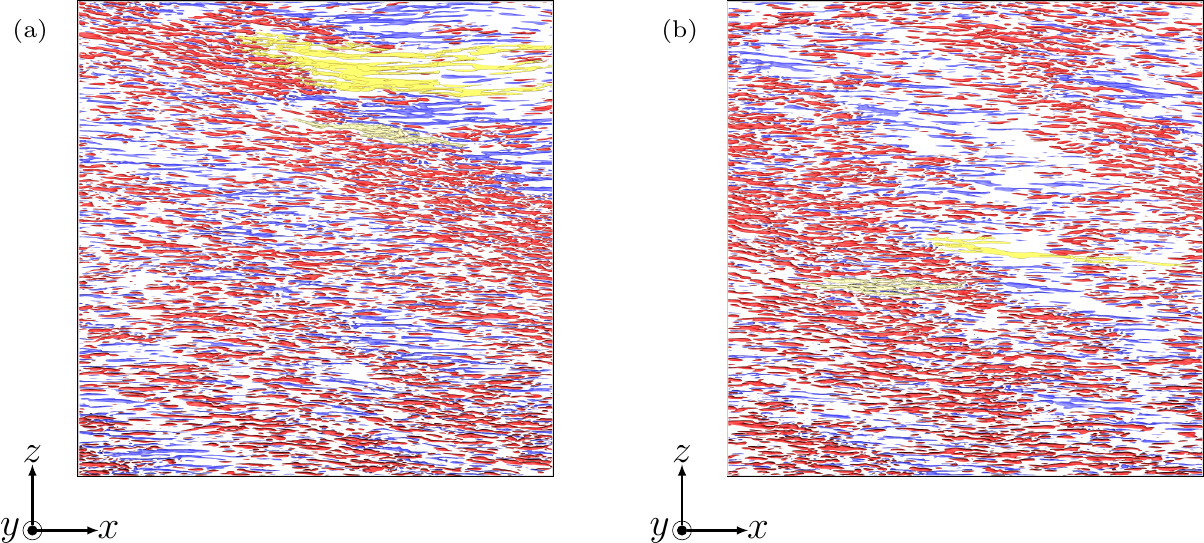}
	\caption{Isosurfaces of $u'$ in the viscous sublayer for grid B. The structure highlighted in light green is the longest low-speed streak within the domain and the one in pale yellow is the longest high-speed streak. (a) corresponds to case S\_2 and (b) to case S\_3. The color specification is according to Table 2 of the main paper where the regions shaded in red correspond to $u' > 0$ and the ones in blue are $u' < 0$.}
\end{figure}

\newpage

\begin{flushleft}
	\noindent C3.2 \textit{Sweeps, ejections and low speed streaks in viscous sublayer}
\end{flushleft}

\begin{figure}[h!]
	\centering
	\includegraphics[width=\linewidth]{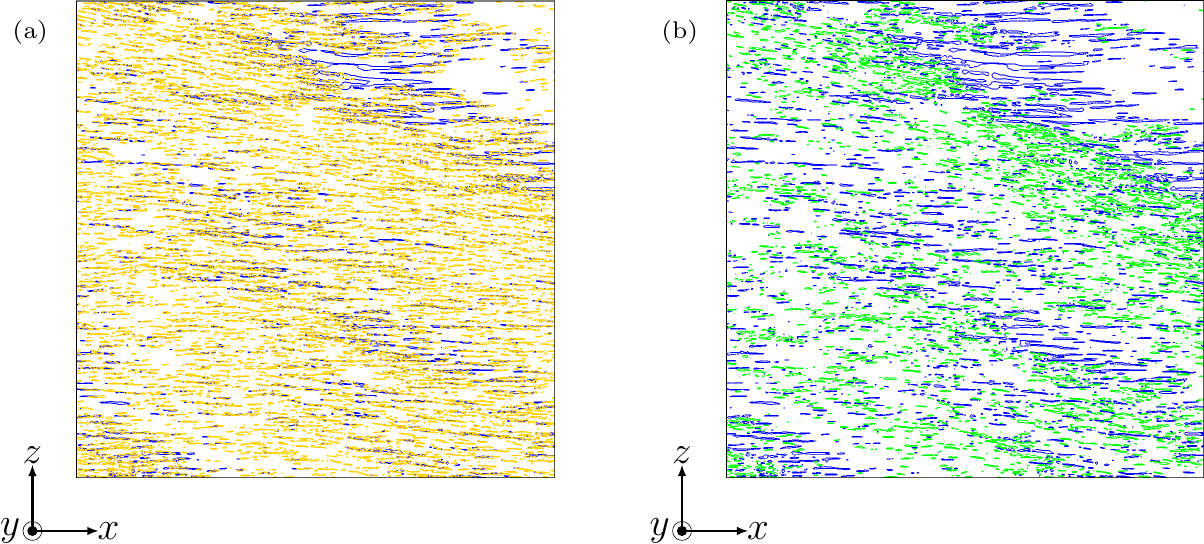}
	\caption{Contour plots of (a) ejections on top of low-speed streaks and (b) sweeps and low-speed streaks at $y^{+} \approx 3.58$ for case S\_2 and grid B. The color specification is according to table 2 of the main paper where low-speed streaks are colored blue, sweeps colored green and ejections colored yellow.}
	\label{fig:sweepFirst}
\end{figure}

\begin{figure}[h!]
	\centering
	\includegraphics[width=\linewidth]{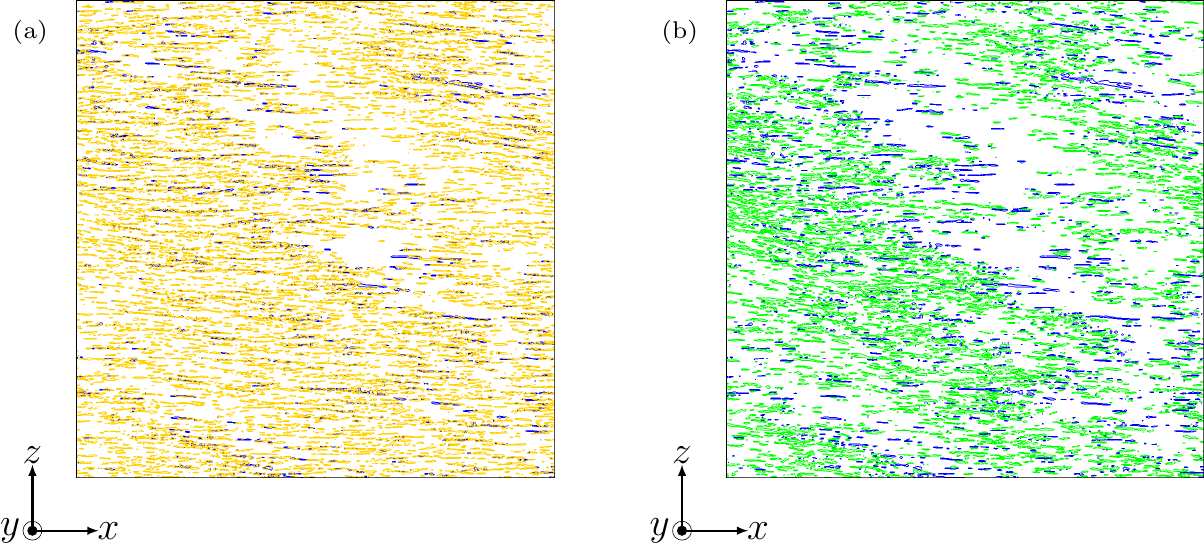}
	\caption{Similar to figure \ref{fig:sweepFirst}, contour plots of (a) ejections on top of low-speed streaks and (b) sweeps and low-speed streaks at $y^{+} \approx 3.58$ are shown for case S\_3 and grid B.}
\end{figure}

\newpage

\begin{flushleft}
	\noindent C3.3 \textit{Pockets with sweeps and ejections in viscous sublayer}
\end{flushleft}

\begin{figure}[h!]
	\centering
	\includegraphics[width=\linewidth]{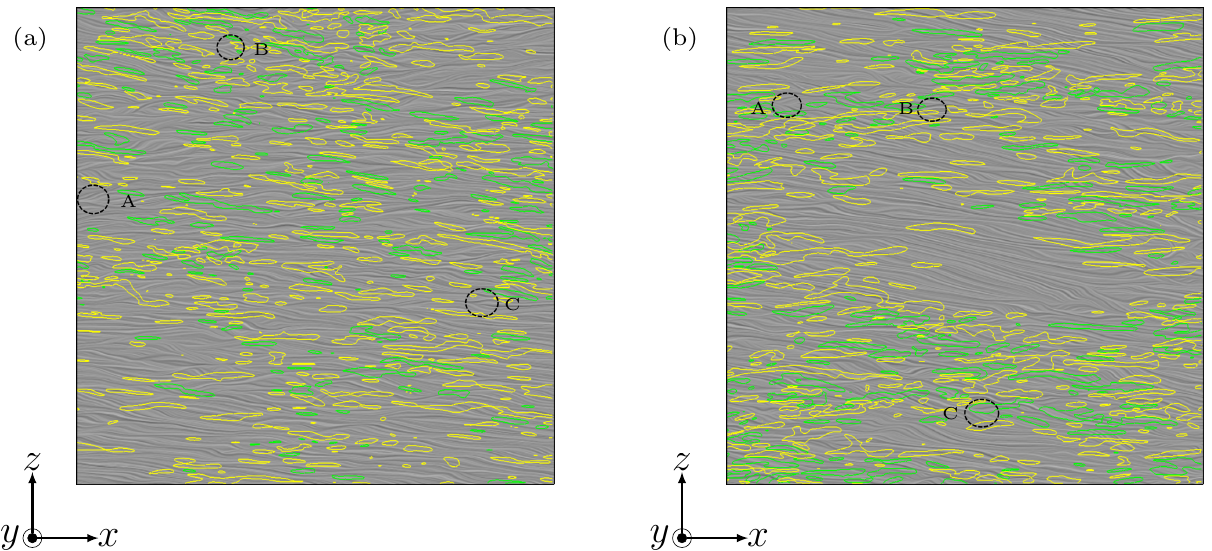}
	\caption{Pockets are shown with diverging streamlines for (a) case S\_2 and (b) case S\_3. Three regions are highlighted in each case which show examples of pocket-like regions. These are overlayed with sweeps and ejections. The combination of pocket-like region and a sweep/ejection can possibly highlight the presence of a vortex above the viscous sublayer. Streamlines are visualized with Line integral convolution for the grid C. The color specification is according to table 2 of the main paper where sweeps are colored green and ejections colored yellow.}
	\label{fig:pocketFirst}
\end{figure}

\newpage

\begin{flushleft}
	\noindent C3.4 \textit{Clustering for sweeps and ejections in the inner layer}
\end{flushleft}
\begin{figure}[h!]
	\minipage{\textwidth}
	\centerline{\includegraphics[width=0.8\linewidth]{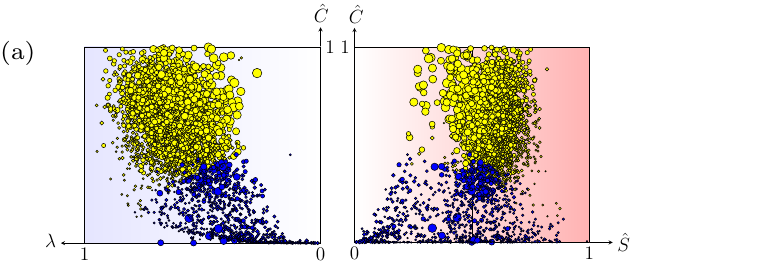}}
	\endminipage\hfill
	\minipage{\textwidth}
	\centerline{\includegraphics[width=0.8\linewidth]{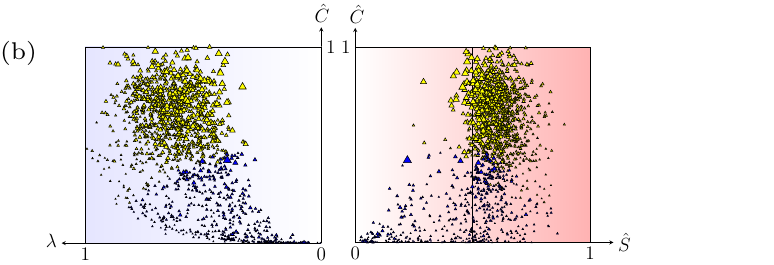}}
	\endminipage\hfill
	\minipage{\textwidth}
	\centerline{\includegraphics[width=0.8\linewidth]{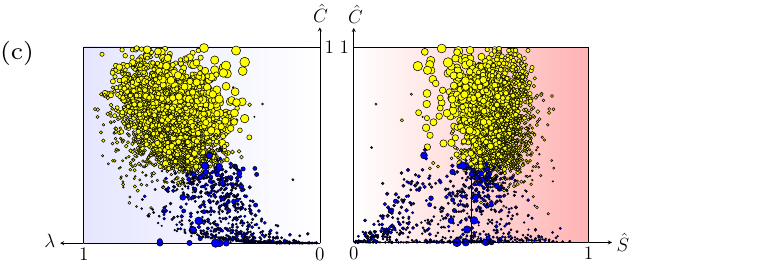}}
	\endminipage\hfill
	\minipage{\textwidth}
	\centerline{\includegraphics[width=0.8\linewidth]{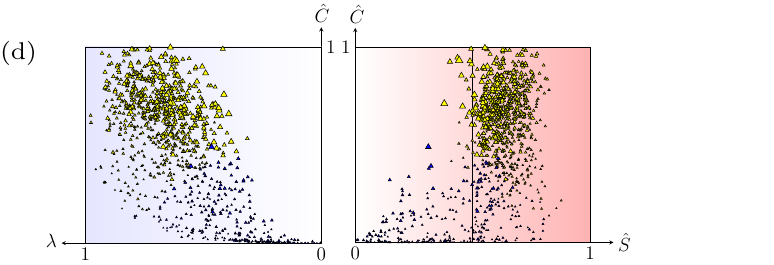}}
	\endminipage\hfill
	\caption{The visualization space for sweeps (a, b) and ejections (c, d) are shown here. Circle markers are used for case N whereas triangles are for case S\_1. Yellow and blue colors correspond to two clusters identified by the K-means algorithm segregating the tube-like and sheet-like structures respectively.}
	\label{fig:ClusteringSweepsEjections}
\end{figure}

\newpage

\begin{flushleft}
	\noindent C3.5 \textit{Visualization of clustered sweeps and ejections in the inner layer}
\end{flushleft}
\begin{figure}[h!]
	\minipage{\textwidth}
	\centerline{\includegraphics[width=0.82\linewidth]{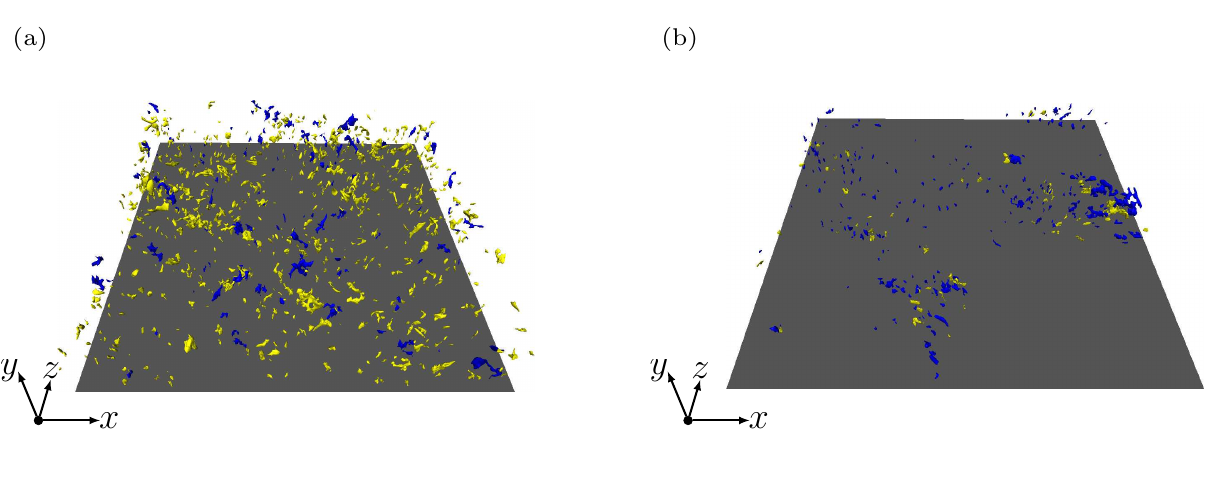}}
	\endminipage\hfill
	\minipage{\textwidth}
	\centerline{\includegraphics[width=0.82\linewidth]{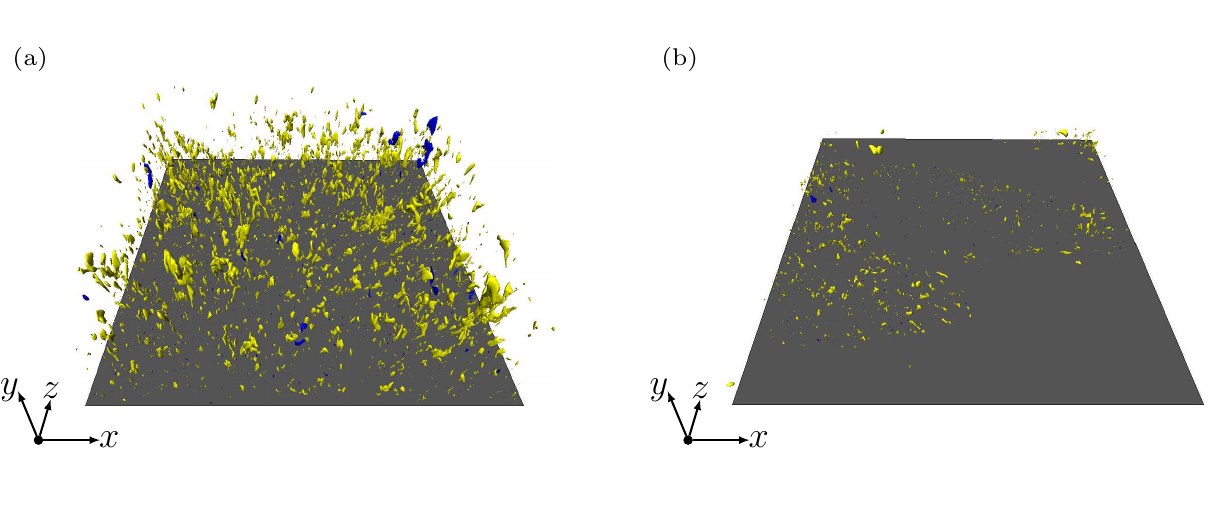}}
	\endminipage\hfill
	\caption{Reconstruction of sweeps (first row) and ejections (second row) shown in the visualization space in section C3.4 for (a) case N and (b) case S\_1. The colors yellow and blue correspond to tube-like and sheet-like structures respectively. }
	\label{fig:VisualizationSweepsEjectionsClustered}
\end{figure}

\newpage

\begin{flushleft}
	\noindent C3.6 \textit{Geometrical characterization for S\_2 and S\_3 in the buffer layer}
\end{flushleft}
\begin{figure}[h!]
	\minipage{\textwidth}
	\centerline{\includegraphics[width=0.7\linewidth]{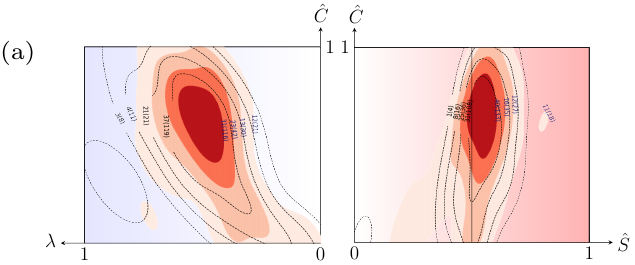}}
	\endminipage\hfill
	\minipage{\textwidth}
	\centerline{\includegraphics[width=0.7\linewidth]{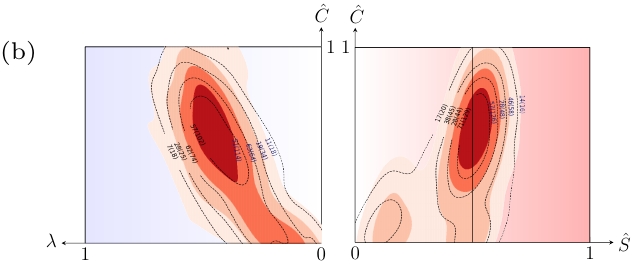}}
	\endminipage\hfill
	\minipage{\textwidth}
	\centerline{\includegraphics[width=0.7\linewidth]{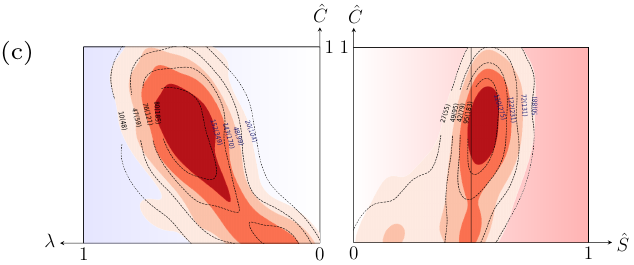}}
	\endminipage\hfill
	\minipage{\textwidth}
	\centerline{\includegraphics[width=0.7\linewidth]{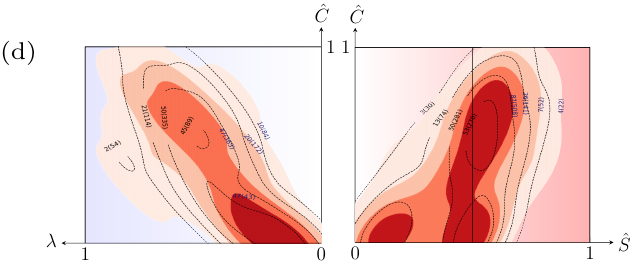}}
	\endminipage\hfill
\end{figure}
\begin{figure}[h!]
	\minipage{\textwidth}
	\centerline{\includegraphics[width=0.7\linewidth]{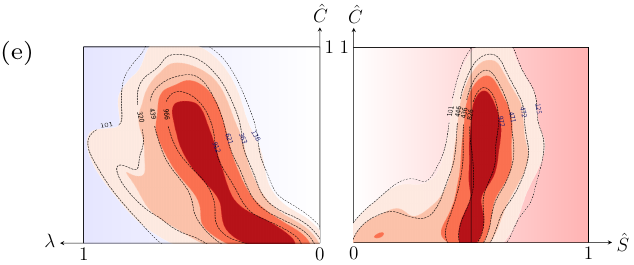}}
	\endminipage\hfill
	\minipage{\textwidth}
	\centerline{\includegraphics[width=0.7\linewidth]{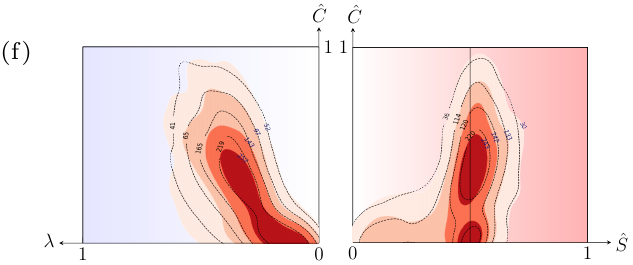}}
	\endminipage\hfill
	\caption{The visualization space for all Robinsion structures except pockets, backs and bulges are shown here with joint pdfs. (a - f) correspond to high-speed streaks, low-speed streaks, sweeps, ejections, vortices and shear layers. Filled contours are used for case S\_2 (with dark shade of red showing the region of high density) whereas unfilled contours with dashed lines are for case S\_3. The number of structures between contours are also indicated - dark blue for case S\_2 and black for case S\_3. Additionally, numbers in parenthesis indicate structures which start within the viscous sublayer and end in the buffer layer. Numbers outside parenthesis indicate structures within the buffer layer itself.}
	\label{fig:GeometryBufferLayerHSSLSS}
\end{figure}

\clearpage
\begin{flushleft}
	\noindent C3.7 \textit{Geometrical characterization for S\_2 and S\_3 in the inner layer}
\end{flushleft}
\begin{figure}[h!]
	\minipage{\textwidth}
	\centerline{\includegraphics[width=0.7\linewidth]{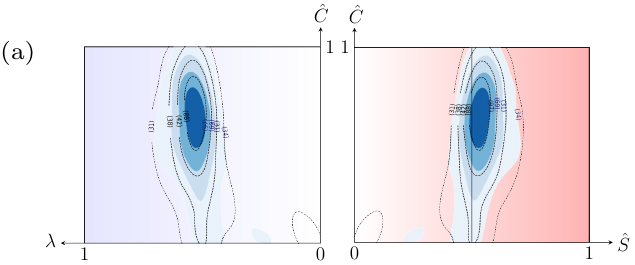}}
	\endminipage\hfill
	\minipage{\textwidth}
	\centerline{\includegraphics[width=0.7\linewidth]{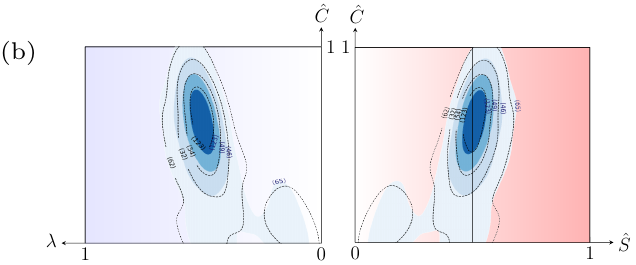}}
	\endminipage\hfill
	\minipage{\textwidth}
	\centerline{\includegraphics[width=0.7\linewidth]{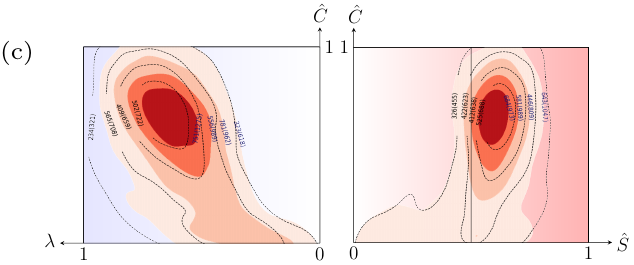}}
	\endminipage\hfill
	\minipage{\textwidth}
	\centerline{\includegraphics[width=0.7\linewidth]{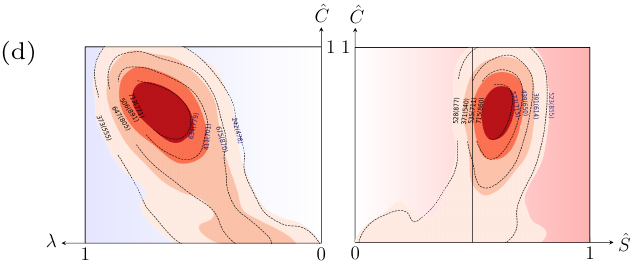}}
	\endminipage\hfill
\end{figure}

\begin{figure}[h!]
	\minipage{\textwidth}
	\centerline{\includegraphics[width=0.7\linewidth]{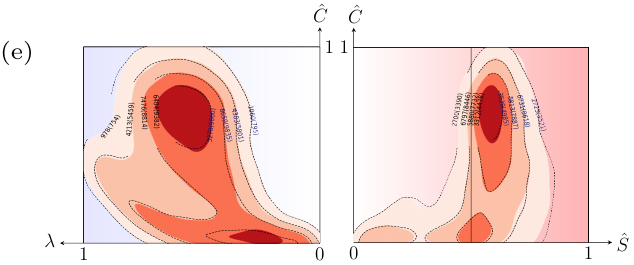}}
	\endminipage\hfill
	\minipage{\textwidth}
	\centerline{\includegraphics[width=0.7\linewidth]{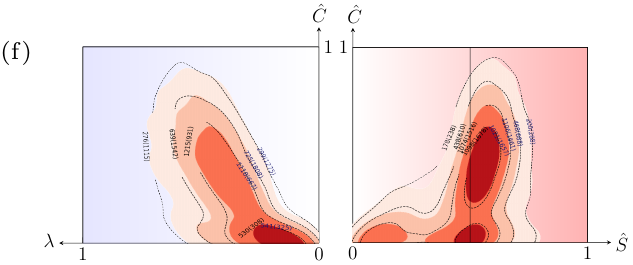}}
	\endminipage\hfill
	\caption{Similar to figure C3.6, the visualization space for high-speed streaks, low-speed streaks, sweeps, ejections, vortices and shear layers (a - f) are shown here. In this case, numbers in parenthesis indicate structures which start within the viscous sublayer and end in the inner layer. Numbers outside parenthesis indicate structures beyond the buffer layer. For (a, b), structures within parenthesis i.e., structures which start within the viscous sublayer and end in the inner layer are shown. This is highlighted with shades of blue.}
	\label{fig:GeometryInnerLayerHSSLSS}
\end{figure}

\clearpage
\begin{flushleft}
	\noindent C3.8 \textit{Geometrical characterization for S\_2 and S\_3 in the outer layer}
\end{flushleft}

\begin{figure}[h!]
	\centering
	\includegraphics[width=0.7\linewidth]{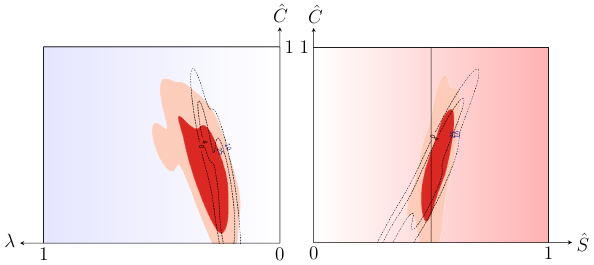}
	\caption{Visualization space for backs for case S\_2 (filled contours) and case S\_3 (unfilled contours). The number of structures between contours are also indicated - dark blue for case S\_2 and black for case S\_3.}
\end{figure}

\begin{figure}[h!]
	\minipage{\textwidth}
	\centerline{\includegraphics[width=1\linewidth]{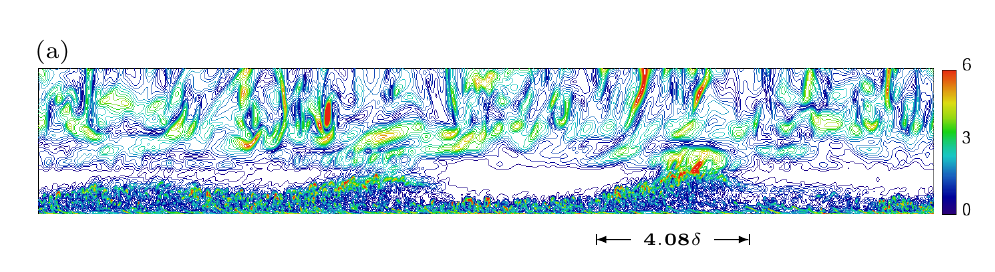}}
	\endminipage\hfill
	\minipage{\textwidth}
	\centerline{\includegraphics[width=1\linewidth]{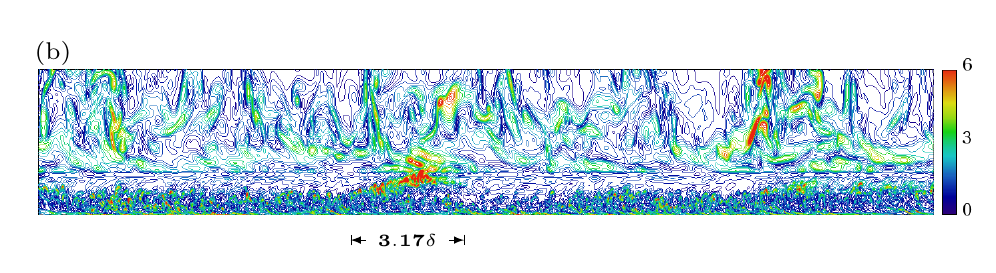}}
	\endminipage\hfill
	\caption{Contours of vorticity magnitude along the XY plane highlighting the presence of $\delta$-scale bulges. Here, the full domain in the streamwise direction is visualized until $y^{+} < 1550$. (a) shows case S\_2 and (b) shows case S\_3. The wall-normal direction, which is $1.75\delta$, has been exaggerated 3 times to show the structures clearly.}
	\label{fig:GeometryOuterLayerBulges}
\end{figure}

\clearpage

\begin{flushleft}
	\noindent C3.9 \textit{Special study: Geometrical characterization of hairpin-like structures}
\end{flushleft}

\begin{figure}[h!]
	\centering
	\includegraphics[width=0.7\linewidth]{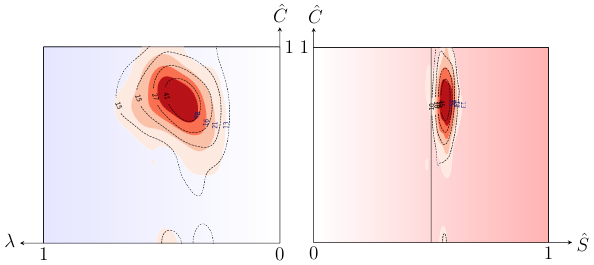}
	\caption{Visualization space with for 100 hairpin-like structures are shown for case S\_2 (filled contours) and case S\_3 (unfilled contours).}
\end{figure}

\clearpage

\begin{flushleft}
	\noindent C3.10 \textit{Visualization of structures in the buffer layer for case N and S\_1}
\end{flushleft}

\begin{figure}[h!]
	\minipage{\textwidth}
	\centerline{\includegraphics[width=0.82\linewidth]{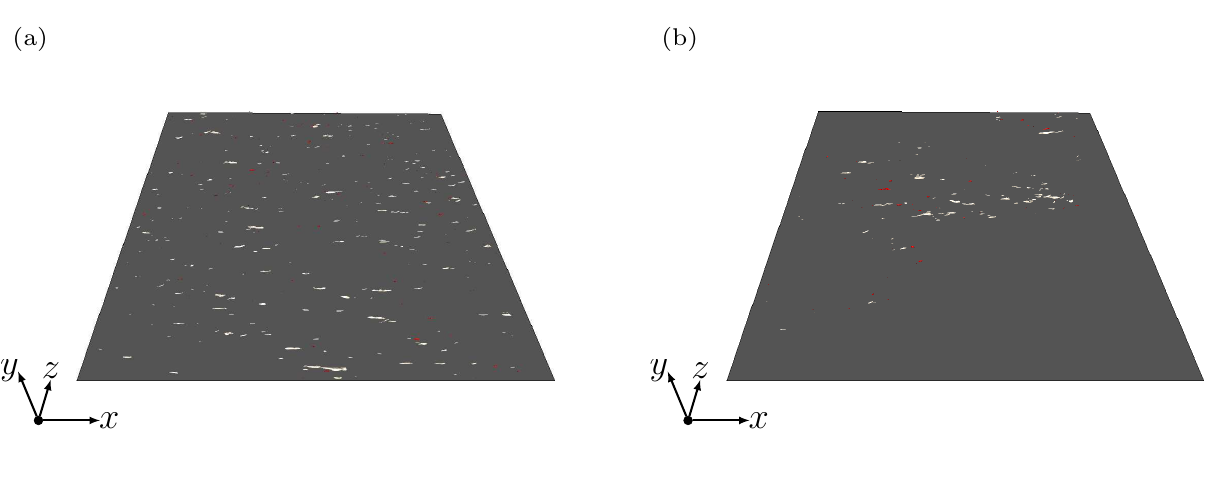}}
	\caption*{Reconstruction of high speed streaks showing structures in the visualization space of figure 10(a) of the main paper for (a) case N and (b) case S\_1.}
	\endminipage\hfill
	\minipage{\textwidth}
	\centerline{\includegraphics[width=0.82\linewidth]{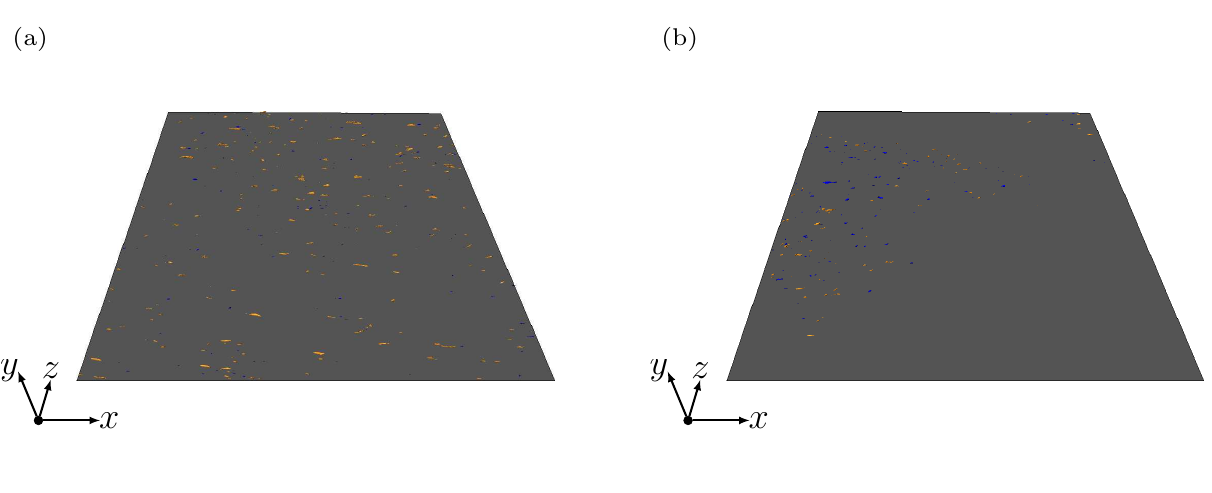}}
	\caption*{Reconstruction of low speed streaks showing structures in the visualization space of figure 10(b) of the main paper for (a) case N and (b) case S\_1.}
	\endminipage\hfill
	\minipage{\textwidth}
	\centerline{\includegraphics[width=0.82\linewidth]{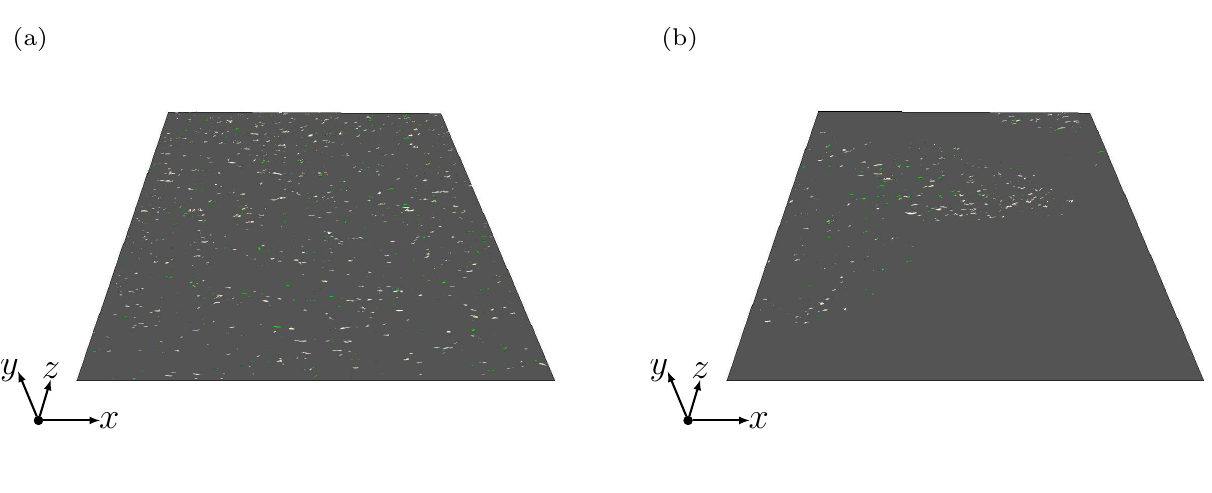}}
	\caption*{Reconstruction of sweeps showing structures in the visualization space of figure 10(c) of the main paper for (a) case N and (b) case S\_1.}
	\endminipage\hfill
\end{figure}

\begin{figure}[h!]
	\minipage{\textwidth}
	\centerline{\includegraphics[width=0.82\linewidth]{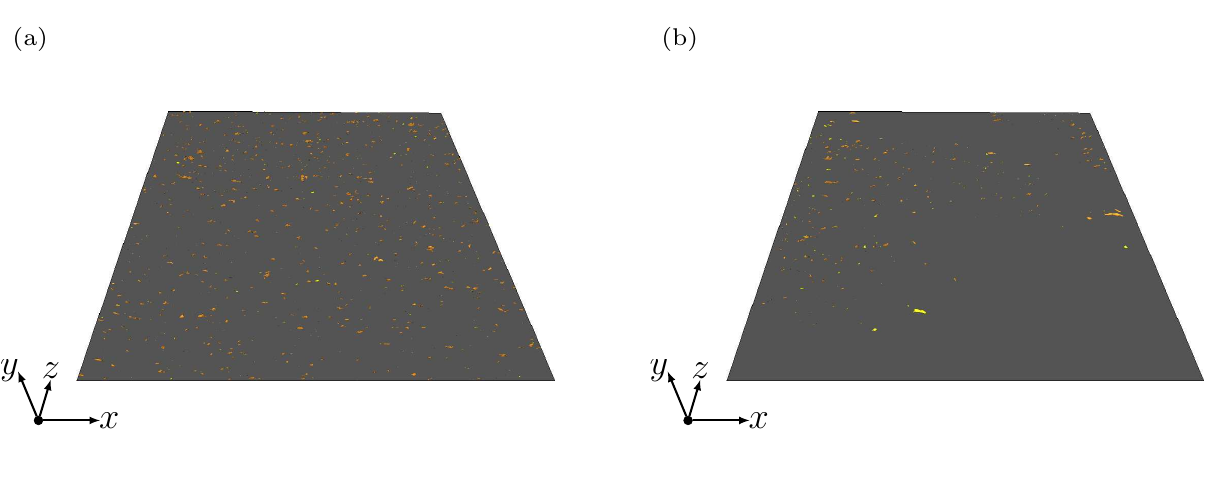}}
	\caption*{Reconstruction of ejections showing structures in the visualization space of figure 10(d) of the main paper for (a) case N and (b) case S\_1.}
	\endminipage\hfill
	\minipage{\textwidth}
	\centerline{\includegraphics[width=0.82\linewidth]{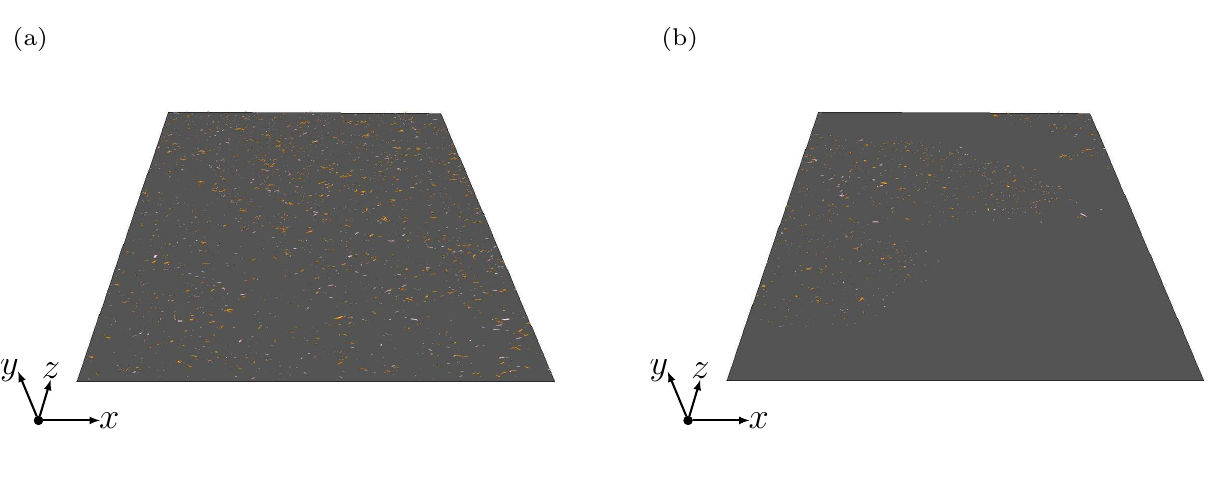}}
	\caption*{Reconstruction of vortices showing structures in the visualization space of figure 10(e) of the main paper for (a) case N and (b) case S\_1.}
	\endminipage\hfill
	\minipage{\textwidth}
	\centerline{\includegraphics[width=0.82\linewidth]{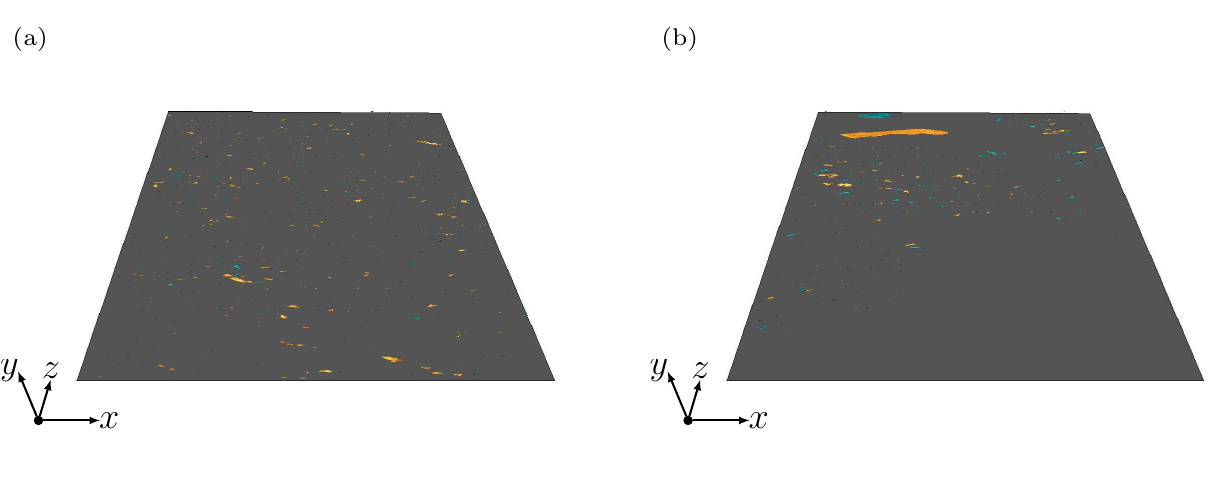}}
	\caption*{Reconstruction of shear layers showing structures in the visualization space of figure 10(f) of the main paper for (a) case N and (b) case S\_1.}
	\endminipage\hfill
	\caption{Structures which start and end within the buffer layer itself are highlighted according to table 2 of the main paper. Other structures which start within the viscous sublayer and end in the buffer layer are shown in orange except for high speed streaks and sweeps are shown in pale yellow.}
\end{figure}

\clearpage

\begin{flushleft}
	\noindent C3.11 \textit{Visualization of structures in the inner layer for case N and S\_1}
\end{flushleft}

\begin{figure}[h!]
	\minipage{\textwidth}
	\centerline{\includegraphics[width=0.82\linewidth]{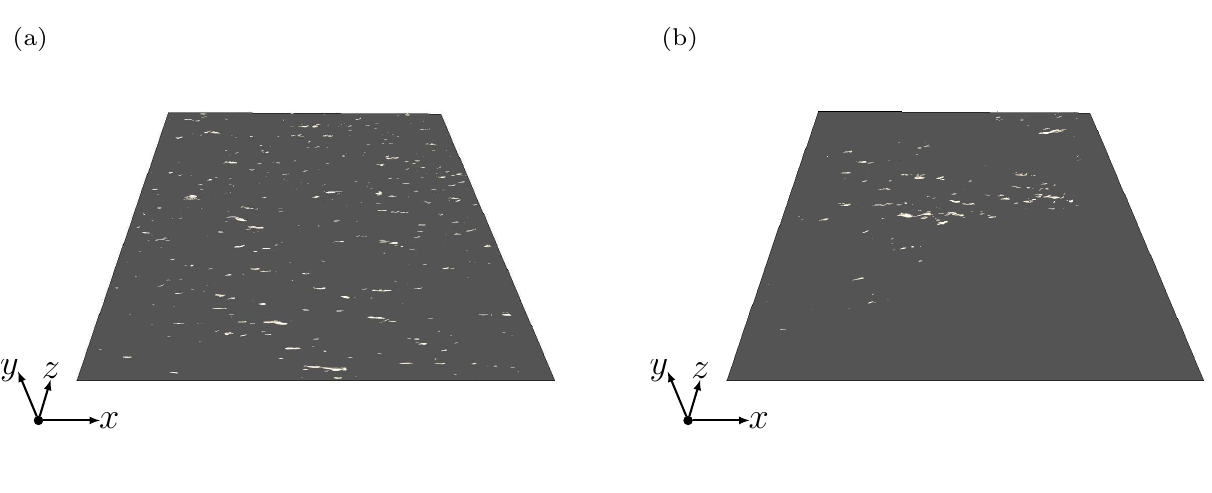}}
	\caption*{Reconstruction of high speed streaks showing structures in the visualization space of figure 11(a) of the main paper for (a) case N and (b) case S\_1.}
	\endminipage\hfill
	\minipage{\textwidth}
	\centerline{\includegraphics[width=0.82\linewidth]{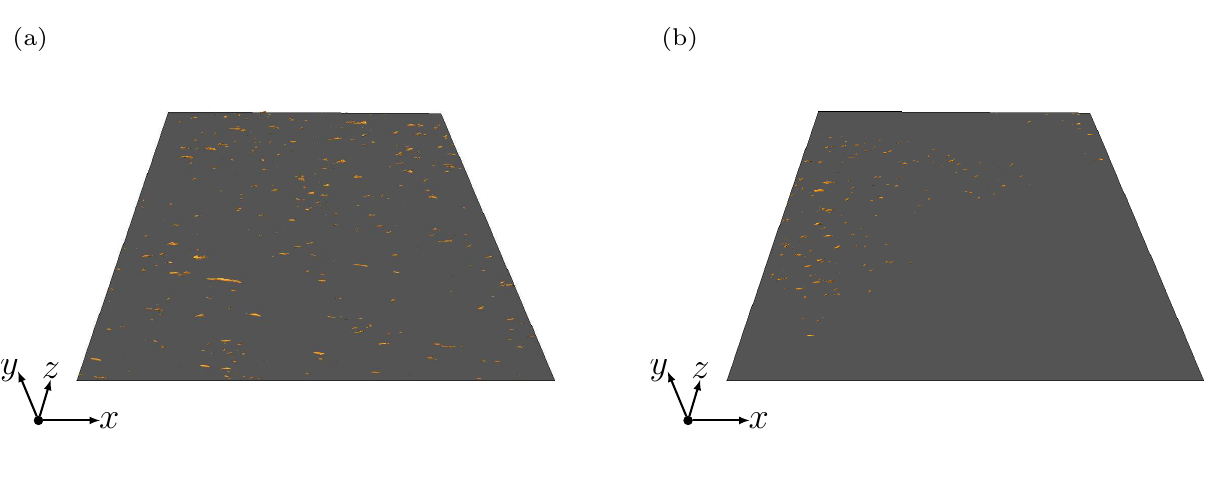}}
	\caption*{Reconstruction of low speed streaks showing structures in the visualization space of figure 11(b) of the main paper for (a) case N and (b) case S\_1.}
	\endminipage\hfill
	\minipage{\textwidth}
	\centerline{\includegraphics[width=0.82\linewidth]{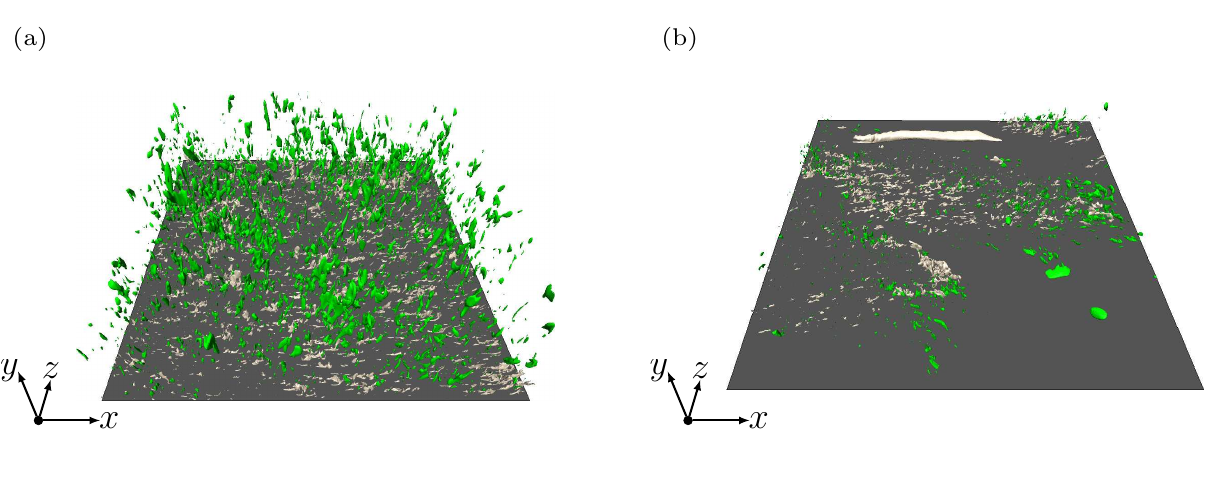}}
	\caption*{Reconstruction of sweeps showing structures in the visualization space of figure 11(c) of the main paper for (a) case N and (b) case S\_1.}
	\endminipage\hfill
\end{figure}

\begin{figure}[h!]
	\minipage{\textwidth}
	\centerline{\includegraphics[width=0.82\linewidth]{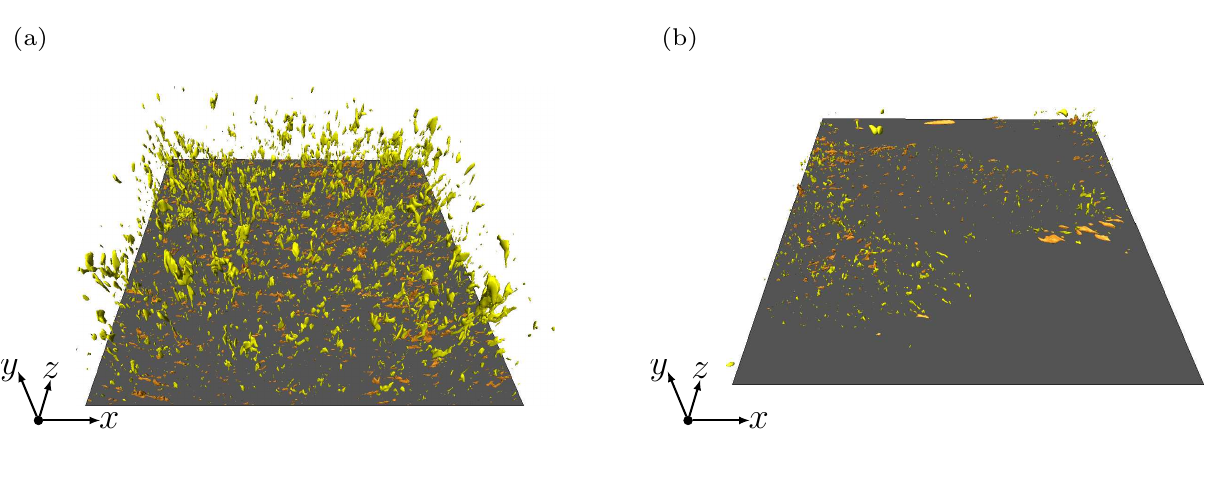}}
	\caption*{Reconstruction of ejections showing structures in the visualization space of figure 11(d) of the main paper for (a) case N and (b) case S\_1.}
	\endminipage\hfill
	\minipage{\textwidth}
	\centerline{\includegraphics[width=0.82\linewidth]{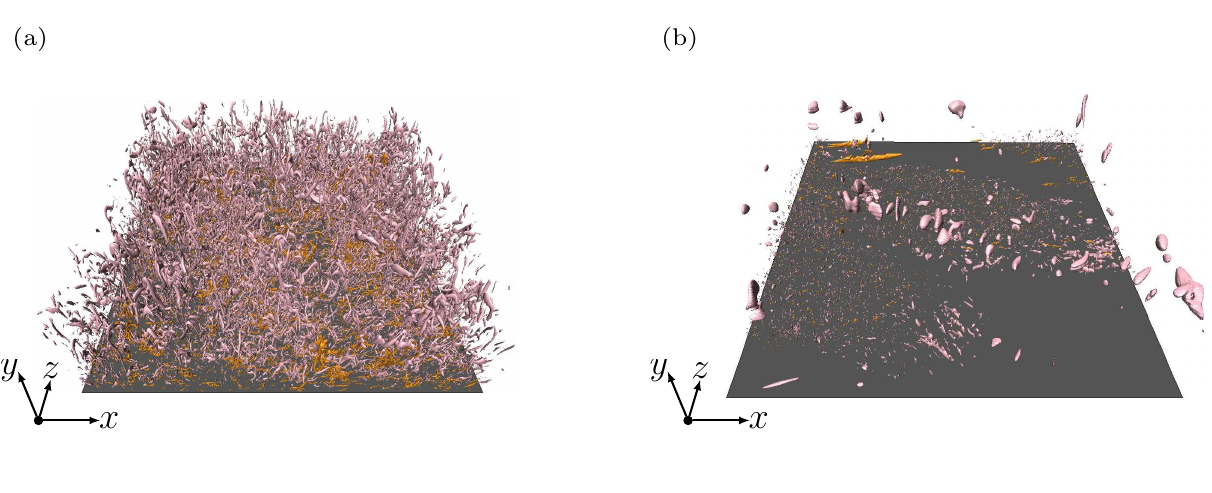}}
	\caption*{Reconstruction of vortices showing structures in the visualization space of figure 11(e) of the main paper for (a) case N and (b) case S\_1.}
	\endminipage\hfill
	\minipage{\textwidth}
	\centerline{\includegraphics[width=0.82\linewidth]{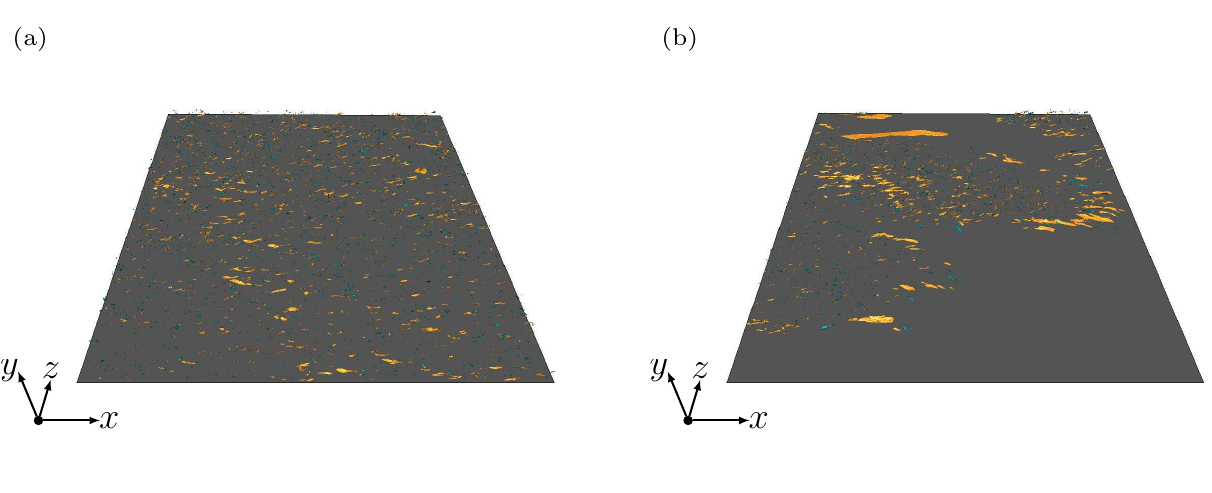}}
	\caption*{Reconstruction of shear layers showing structures in the visualization space of figure 11(f) of the main paper for (a) case N and (b) case S\_1.}
	\endminipage\hfill
	\caption{Structures which start and end within the inner layer itself are highlighted according to table 2 of the main paper. Other structures which start within the viscous sublayer and end in the buffer layer are shown in orange except for high speed streaks and sweeps are shown in pale yellow.}
\end{figure}

\clearpage

\begin{flushleft}
	\noindent C3.12 \textit{Visualization of structures in the buffer layer for case S\_2 and S\_3}
\end{flushleft}

\begin{figure}[h!]
	\minipage{\textwidth}
	\centerline{\includegraphics[width=0.82\linewidth]{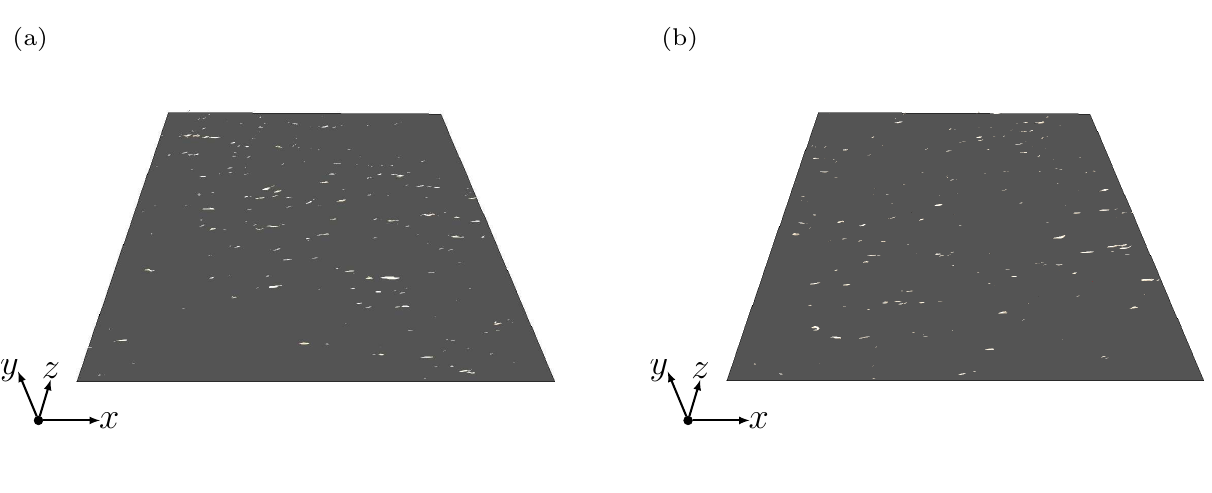}}
	\caption*{Reconstruction of high speed streaks showing structures in the visualization space C3.6(a) for (a) case S\_2 and (b) case S\_3.}
	\endminipage\hfill
	\minipage{\textwidth}
	\centerline{\includegraphics[width=0.82\linewidth]{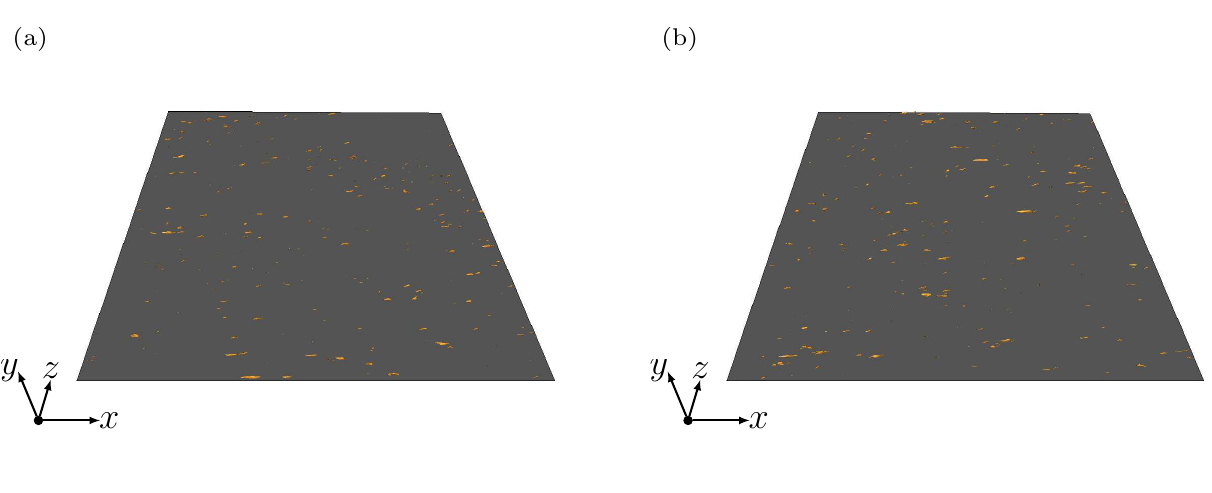}}
	\caption*{Reconstruction of low speed streaks showing structures in the visualization space C3.6(b) for (a) case S\_2 and (b) case S\_3.}
	\endminipage\hfill
	\minipage{\textwidth}
	\centerline{\includegraphics[width=0.82\linewidth]{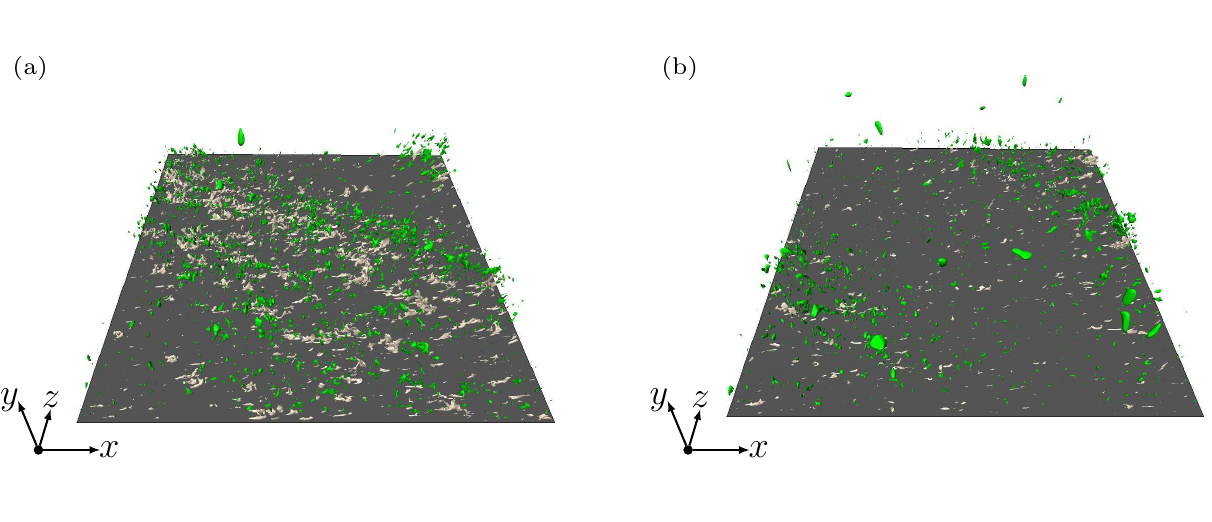}}
	\caption*{Reconstruction of sweeps showing structures in the visualization space C3.6(c) for (a) case S\_2 and (b) case S\_3.}
	\endminipage\hfill
\end{figure}

\begin{figure}[h!]
	\minipage{\textwidth}
	\centerline{\includegraphics[width=0.82\linewidth]{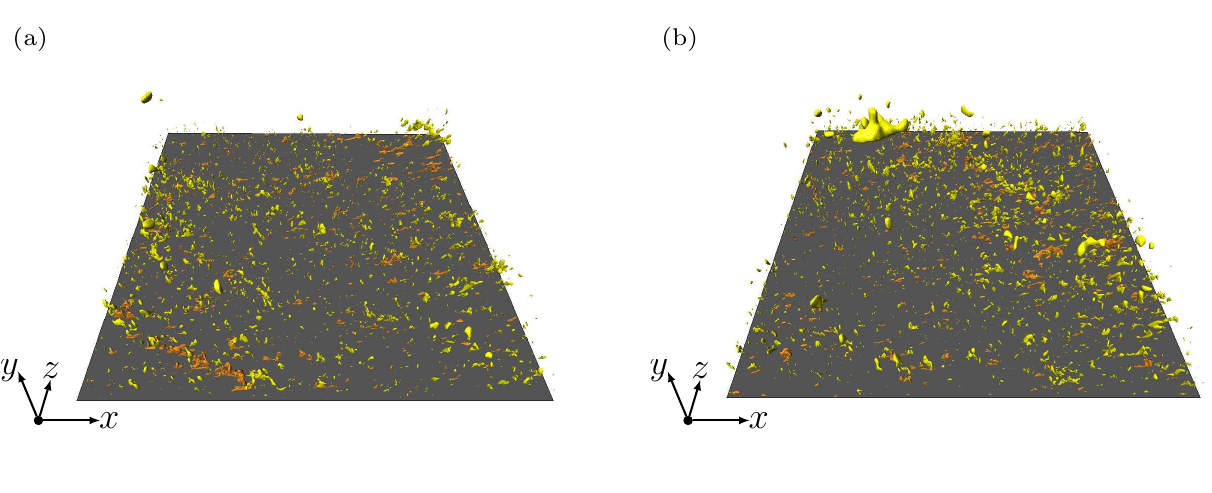}}
	\caption*{Reconstruction of ejections showing structures in the visualization space C3.6(d) for (a) case S\_2 and (b) case S\_3.}
	\endminipage\hfill
	\minipage{\textwidth}
	\centerline{\includegraphics[width=0.82\linewidth]{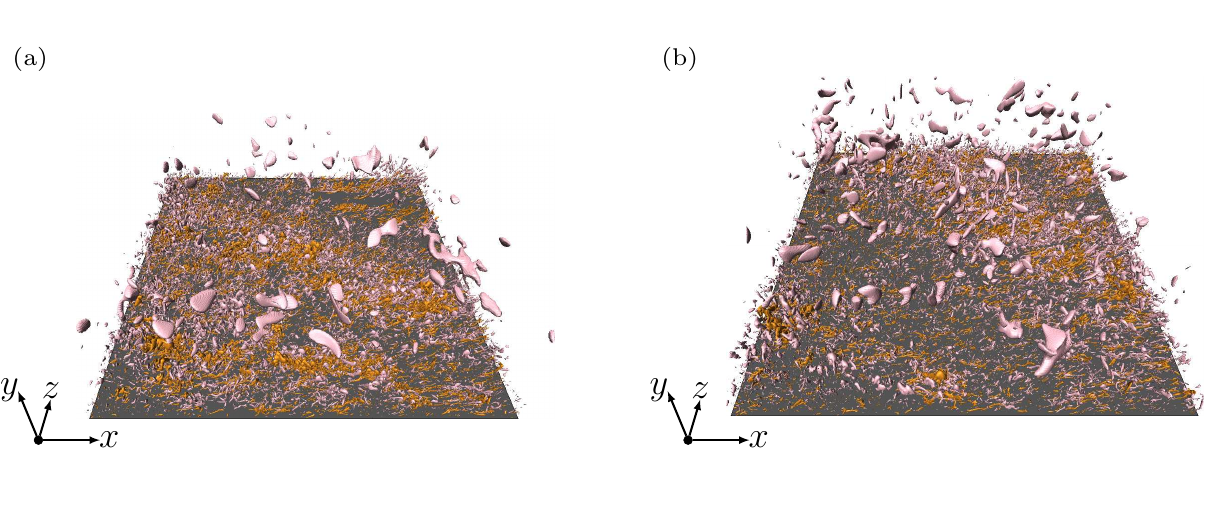}}
	\caption*{Reconstruction of vortices showing structures in the visualization space C3.6(e) for (a) case S\_2 and (b) case S\_3.}
	\endminipage\hfill
	\minipage{\textwidth}
	\centerline{\includegraphics[width=0.82\linewidth]{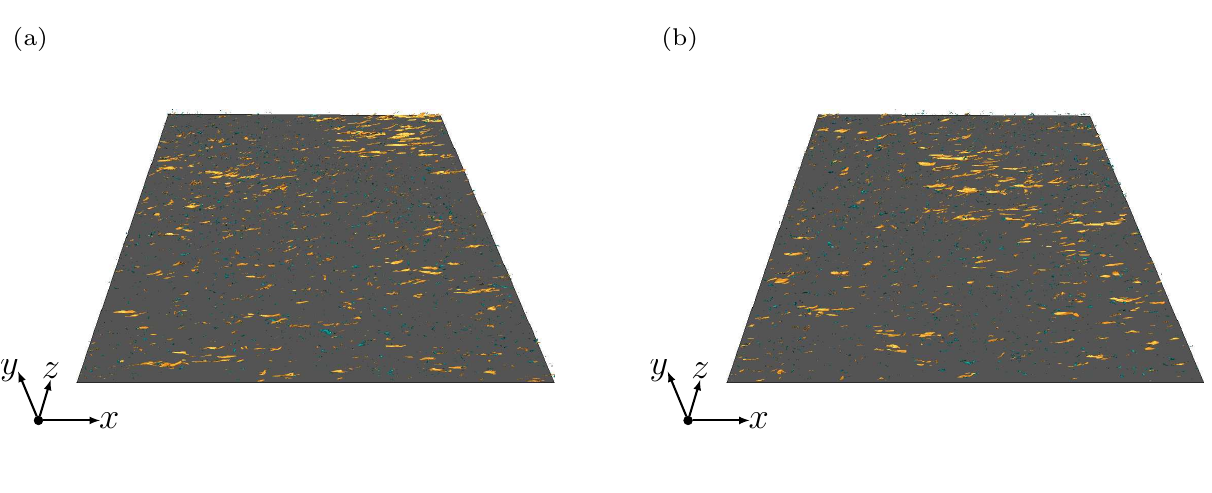}}
	\caption*{Reconstruction of shear layers showing structures in the visualization space C3.6(f) for (a) case S\_2 and (b) case S\_3.}
	\endminipage\hfill
	\caption{Structures which start and end within the inner layer itself are highlighted according to table 2 of the main paper. Other structures which start within the viscous sublayer and end in the buffer layer are shown in orange except for high speed streaks and sweeps are shown in pale yellow.}
\end{figure}

\clearpage

\begin{flushleft}
	\noindent C3.13 \textit{Visualization of structures in the inner layer for case S\_2 and S\_3}
\end{flushleft}

\begin{figure}[h!]
	\minipage{\textwidth}
	\centerline{\includegraphics[width=0.82\linewidth]{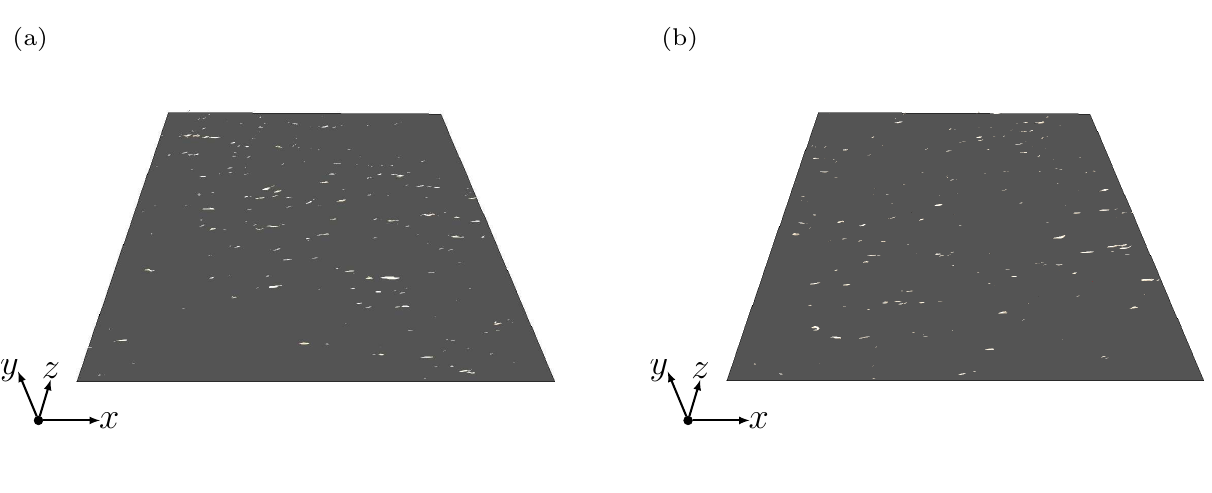}}
	\caption*{Reconstruction of high speed streaks showing structures in the visualization space C3.7(a) for (a) case S\_2 and (b) case S\_3.}
	\endminipage\hfill
	\minipage{\textwidth}
	\centerline{\includegraphics[width=0.82\linewidth]{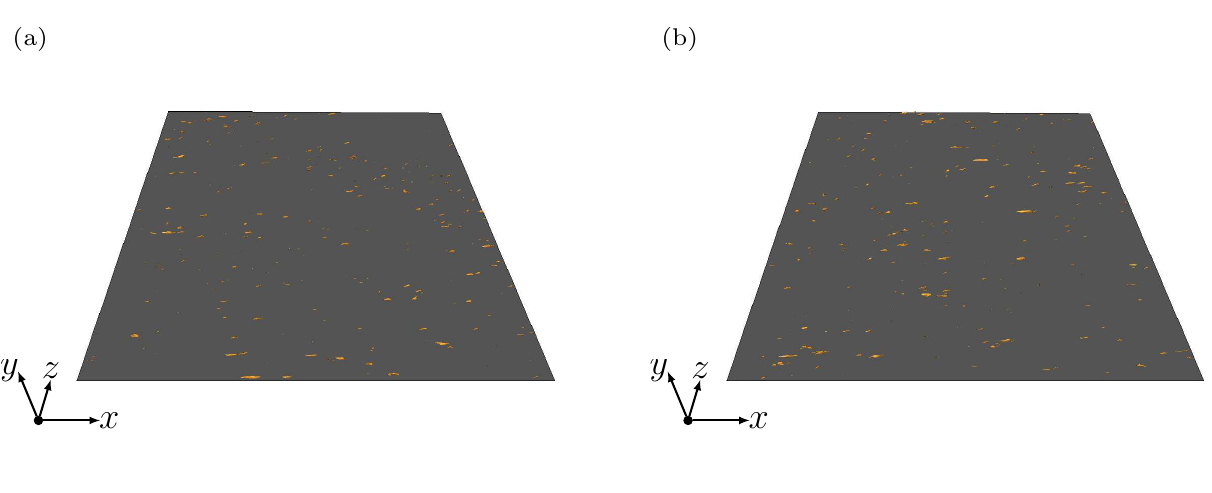}}
	\caption*{Reconstruction of low speed streaks showing structures in the visualization space C3.7(b) for (a) case S\_2 and (b) case S\_3.}
	\endminipage\hfill
	\minipage{\textwidth}
	\centerline{\includegraphics[width=0.82\linewidth]{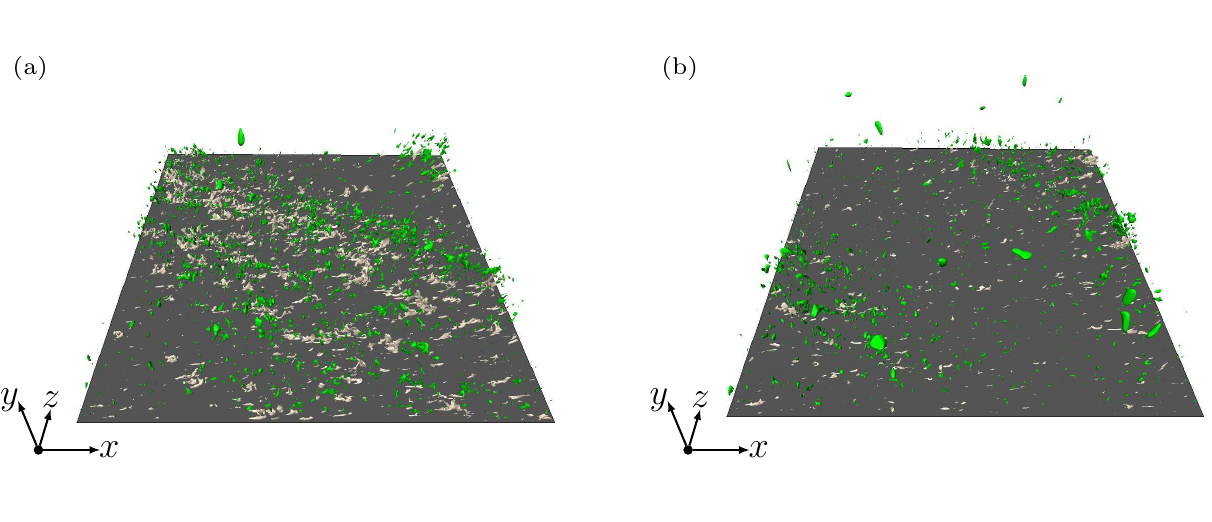}}
	\caption*{Reconstruction of sweeps showing structures in the visualization space C3.7(c) for (a) case S\_2 and (b) case S\_3.}
	\endminipage\hfill
\end{figure}

\begin{figure}[h!]
	\minipage{\textwidth}
	\centerline{\includegraphics[width=0.82\linewidth]{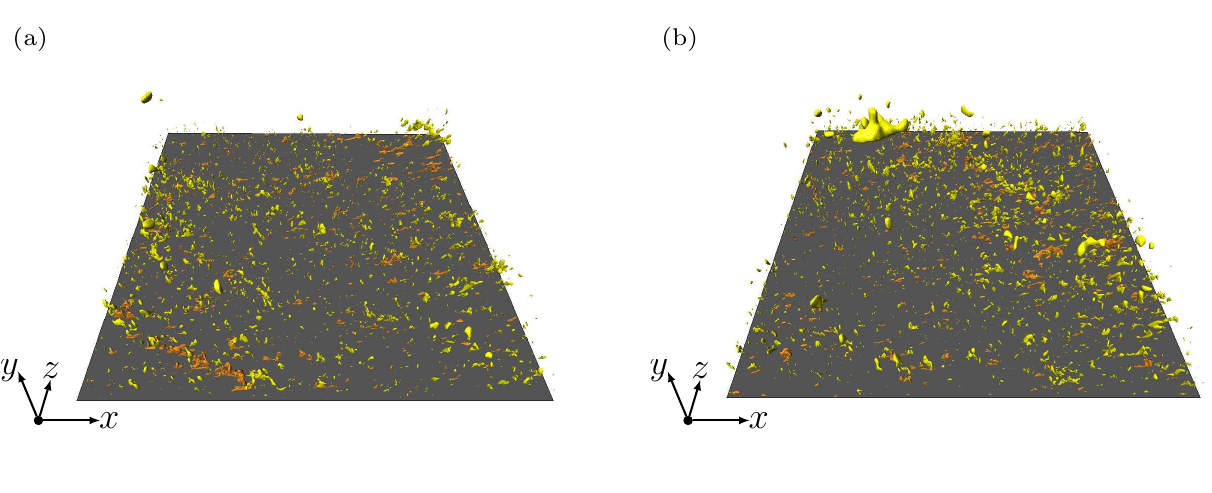}}
	\caption*{Reconstruction of ejections showing structures in the visualization space C3.7(d) for (a) case S\_2 and (b) case S\_3.}
	\endminipage\hfill
	\minipage{\textwidth}
	\centerline{\includegraphics[width=0.82\linewidth]{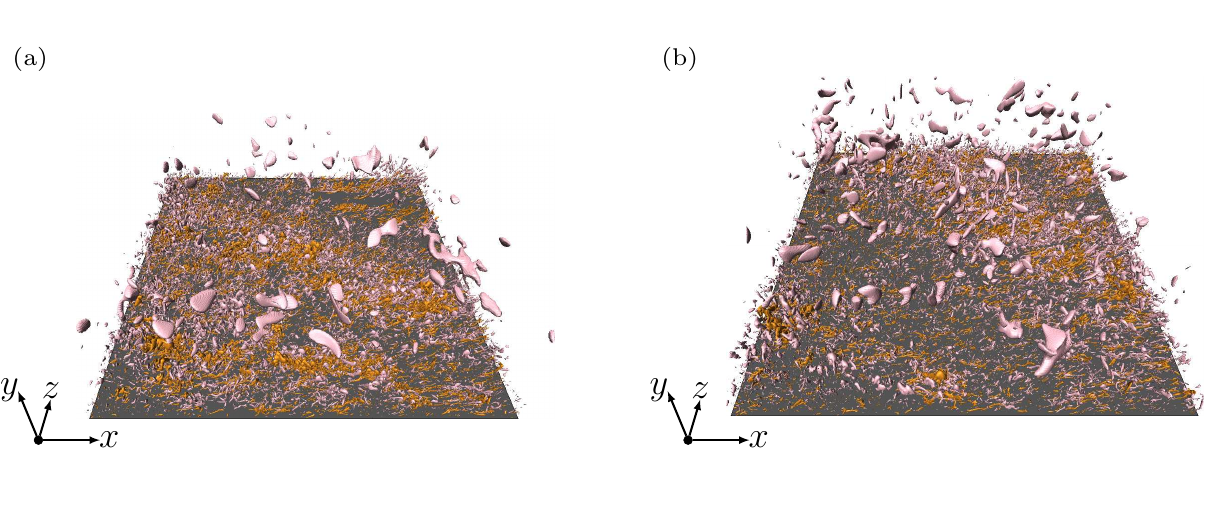}}
	\caption*{Reconstruction of vortices showing structures in the visualization space C3.7(e) for (a) case S\_2 and (b) case S\_3.}
	\endminipage\hfill
	\minipage{\textwidth}
	\centerline{\includegraphics[width=0.82\linewidth]{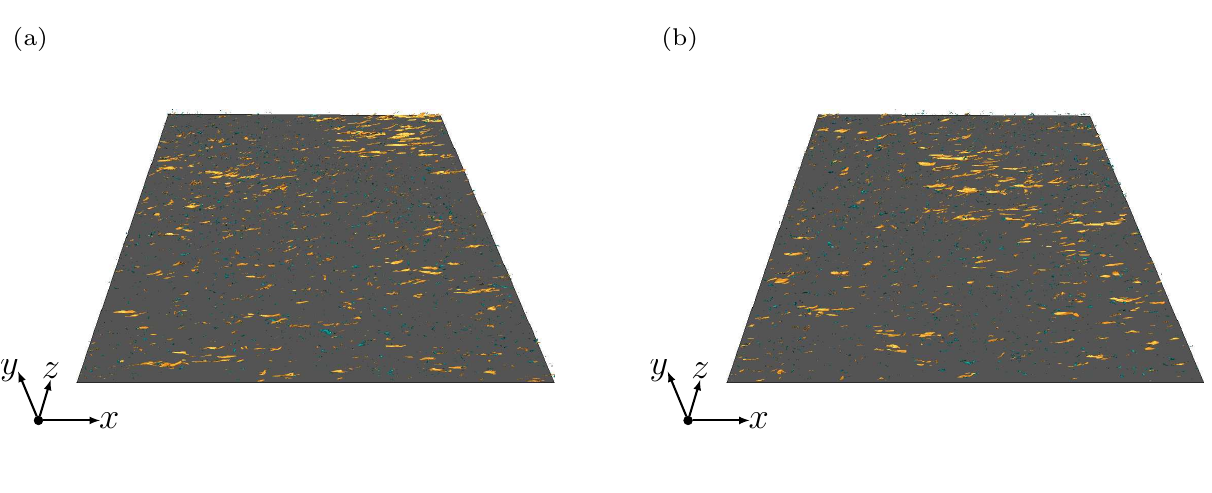}}
	\caption*{Reconstruction of shear layers showing structures in the visualization space C3.7(f) for (a) case S\_2 and (b) case S\_3.}
	\endminipage\hfill
	\caption{Structures which start and end within the inner layer itself are highlighted according to table 2 of the main paper. Other structures which start within the viscous sublayer and end in the buffer layer are shown in orange except for high speed streaks and sweeps are shown in pale yellow.}
\end{figure}

\clearpage

\noindent \Large C4. Conditional one-point statistics

\normalsize

\begin{flushleft}
	\noindent C4.1 \textit{Vertical profiles for case S\_2, S\_3 and N.}
\end{flushleft}

\begin{figure}[h!]
	\minipage{\textwidth}
	\centerline{\includegraphics[width=\linewidth]{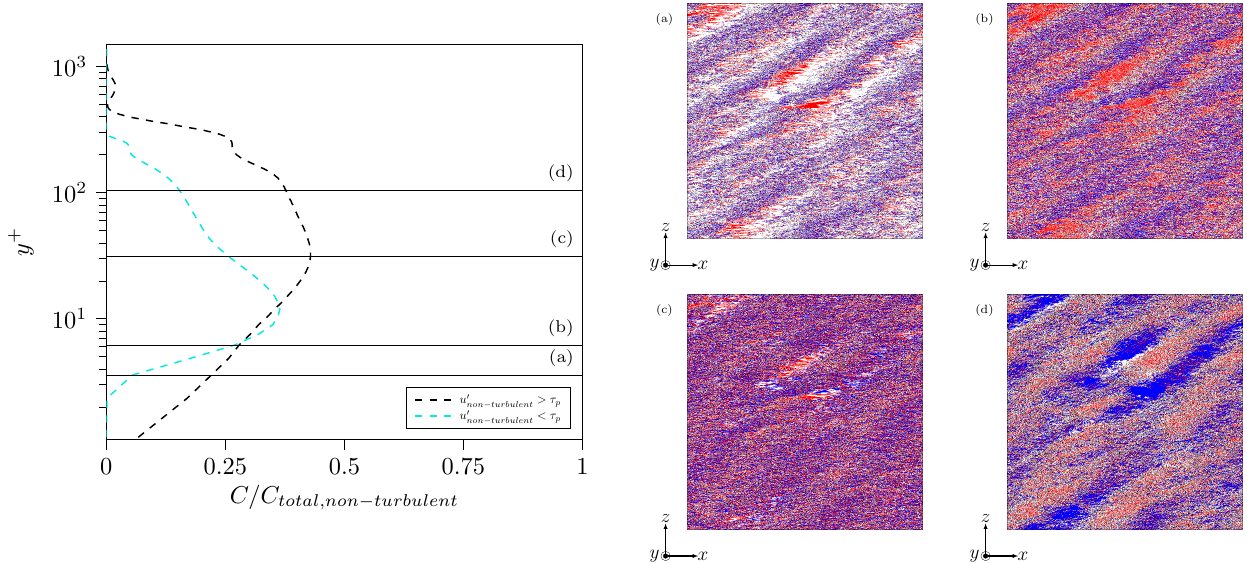}}
	\caption*{(a) $y^+ \approx 3.58$, (b) $y^+ \approx 6.18$, (c) $y^+ \approx 31.26$ (d) $y^+ \approx 104$.}
	\endminipage\hfill
	\minipage{\textwidth}
	\centerline{\includegraphics[width=\linewidth]{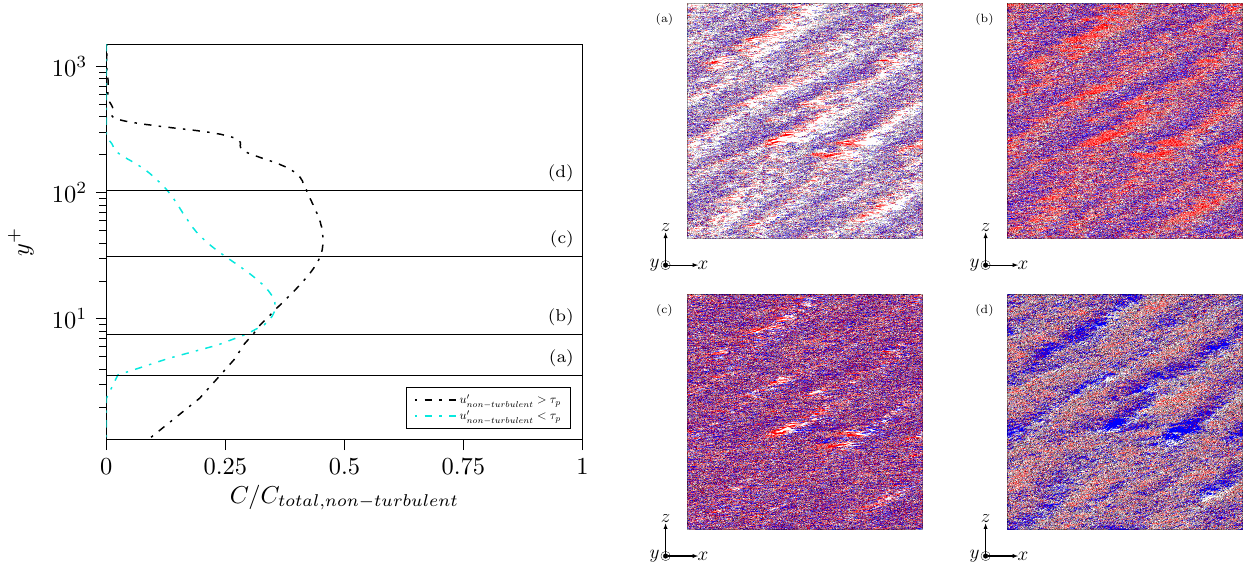}}
	\caption*{(a) $y^+ \approx 3.58$, (b) $y^+ \approx 7.55$, (c) $y^+ \approx 31.26$ (d) $y^+ \approx 104$.}
	\endminipage\hfill
\end{figure}

\clearpage

\begin{figure}[h!]
	\minipage{\textwidth}
	\centerline{\includegraphics[width=\linewidth]{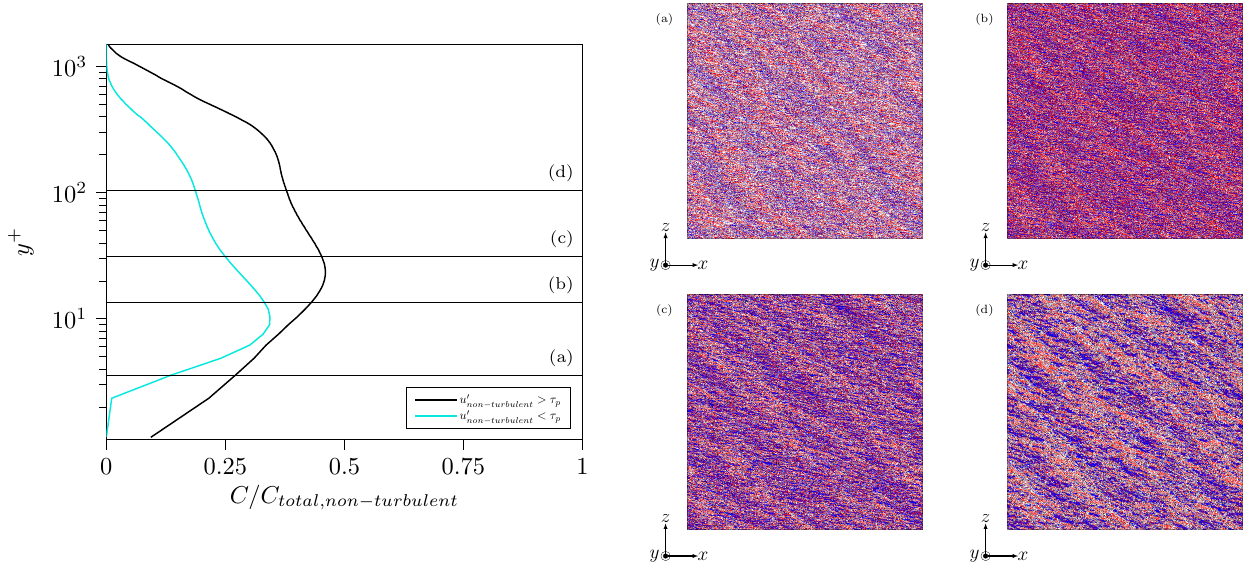}}
	\caption*{(a) $y^+ \approx 3.58$, (b) $y^+ \approx 13.45$, (c) $y^+ \approx 31.26$ (d) $y^+ \approx 104$.}
	\endminipage\hfill
	\caption{For case S\_2 (panel 1), S\_3 (panel 2) and N (panel 3), vertical profiles indicating the number of points $C$ satisfying the condition $\langle u'_{nonturbulent} ´> 0 \rangle$ and $\langle u'_{nonturbulent} < 0 \rangle$ over the total nonturbulent points $C_{total, nonturbulent}$ are shown on the left plot. On the right, horizontal slices of $\langle u'_{nonturbulent} > 0 \rangle$ (red), $\langle u'_{nonturbulent} < 0 \rangle$ (blue) and $Q-$criterion (black) are shown. Line specification is according to table 1 of the main paper.}
\end{figure}